\documentclass[10pt,pre,twocolumn,tightenlines,longbibliography,notitlepage,superscriptaddress,nofootinbib]{revtex4-2}
\pdfoutput=1

\usepackage{xcolor}
\usepackage{verbatim}
\usepackage{graphicx}
\usepackage{natbib}
\usepackage{siunitx}
\usepackage{float}

\newcommand*{\rom}[1]{\uppercase\expandafter{\romannumeral #1\relax}}

\begin{document}

\title{Geometrically programmed self-limited assembly of tubules using DNA origami colloids}% as subunits}

% Use letters for affiliations, numbers to show equal authorship (if applicable) and to indicate the corresponding author
\author{Daichi Hayakawa}
\affiliation{Martin A. Fisher School of Physics, Brandeis University, Waltham, Massachusetts 02453, USA}
\author{Thomas E. Videb\ae k}
\affiliation{Martin A. Fisher School of Physics, Brandeis University, Waltham, Massachusetts 02453, USA}
\author{Douglas M. Hall}
\affiliation{Department of Polymer Science and Engineering, University of Massachusetts, Amherst, Massachusetts 01003, USA}
\author{Huang Fang}
%\affiliation{Martin A. Fisher School of Physics, Brandeis University, Waltham, Massachusetts 02453, USA}
\author{Christian Sigl}
\affiliation{Department of Physics, Technical University of Munich, Munich, Germany}
\author{Elija Feigl}
\affiliation{Department of Physics, Technical University of Munich, Munich, Germany}
\author{Hendrik Dietz}
\affiliation{Department of Physics, Technical University of Munich, Munich, Germany}
\author{Seth Fraden}
\affiliation{Martin A. Fisher School of Physics, Brandeis University, Waltham, Massachusetts 02453, USA}
\author{Michael F. Hagan}
\affiliation{Martin A. Fisher School of Physics, Brandeis University, Waltham, Massachusetts 02453, USA}
\author{Gregory M. Grason}
\affiliation{Department of Polymer Science and Engineering, University of Massachusetts, Amherst, Massachusetts 01003, USA}
\author{W. Benjamin Rogers}
\email{wrogers@brandeis.edu}
\affiliation{Martin A. Fisher School of Physics, Brandeis University, Waltham, Massachusetts 02453, USA}

\begin{abstract}
 Self-assembly is one of the most promising strategies for making functional materials at the nanoscale, yet new design principles for making self-limiting architectures, rather than spatially unlimited periodic lattice structures, are needed. To address this challenge, we explore the trade-offs between addressable assembly and self-closing assembly of a specific class of self-limiting structures: cylindrical tubules. We make triangular subunits using DNA origami that have specific, valence-limited interactions and designed binding angles, and study their assembly into tubules that have a self-limited width that is much larger than the size of an individual subunit. In the simplest case, the tubules are assembled from a single component by geometrically programming the dihedral angles between neighboring subunits. We show that the tubules can reach many micrometers in length and that their average width can be prescribed through the dihedral angles. We find that there is a distribution in the width and the chirality of the tubules, which we rationalize by developing a model that considers the finite bending rigidity of the assembled structure as well as the mechanism of self-closure. Finally, we demonstrate that the distributions of tubules can be further sculpted by increasing the number of subunit species, thereby increasing the assembly complexity, and demonstrate that using two subunit species successfully reduces the number of available end states by half. These results help to shed light on the roles of assembly complexity and geometry in self-limited assembly and could be extended to other self-limiting architectures, such as shells, toroids, or triply-periodic frameworks.
\end{abstract}

\maketitle

Self-assembly is a fundamental building principle used by Nature to make functional materials, including virus capsids for encapsulation and delivery \cite{caspar_physical_1962}, cytoskeletal filaments for transport \cite{sui_structural_2010}, and macromolecular machines with diverse roles, like the ribosome for protein synthesis \cite{wimberly_structure_2000}.  Recently, there has been considerable effort aimed at mimicking biological self-assembly to synthesize user-prescribed structures from synthetic nanometer- and micrometer-scale particles. For example, by encoding short-range specific interactions, DNA-grafted colloidal particles can be programmed to assemble into a variety of two- and three-dimensional crystal phases with prescribed symmetry groups and lattice constants \cite{macfarlane_nanoparticle_2011, rogers2016using, fang2020two,he_colloidal_2020}. However, rather than unbounded crystal phases, the aforementioned biological functionalities---encapsulation, motility, and protein synthesis---arise from self-limiting structures that have one or more self-limited length scales. 

How does one go beyond periodic lattice structures with macroscopically uncontrolled dimensions, to program the assembly of self-limiting architectures that have self-limited length scales that are arbitrarily large with respect to the size of the individual subunits \cite{Hagan2021Jun}? There are two prominent paradigms for prescribing self-limited assembly: 1) addressable assembly and 2) self-closing assembly. In addressable assembly, every component of a multi-species ensemble is distinct and is programmed to occupy a specific location within a target structure \cite{zeravcic2014size,jacobs2015rational,ke_dna_2014}. Therefore, increasing the self-limited length scale necessitates increasing the assembly complexity in terms of the number and interaction specificity of the subunits. In self-closing assembly, anisotropic interactions give rise to the accumulation of curvature during growth that causes the structure to self-close at a finite size \cite{Hagan2021Jun}. In contrast to addressable assembly, the self-limited length scale in self-closing assembly is therefore controlled by the binding angles between neighboring subunits and is programmed geometrically rather than through specific interactions, and as such, may require only one or relatively few subunits to target a specific self-closing size. But which strategy should one select for a particular target geometry, and which approach is more accurate, precise, or economical? Because the vast majority of examples either address only one paradigm or do not distinguish between the two, as in DNA-coated colloids \cite{halverson2013dna}, DNA tiles \cite{rothemund_design_2004, yin_programming_2008, wei_complex_2012, shen_novo_2018,mohammed2013directing}, hierarchical assembly of DNA origami \cite{tikhomirov_programmable_2017,wagenbauer_gigadalton-scale_2017,sigl_programmable_2021} and designed proteins \cite{bale_accurate_2016,shen_novo_2018}, direct comparisons between addressability and geometry in prescribing self-limited assembly are obscured.

In this article, we create an experimental platform using DNA origami for examining the roles of geometric and interaction specificity in arguably the simplest system that could in principle be programmed by geometry alone: the assembly of cylindrical tubules from rigid triangular monomers. The tubule is the ideal target geometry because it represents an infinitely large class of structures, each of which can be assembled from a {\it single subunit} species by controlling the dihedral angles between neighboring subunits. As anticipated, we show that we can vary the self-limited width of the tubules by tuning the dihedral angles between neighboring subunits without changing the interaction complexity. However, we find that the width of the assembled tubules takes a range of values and that the distribution broadens upon increasing the mean tubule width. We discuss how this distribution of the self-limited width is a generic feature of self-closing assembly when the self-limited length scale is sufficiently large compared to the size of the building block \cite{videbaek2021tiling}, which we understand using a simple theory. We conclude by exploring how increasing the assembly complexity enables us to constrain the width distribution by using multiple species of triangles with increased numbers of specific interactions. For a binary mixture of triangle species, we validate that the allowed states of the assembled tubules are reduced by half, as expected from simple geometrical arguments.  

\section*{Results and Discussion}
\subsection{Design principles}
\begin{figure}[t]
 \centering
 \includegraphics[width=.99\linewidth]{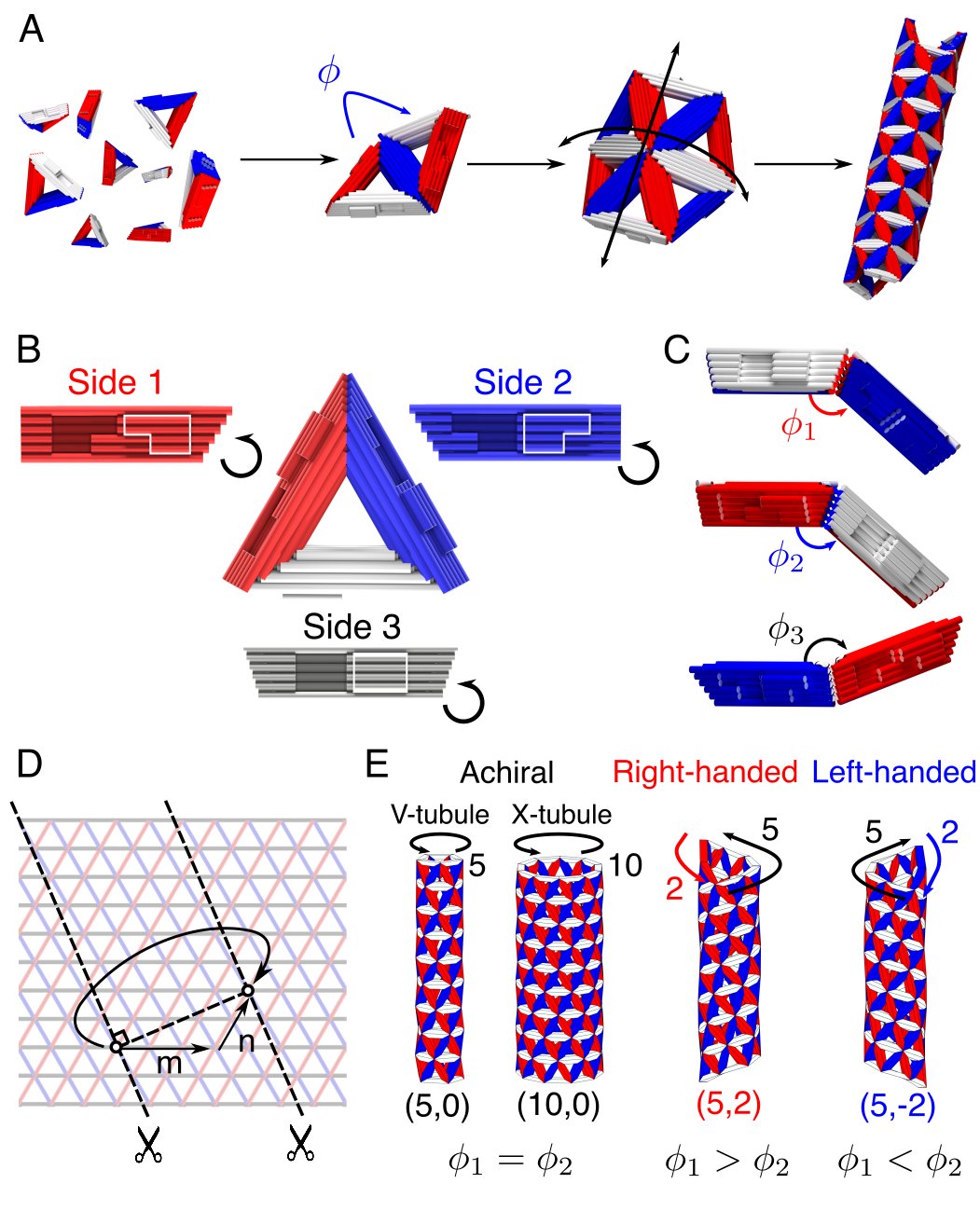}
 \caption{\textbf{Overview of the design rules for tubule assembly.}
(A) Triangular subunits bind at the edges with dihedral angles $\phi$ that specify the principal curvatures of the assembly to favor the formation of cylindrical tubules. (B--C) Design principles of triangular subunits assembling into tubules. (B) 
Each side has self-complementary lock-and-key interactions encoded in shape-complementary protrusions and recesses. Holes on the faces are shaded, while protrusions are highlighted with white. Black arrows indicate the sides that bind together. In this case, each side binds to itself. (C) The sides are also beveled so that when two triangles bind, they form a unique dihedral angle $\phi$ that is determined by the bevel angle. (D) A triangular lattice that is cut along two parallel lines can be rolled into a tubule, with periodicity determined by indices $m$ and $n$. $m$ denotes the number of steps along side 3, while $n$ is the number of steps along side 1 or 2 (if $n>0$, it is side 1, if $n<0$, it is side 2). (E) Any tubule is identified by an index pair, $(m,n)$. V-tubule and X-tubule, corresponding to (5,0) and (10,0), are achiral tubules with 5 and 10 triangles in circumference, respectively. Tubules of varying width and chirality can be assembled by tuning the magnitudes of the dihedral angles $\phi_1$, $\phi_2$, and $\phi_3$.
}
 \label{fig:1}
\end{figure}

Our system consists of rigid triangular subunits made by DNA origami~\cite{sigl_programmable_2021} that encode all of the information necessary to self-assemble tubules with user-prescribed geometries (Fig.~\ref{fig:1}A). Assembling tubules from triangular subunits requires that we specify two types of information: 1) the interaction specificity between the edges and 2) the local curvature between neighboring subunits. We encode the specific interactions by using unique protrusions and recesses that give rise to shape-complementary lock-and-key interactions whose attraction originates from blunt-end stacking (Fig.~\ref{fig:1}B) \cite{gerling_dynamic_nodate}. At a minimum, we require three specific interactions---one per side---which are each homophilic. These three specific interactions allow the triangular subunits to assemble into a deterministic triangular lattice, where each side aligns with one of the three lattice directions.

%figure 1

Additionally, we encode the local curvature by specifying the dihedral angles between neighboring subunits. Specifically,  we design the bevel angles of each side of the triangular subunit such that it forms the three dihedral angles with its three neighbors corresponding to a target tubule geometry (Fig.~\ref{fig:1}C). For example, the subunit in Fig.~\ref{fig:1}B--C, which assembles into a tubule with five subunit edges in circumference, has dihedral angles of 138.2, 138.2, and -158.4 degrees for sides 1, 2, and 3, respectively. The accumulation of curvature from the dihedral angles upon assembly results in the formation of a curved triangular lattice. For appropriately chosen angles, the lattice will close upon itself to form a tubule (Fig.~\ref{fig:1}D), which can be chiral or achiral depending on how the tubule closes (Fig.~\ref{fig:1}E). See Supplementary Information II for more details of the tubule geometry.

Because the curvature originates from the combination of three bevel angles, the width, handedness, and pitch of the resulting tubules can be programmed using these three angles alone. Indeed, the final tubule structure is equivalent to the equilateral Yoshimura pattern of a buckled cylindrical shell \cite{yoshimura1955mechanism}, which constitutes a class of origami tilings of cylinders via a single triangular facet with a single set of fold angles on its three edges. In the case of tubules, larger dihedral angles in the direction perpendicular to the tubule axis produce wider tubules. Furthermore, setting the dihedral angles of sides 1 and 2 to different values produces chiral tubules with different pitch and handedness (Fig.~\ref{fig:1}E). The relative magnitude of the two dihedral angles determines the handedness and the difference between the two angles determines the pitch, with a larger difference leading to a larger pitch.  Thus, by tuning the dihedral angles of the three sides, we can assemble a variety of different tubule structures. We note that the ability to program any tubule geometry using a single triangular facet is in contrast to the ability to program assembly of an icosahedral shell, for example, which requires a larger and larger number of unique facets upon increasing the shell diameter \cite{caspar_physical_1962,sigl_programmable_2021}.

The resulting tubules can be uniquely classified using a pair of lattice indices, $(m,n)$~\cite{Lee2009Jan}. Here, $m$ refers to vector along side 3 of the triangle and $n$ refers to the vector along side 1 ($n>0$) or side 2 ($n<0$). Therefore, $(m,n)$ defines the shortest distance along the triangular lattice to go around the tubule and come back to the same vertex. Note that we only consider the case $m\geq{|n|}$. With our convention, we use the positive index $n$ to define right-handed tubules and the negative index $n$ for left-handed tubules. Each tubule type is associated with a different set of dihedral angles for the three sides.

\subsection{Tubule assembly}
\begin{figure*}[!t]
 \centering
 \includegraphics[width=.99\linewidth]{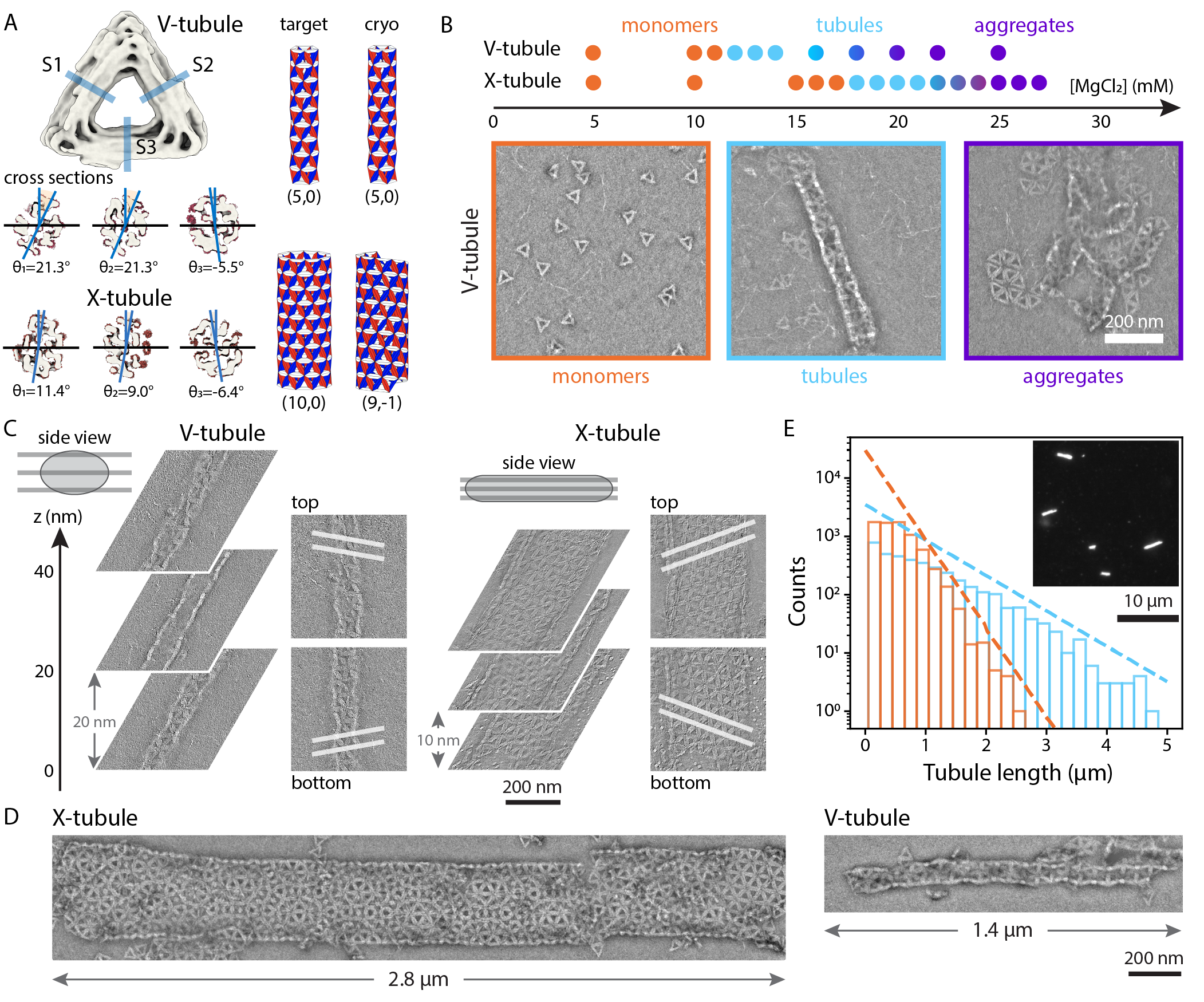}
 \caption{\textbf{Self-assembly of tubules.}
(A) Cryo-EM single-particle reconstructions for V-triangle and X-triangle. The figure shows cross-sections of three sides of the triangles (gray) and the estimated bevel angles, $\theta$, from fitting the cryo-EM reconstruction to a pseudo-atomic model (maroon) (see Supplementary Information V). Tubule illustrations show the target geometry (labeled target) and the expected tubule geometry based on our cryo-EM measurements (labeled cryo) for the V-tubule and X-tubule.
(B) Negative-stain transmission electron micrographs of V-triangles after incubation at different $\text{MgCl}_{\text{2}}$ concentrations for a week at 40 $^{\circ}$C. The points above the images show the $\text{MgCl}_{\text{2}}$ concentrations that lead to assembly of monomers (orange), tubules (cyan), or aggregates (purple). 
(C) Tomography reconstructions of the V- and X-tubules, showing different $z$ positions. The orientation of side 3 is illustrated with white lines. (D) Images of tubules observed with V-triangles (right) and X-triangles (left). For the X-tubule one can see a Moir\'e pattern where the tubule has closed on itself. (E) Histograms of the tubule length observed with epi-fluorescence for V-tubules (orange) and X-tubules (cyan). Dashed lines are guides for the eye to highlight the exponential decay in length (see Supplementary Information VII A). The inset shows an example fluorescence image.
}
 \label{fig:2}
\end{figure*}

To demonstrate the utility of our experimental approach, we design two different monomers that assemble into tubules with different widths: V-triangle, which targets a (5,0) tubule, and X-triangle\footnote[1]{Roman numerals V and X are used as shorthand for the target tubule types (5,0) and (10,0) respectively.}, 
which targets a (10,0) tubule (Fig.~\ref{fig:2}A). Each side of the triangle has a cross-section that is 4-by-6 double helices arranged on a square lattice (Fig.~S3A). 
We specify the three unique bevel angles required for a particular tubule geometry by varying the relative lengths of the double helices (Fig.~S3B). We encode the specific shapes of the protrusions and recesses by designing the scaffold routing, which we choose to disallow off-target binding and to enforce the correct relative orientations of neighboring subunits. The size of the subunits is set by the length of the DNA scaffold, which is 8064-nucleotides long. See Supplementary Information III for details of the subunit design.

We fold, purify, and assemble the subunits using standard DNA origami protocols. In brief, we fold the origami using a slow, linear temperature ramp, purify the resulting monomers by gel extraction, and then assemble them at constant temperature for one week in a rotating incubator~\cite{wagenbauer_how_2017} (see Supplementary Information V). For a given design, we perform multiple assembly experiments at different concentrations of MgCl$_2$ to tune the strength of the intersubunit attraction. Finally, we characterize the structures of the individual monomers using single-particle cryogenic electron microscopy (cryo-EM)~\cite{zivanov2018RELION3} and the entire assemblies using negative-stain transmission electron microscopy (TEM).

The bevel angles of the folded monomers are close to the target values but do not match them exactly. By fitting a pseudo-atomic model to our cryo-EM reconstructions~\cite{kube2020revealing} we make an estimate of the bevel angles of each of the three sides for both of our monomers (Fig.~\ref{fig:2}A). We find that for the V-triangle the three sides have angles of ($21.3\pm0.1,\ 21.3\pm0.1,\ -5.5\pm0.1$) compared with the target angles of ($20.9,\ 20.9,\ -10.8$). Comparing these angles to the angles of different tubule types, this monomer is closest to a ($5,0$) tubule, as designed. In contrast, for the X-triangle, we see monomer bevel angles of ($11.4\pm0.1,\ 9.0\pm0.1,\ -6.4\pm0.1$) compared with ($10.4,\ 10.4,\ -5.3$), yielding a closest state of a ($9,-1$) tubule. While the ($9,-1$) tubule has a similar diameter to the ($10,0$) one, it is chiral and left-handed rather than achiral, as in the designed monomer. See Supplementary Information VIII for more detail on the cryo-EM reconstructions.

The assembly of the triangular subunits into self-limiting structures depends on the intersubunit attraction, which can be controlled by varying the concentration of MgCl$_2$.
For both designs, we observe the same sequence of outcomes upon increasing the Mg$^{2+}$ concentration (Fig.~\ref{fig:2}B). At low Mg$^{2+}$ concentration, we observe only monomers and oligomers containing a few subunits. At intermediate Mg$^{2+}$ concentration, we see the assembly of filament-like, self-limited structures in coexistence with monomers. The fact that we observe the coexistence of assemblies and monomers suggests that assembly occurred near to equilibrium and that the assembled filaments likely grew by consuming monomers rather than the merging of clusters. At higher Mg$^{2+}$ concentrations, we observe large disordered aggregates and rarely observe monomers or small clusters. Therefore, we hypothesize that these large aggregates are likely due to kinetic arrest \cite{Whitelam2015Apr}. These results are consistent with previous measurements of the stacking interactions between blunt ends, which show that the interactions become stronger upon increasing Mg$^{2+}$ concentration~\cite{kilchherr_single-molecule_2016}.

TEM tomography confirms that the filament-like structures are indeed tubules. First, we observe a single filament assembled from V-triangles (Fig.~\ref{fig:2}C). Looking at different slices through the filament in the direction normal to the EM grid, or $z$-stack, we see that the structure clearly shows two triangular lattices at different $z$ positions, separated by a hollow core. The distance between the top and bottom lattices is around 40 nm, which is shorter than the expected diameter of 88 nm, but longer than 20 nm, or twice the thickness of a triangle. Therefore, we hypothesize that the cross-section of the tubule deposited on grid is elliptical. Finally, we also see that the top and bottom lattices have different orientations of the triangular lattice with respect to one another, as would be expected for a chiral tubule. Taken together, these observations confirm that the filament is indeed a tubule. Going further, the lattice orientation can be tracked around the tubule, yielding a (4,1) tubule for this case. While this determination only gives us the magnitude of \textit{n}, we also determine the handedness by comparing our reconstructions to reconstructions of a DNA origami nanohelix labeled with gold nanoparticles in a right-handed, chiral arrangement
~(see Supplementary Information VII B)~\cite{kuzyk_dna-based_2012,briegel_challenge_2013}.

Similarly, we find that the wider filaments formed from X-triangles are also tubules. As with the V-triangles, the tomograms of the wide filaments show two triangular lattices, which are mirror images of one another, and spaced apart along the $z$ direction. In this case, the distance between the bottom and the top lattices is around 20 nm, which matches the height of two triangles. We suspect that this spacing is due to the flattening of the X-tubule during the sample preparation. This hypothesis is further supported by the observation that the two triangular lattices appear to be planar, as expected for a flattened tubule.  Finally, similar to the example V-tubule, the specific X-tubule shown in Fig.~\ref{fig:2}C is also chiral and right-handed, as seen in the mirror reflection of the top and bottom lattices (see Supplementary Information VII B). 

Remarkably, we find that both V-tubules and X-tubules can grow to micrometers in length and exhibit length distributions that are characteristic of equilibrium one-dimensional growth. Figure~\ref{fig:2}D shows examples of some of the longest tubules that we observe in electron microscopy for both designs. The V-tubule is roughly 1200 nanometers in length and is made from about 210 subunits. The X-tubule is nearly 3 micrometers long and is assembled from about 1400 subunits. To complement our EM observations, we also perform epi-fluorescence experiments to get a more complete view of the tubule lengths (Fig.~\ref{fig:2}E, see Supplementary Information VII A for details). We find that both V- and X-tubules are characterized by length distributions that decay exponentially, with means of 0.5 micrometers and 0.9 micrometers, respectively, which is consistent with expectations for equilibrium one-dimensional assembly \cite{phillips2012physical}. In the extremes, we find assemblies reaching up to 5 micrometers in length. Comparing our X-tubule results to earlier attempts to assemble tubules of similar width from DNA origami subunits~\cite{wagenbauer_gigadalton-scale_2017} shows that our approach yields assemblies that are roughly an order of magnitude longer and contain about five times the number of subunits. Furthermore, our results show that although the width of the tubule is self-limited, the tubule lengths are unconstrained.

\subsection{Tubule distributions}

\begin{figure}[t]
 \centering
 \includegraphics[width=\linewidth]{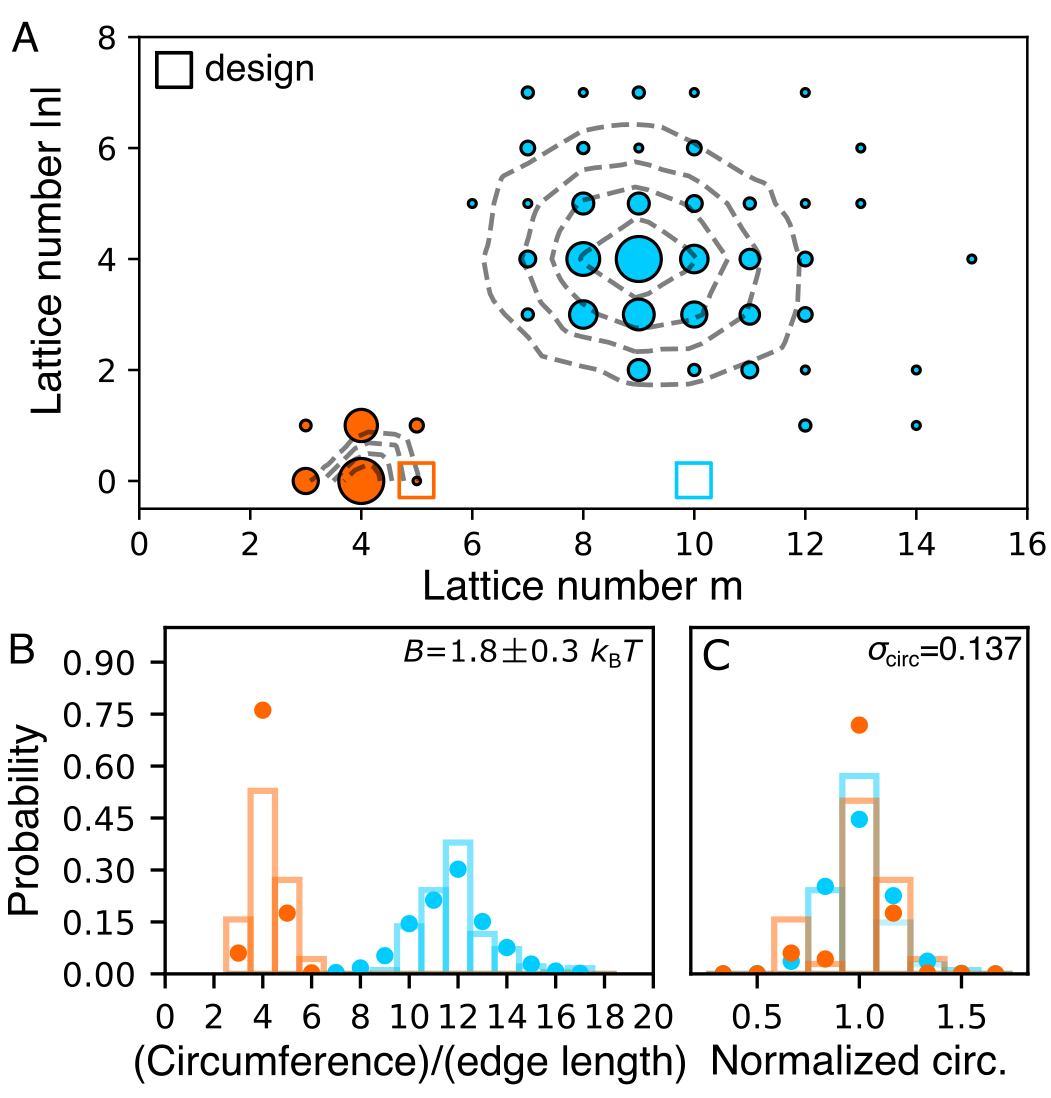}
 \caption{\textbf{Measured distributions of the lattice numbers of observed tubules}
(A) Lattice number distribution of the two tubules. The size of each circle is proportional to the number of tubules found for that lattice number. Orange represents data for the V-tubule, while cyan represents data for the X-tubule. The square symbols denote the target states. Contours show the expected spread based on model predictions, described in the text, where the spacing between contours is 20\% of the peak probability. The plot shows distribution from N=70 and N=182 tubules for V- and X-tubules, respectively. (B) Histograms of the measured circumferences of the tubules. Points come from a model prediction using a bending rigidity of $B=1.8\pm 0.3\ k_\mathrm{B}T$; bars show experimental data. (C) The corresponding distributions of the circumferences normalized by the mean, whose variance, $\sigma_\mathrm{circ}^2$, is used to estimate $B$. $\sigma_\mathrm{circ}=0.137$ is the average of the standard deviations obtained from the V- and X- tubule distributions.
}
 \label{fig:3}
\end{figure}

Whereas the results above show the structure of single tubules, our assembly experiments yield an ensemble of tubules that exhibits a variety of widths and lattice orientations, despite being formed from a single monomer type. For both the V- and X-tubule designs, we identify the pair of indices ($m,n$) that classifies the structure for each of hundreds of individual tubules, and create a distribution (Fig.~\ref{fig:3}A). The distribution shows that the most probable tubule types are (4,0) and (9,4) for the V- and X-tubules respectively, with the probability of different states falling off with distance from this tubule type. Furthermore, we find that the breadth of the X-tubule distribution is larger than that of the V-tubule. Although we cannot easily determine the chirality of every tubule, we examined 13 chiral V-tubules and 15 chiral X-tubules using tomography. In both cases, all of the tubules that we examined were right-handed. See the Supplementary Information VII B and VII C for a detailed description of the tubule classification.

The observed distributions beg two important questions: 1) Why are the tubule distributions not centered around the target structures? and 2) What determines the breadth of the tubule distribution around the most probable tubule type? We address the first question by returning to our single-particle reconstructions of the monomers from cryo-EM (Fig.~\ref{fig:2}A).  While the V-triangle estimates a state close to the peak of the experimental distribution, the X-triangle estimate is 13\% narrower in width than the (9,4) peak that we observe. We note that the cryo-EM map of the X-triangle shows that some parts of the lock-and-key design protrude from the structure due to a missed crossover in the origami design (Fig.~S16). Therefore, we hypothesize that this aspect of the structure causes a poor fit for the lock-and-key shapes of the interaction, leading to unintended torques that could skew the dihedral angles to a different value. This misfit would cause a shift in the mean of the distribution away from what is expected from the geometry of the monomer alone.

Next, we tackle the origin of the breadth of the distribution. We start by noting that tubules with neighboring lattice numbers only vary by small changes in their dihedral angles. Therefore, if the bending rigidity of the assembly is sufficiently small, thermal fluctuations will cause the dihedral angles between neighboring triangles to explore a range of possible values, leading to tubules with larger or smaller diameters. 

This idea can be captured by considering the Helfrich energy of an elastic sheet, $E_\mathrm{H}=\frac{1}{2}BA(\Delta\kappa_{\perp})^2$ where $A$ is the surface area, $B$ is the bending rigidity, and $\Delta\kappa_{\perp}$ is the fluctuation of the sheet's curvature in the circumferential direction \cite{helfrich_intrinsic_1988}. We assume that as the assemblies grow, they must first form a patch-like circular sheet before closing into a tubule. Thermal fluctuations will cause this sheet to accommodate different curvatures from its preferred value, but once it grows large enough to close into a tubule we assume that it can not open into a sheet again, thereby locking in a specific tubule circumference. If the growth rate is faster than the dissociation of a subunit-subunit bond, then once a tubule forms and starts to grow, the possibility of opening a large number of bonds to allow the tubule to reform into a different type becomes increasingly unlikely. Therefore, we estimate the size of assemblies at this closure point as a disk with a diameter that corresponds to the circumference of the closed tubule. The fluctuations of the sheet's curvature at the point of closure will inherently lead to a distribution of tubule circumferences. Under these assumptions, we can write the Helfrich energy for a tubule at closure as
\begin{equation}
    E_\mathrm{H} = \frac{1}{2}B\pi^3 \left(\frac{\Delta C}{C}\right)^2
\end{equation}
where $C$ is the circumference of a tubule. Assuming that the circumferences follow a Boltzmann distribution, $P \sim \exp(-E_\mathrm{H}/k_\mathrm{B}T)$, we can relate the spread of the widths to the bending rigidity (\cite{videbaek2021tiling}, see Supplementary Information IX for details).

The predictions from the Helfrich energy are consistent with our experimental observations, suggesting that the width distribution is determined by the bending rigidity of the growing assembly and the irreversibility of closure. Figure~\ref{fig:3}B shows the distribution of the tubule circumference for both V- and X-tubules. Following the insight we gained from considering the Helfrich energy, we rescale the circumference by the mean circumference of each distribution. Figure~\ref{fig:3}C shows that for both V- and X-tubules the scaled distributions have similar breadths, with standard deviations of 0.149 and 0.125, respectively. The importance of this observation is that it is an inherent feature of self-closing assemblies at finite temperature: Whenever the assembly has a finite bending rigidity, the system will form a distribution of end states with a breadth that depends on the self-limited length scale relative to the size of the subunit.

Using the variance of the width distributions, we make an estimate of the bending rigidity of the V- and X-tubules. Assuming both assemblies have the same bending rigidity, we find $B = 1.8 \pm 0.3$ $k_\mathrm{B}T$. To our knowledge, this is the first estimate of bending rigidity of shape-complementary interactions in DNA origami. Moreover, we can use this bending rigidity as an input to perform more detailed energetics calculations (see Supplementary Information IX) to get a complete ($m,n$) distribution for the expected tubules~\cite{videbaek2021tiling}, which is shown in the contours of Fig.~\ref{fig:3}A. Similarly, we can compute the circumference distributions for the model. In both cases, we find that our model predictions match the experimental data well (Fig.~\ref{fig:3}B), further supporting the idea that the distribution of tubules we observe in the experiment is due to finite bending rigidity and is therefore expected for self-closing structures.

\subsection{Pruning the tubule distribution by increasing the assembly complexity}

Our observations of the X-tubule assembly highlight an inherent challenge in using self-closing assembly alone to target finite-sized soft materials: As we target larger self-closing lengths, the distribution of the self-limited dimension also gets broader. So how do we overcome this challenge to assemble wider tubules without compromising the accuracy of the assembly? 

Here we turn our attention toward the other paradigm for encoding self-limitation: addressable assembly. The rationale of this approach is relatively easy to understand. By increasing the number of particle species per structure, and therefore the total number of specific interactions, the location of any given particle within that structure becomes more precisely defined. In the fully-addressable limit, each particle species can only occupy a single position within the final structure while simultaneously maximizing the number of its favorable interactions. Therefore, by eliminating the other ways in which the particles can be arranged, the yield of the target assembly can be increased.

\begin{figure}[t]
 \centering
 \includegraphics[width=\linewidth]{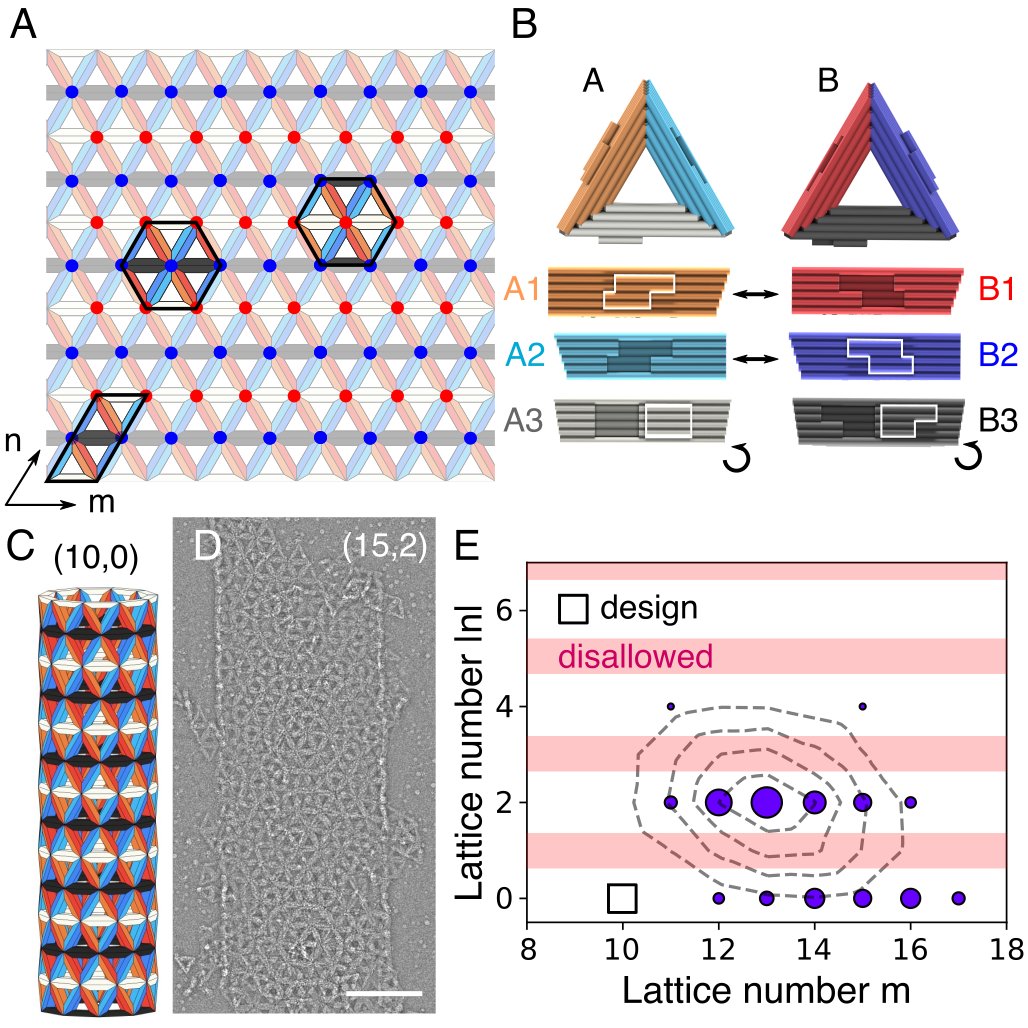}
 \caption{\textbf{Schematics for binary species tubules and their lattice number distributions}
(A) Triangular tiling using two types of monomers. The highlighted region in the lower left shows the primitive cell of the tiling. Red and blue dots denote the two distinct vertices present in the pattern, which are highlighted as well. (B) Schematic representation of the two types of monomers we use to construct the tubule. Black arrows represent the binding pairs of the lock-and-key interactions. (C) Diagram of the intended (10,0) tubule structure. (D) Negative-stain TEM micrograph of a tubule made from the two species, whose tubule type is (15,2). The scale bar is 200 nm. (E) Tubule type distribution of the binary mixture. The size of each circle is proportional to the number of tubules found for that pair of lattice indices. The square symbol denotes the target state. Contours show the expected spread based on model predictions with $B=3.1\ k_\mathrm{B}T$, where the spacing between contours is 20\% of the peak probability. This $B$ value is estimated in the same manner as described previously. Red regions are disallowed due to the tiling pattern of the binary species, as seen in the experiments. The plot shows distribution from N=113 tubules.
}
 \label{fig:4}
\end{figure}

We explore this approach by assembling a (10,0) tubule from two unique species of triangular subunits. Figure~\ref{fig:4} shows our specific realization of this concept using a binary tiling that is compatible with tubule self-assembly (Fig.~\ref{fig:4}A, see Supplementary Information IV). The two species are placed periodically and unidirectionally in the pattern. Since each side encodes a specific dihedral angle, the sides must always remain in the same orientation. As before, a triangular lattice of these two species can be rolled up into a (10,0) tubule. However, unlike before, the binary lattice contains two distinct types of vertices. As a result, the assembled tubules have an additional constraint: When the triangular lattice closes to form a tubule, the closing vertices must match. Therefore, for our specific case, we expect that only tubules with an even lattice number $n$ can assemble. The specific implementation of the lock-and-key geometries and their interactions is shown in Fig.~\ref{fig:4}B and the expected assembly is shown in Fig.~\ref{fig:4}C. 

The distribution of assembled tubules demonstrates that increasing the assembly complexity limits the accessible states. Figure~\ref{fig:4}D shows an example tubule and Fig.~\ref{fig:4}E is the tubule distribution that we measure. We find that the distribution is maximal at (13,2), with some tubules being chiral and others achiral. All 7 tomography reconstructions of chiral tubules were identified to be right-handed (see Supplementary Information VII B). Most importantly, we do not find a single tubule classified by an odd value of $n$, as we intended by our design. We also find that the probability of the most common state is increased compared to what it would be with a single particle type. The binary assembly has a larger average circumference than the X-tubule, which would suggest that the probability for the most common state would be comparatively lower. However, we find that the most populated state for the binary assembly is 22.1\% compared to 17.6\% for the X-tubule. This increase in probability is a result of using more species of triangles for the assembly since some fraction of tubules that might have had an odd $n$ are distributed to nearby even $n$ states.

As before, although the distribution is not centered around the target state, it has a width that is consistent with our energetics model. We hypothesize that the shift in the center of the distribution is again due to the experimental design challenge of accurately encoding the dihedral angles in our DNA origami subunits. The bevel angles of the triangles are designed to target X-tubules, similar to the previous case. However, cryo-EM reconstructions of the individual subunits reveal bevel angles that would target a (13,-1) tubule (see Supplementary Information VIII).

\section*{Conclusion}

In summary, we devised a new class of colloidal particles using DNA origami and studied the roles of geometry and interaction complexity in their self-limited assembly into tubules. Our DNA origami design scheme allows for control over the valence, interaction specificity, and local curvature of the subunits, independently. We demonstrate that the new DNA origami colloids enable the assembly of complex structures, such as tubules, with self-limited dimensions that are much larger than the individual subunits. Using this design principle, we designed and assembled tubules of different widths, demonstrating that information encoded in the geometry of individual subunits can be used to program the geometry of the entire assembly. However, due to the intrinsic flexibility of the binding, tubules with a variety of widths and chiralities assembled from the same subunits, which was especially prominent for wider tubules. Such a distribution of assembled tubules for a single monomer type is an inherent feature of bending fluctuations in curvature-limited assemblies. 

Here, we experimentally demonstrated one path toward removing off-target structures to focus the distribution on the target of interest. The precision in reaching a specific state can be increased by combining {\it interaction specificity} with {\it geometric specificity} in a multi-species design. As one specific realization of this concept, we assembled tubules using two species of subunits, which effectively reduced the number of accessible states by half. Therefore, as more species of subunits are added to the system, we anticipate that the yield of the target state will increase, though likely at the cost of longer timescales for assembly~\cite{videbaek2021tiling}. Another possibility other than {\it interaction specificity} that follows from our observations is that one can exploit the {\it geometric specificity} of the system to mitigate the formation of off-target structures. For example, designing ever more rigid binding sites should increase the energetic cost to deform the structure away from the target geometry. While it is not clear how much control there is over this aspect given the material properties of the subunits themselves, one might expect the bending rigidity to vary with the particle aspect ratio (i.e. thickness to width) based on elastic considerations alone. Finally, Nature, confronted with the same design challenge of assembling tubules from few components, has evolved a third strategy for eliminating off-target states: employing seeded nucleation, as seen with the \textit{in vivo} assembly of microtubules~\cite{chretien_new_1991,sui_structural_2010,roostalu_microtubule_2017}. Thus, if one can tune the supersaturation level to avoid spurious nucleation while still allowing for growth by monomer addition, having templates off of which tubules grow would improve the specificity of a given target structure without impeding the kinetics~\cite{mohammed2013directing}.

A somewhat surprising result of the origami design is how different the bevel angles of the three different versions of the X-triangle are (see Supplementary Information~Table IV). Even though all three of these structures were constructed using the same design principle, resulting in the same length of scaffold strands for each helix, this approach did not result in the same bevel angles for the different sides. One main difference is that the crossover connections between helices had to be placed in different locations  to accommodate the different lock-and-key designs. This subtle change may cause unintended torsion within the sides that impact the relative angles of adjacent sides. One avenue to address this issue would be to use interior supporting struts to add additional length constraints to the system. Another possible issue is that the short (one to three base pairs) single-stranded DNA segments connecting helices at the vertices might over constrain the vertices and add stress to the subunit. A  design modification that might avoid this issue is to only connect the sides at the vertices in a few locations. Lastly, it seems prudent to use existing simulation software, such as oxDNA~\cite{ouldridge2011structural}, ENRG MD~\cite{Maffeo2016Apr}, and mrdna~\cite{maffeo2020mrdna} to screen different arrangements of crossovers to find ones that yield the desired bevel angles.

Overall, our DNA origami colloids represent a powerful platform for programming the assembly of self-limiting architectures.  We argue that our ability to program the local curvature with the precision of a few degrees per subunit opens up new directions in materials design that surpass what is currently possible. Whereas tubules assembled from DNA tiles are made from `floppy' components and therefore prescribed mainly by the {\it interaction specificity} encoded in the tile sequences \cite{rothemund_design_2004, yin_programming_2008, wei_complex_2012, shen_novo_2018}, our tubule structures can be programmed by both {\it interaction specificity} and {\it geometrical specificity}. As we showed, these two paradigms can play important complementary roles in self-limited assembly. Geometric specificity can enable economical designs that require only a few subunit species and interaction specificity can improve the accuracy of assembly by eliminating off-target structures. Therefore, going forward, we anticipate that being able to prescribe both mechanisms of self-limited assembly will allow access to a new library of self-assembled structures with interesting material applications. Examples include 2D lattice-like membranes for patterning or separations \cite{millan_self-assembly_2014,tikhomirov_programmable_2017}, spherical shells for encapsulation and delivery \cite{sigl_programmable_2021}, fibers or length-limited tubules through geometrical frustration \cite{grason_perspective_2016,tyukodi_thermodynamic_2021}, and three-dimensional3D periodic structures for structural coloration \cite{michielsen_gyroid_2008}.    

\section*{Materials and Methods}

Brief descriptions of our experimental methods are provided below. For more detailed methods see the Supplementary Information I.

\subsection*{Assembly of triangular subunits}
DNA origami subunits are assembled in a one-pot reaction with 50 nM of p8064 scaffold DNA (Tilibit) and 200 nM of each staple strand (IDT, see Supplementary Information for sequences) in a standard `folding buffer'. Standard `folding buffers', described previously~\cite{wagenbauer_how_2017}, contain $X$ mM MgCl$_2$, 5 mM Tris base, 1 mM EDTA, and 5 mM NaCl (FoB$X$). Reaction solutions are subjected to a thermal annealing cycle in a Tetrad (Bio-Rad) thermocycler. Optimal MgCl$_2$ concentrations and annealing protocols are described in Supplementary Information Table S1.

\subsection*{Purification of subunits}
All origami subunits are purified by gel extraction and concentrated by ultrafiltration. We use a 1.5\% agarose gel with 0.5XTBE buffer, 5.5 mM MgCl$_2$, and 0.5x SYBR-safe (Invitrogen). Custom-made gel combs that can hold 0.2 ml per well were used to increase the throughput. The folded solution is mixed 5:1 with loading dye (30\% w/v Ficoll 400, 0.1\% w/v bromophenol blue, 3xTBE) and run at 110V for 2 hours. We remove the monomer band with a razor blade and dice it into small pieces. Gel pieces are placed in a Freeze 'N Squeeze spin column (Bio-Rad) and kept at -20$^\circ$ C for 5 minutes, then spun down for 5 minutes at 13 krcf. The subnatant is concentrated by ultrafiltration with 0.5 ml Amicon 100 kDa filters. Amicon filters are first equilibrated by centrifuging down 0.5 ml of 1xFoB5 at 5 krcf for 7 minutes, after which the flow-through is removed. The DNA origami solution is added up to 0.5 ml and centrifuged at 14 krcf for 15 minutes, then the flow-through is removed. This process is repeated until all of the DNA origami solution has been filtered. Finally, we place the filter upside down over a new Amicon tube and centrifuge at 1 krcf for 2 minutes. The DNA origami concentration of the final solution is measured using a Nanodrop (Thermo Scientific).

\subsection*{Self-assembly of tubules}
Purified subunits are assembled in 50 $\mu$l mixtures of 1xFoB$X$ with a monomer concentration of 10 nM. The MgCl$_2$ concentration ($X$) is varied from 5 to 30 mM. The assembly solution is pipetted into a capped 0.1 ml strip tube (Rotor-Gene), which is subsequently place into a 0.2 ml strip tube (Corning) to suppress evaporation and condensation within the tube. Tubes are loaded into a rotating incubator (Roto-Therm, Benchmark Scientific) at 40$^\circ$ C for a week.

\subsection*{Negative stain TEM}
Assembly samples are incubated on glow-discharged FCF400-Cu TEM grids (Electron Microscopy Sciences) for 60--120~s. Grids are then stained with 2\% aqueous uranyl formate solution with 20 mM NaOH for up to 30~s before blotting on filter paper and using vacuum suction to remove excess fluid. Images of the grids are acquired on an FEI Morgagni TEM operated at 80 kV with a Nanosprint5 CMOS camera (AMT) at magnifications between x8000 and x20000. Tomograms of grid samples are acquired on a Tecnai F20 TEM with an FEG run at 200kV with a Gatan Ultrascan 4kx4k CCD camera. Tilt series were observed at a magnification of x32000 from -50$^\circ$ to 50$^\circ$  in 2$^\circ$  increments. Subsequent analysis is performed using Etomo (IMOD~\cite{kremer_computer_1996}).

\subsection*{Cryo-electron microscopy}
Higher concentrations of DNA origami are used to prepare cryo-EM grids, summarized in SI Table S3. Samples are placed on glow-discharged C-flat 1.2/1.3 400 mesh grids (Protochip). Plunge-freezing of the grids in liquid ethane is performed with an FEI Vitrobot with sample volumes of 3 $\mu$l, blot times of 5--8~s, a blot force of -1, and a drain time of 0~s at 20$^\circ$ C and 95\% humidity. All cryo-EM images were acquired with a Tecnai F30 TEM with the field emission gun electron source operated at 300kV and equipped with an FEI Falcon II direct electron detector at a magnification of x39000. Single-particle acquisition was performed with SerialEM. The defocus was set to -2~$\mathrm{\mu}$m for all acquisitions with a pixel size of 2.87~Angstrom.

Image processing was performed using RELION-3~\cite{zivanov2018RELION3}. Contrast-transfer-function (CTF) estimation was performed using CTFFIND4.1~\cite{ctffind}. After picking single particles we performed a reference-free 2D classification from which the best 2D class averages were selected for processing, estimated by visual inspection. The particles in these 2D class averages were used to calculate an initial 3D model. A single round of 3D classification was used to remove heterogeneous monomers and the remaining particles were used for 3D auto-refinement and post-processing. A summary of the cryo-EM reconstructions is shown in SI Table S3.% All post-processed maps are deposited in the Electron Microscopy Data Bank.

\subsection*{Epi-fluorescence imaging of tubules}
To dye our samples, we incubate our assemblies with YOYO-1 dye (Invitrogen) at room temperature for at least half an hour in a solution of 5 nM DNA origami, 500 nM YOYO-1 dye, and 1xFoB, at the same MgCl$_2$ concentration as that of the assembly. We pipette 1.6 $\mu$l of the solution onto a microscope slide that has been cleaned with Alconox, ethanol (90\%), acetone, deionized water, and subsequently plasma-cleaned. After deposition, a plasma-cleaned coverslip is placed on top to create a thin liquid layer. Samples are then imaged on a TE2000 Nikon inverted microscope with a Blackfly USB3 (FLIR) camera.

\begin{acknowledgments}
We acknowledge Don Caspar for sparking an interest in the self-assembly of cylindrical tubules. We thank Botond Tyukodi and Farzaneh Mohajerani for helpful discussions, Thomas Gerling for experimental support, Ali Aghvami, Mike Rigney, and Berith Isaac, for technical support with electron microscopy. TEM images were prepared and imaged at the Brandeis Electron Microscopy facility. This work is supported by the Brandeis University Materials Research Science and Engineering Center, which is funded by the National Science Foundation under award number DMR-2011846. D.H. acknowledges support from the Masason Foundation. D.H., D.M.H., S.F., M.F.H., G.M.G., and W.B.R. conceived the research. D.H. and T.E.V. performed experiments. D.H., T.E.V, H.F., and E.F. analyzed the data. D.H., D.M.H., C.S., and H.D. designed the particles. All authors contributed to writing the manuscript.
\end{acknowledgments}

% Bibliography
\bibliography{main.bib}

\end{document}

\title{Supplemental Information for ``Geometrically programmed self-assembly of tubules using DNA-origami colloids"}% as subunits}
\author{Daichi Hayakawa, Thomas E. Videb\ae k, Douglas M. Hall, Huang Fang, Christian Sigl, Elija Feigl, Hendrik Dietz, Seth Fraden, Michael F. Hagan, Gregory M. Grason, W. Benjamin Rogers}

\maketitle

\section{Experimental methods}\label{sec:methods}
%\todo{identify the correct name and company of all the materials used}

\subsection{Folding DNA origami}\label{subsec:folding}
Each DNA origami particle is folded by mixing 50 nM of p8064 scaffold DNA (Tilibit) and 200 nM each of staple strands (IDT, see SI Tables~\ref{vseq} --~\ref{bseq} for the sequences) with folding buffer and run through an individually optimized annealing protocol. Our folding buffer, or FoB$X$, contains 5 mM Tris Base, 1 mM EDTA, 5 mM NaCl, and $X$ mM MgCl$_2$. We use a Tetrad (Bio-Rad) thermocycler for annealing the solutions. 

To find the optimal folding conditions, we screen a range of temperature ramps and MgCl$_2$ concentrations. We begin with a coarse screening over a wide range of temperature ramps and MgCl$_2$ concentrations: 20 mM MgCl$_2$ mixtures are folded with 4 hour temperature ramps for the eight temperature ranges 50-47$^\circ$C to 64-61$^\circ$C with 2$^\circ$C increments in the starting temperature and stepping by 1$^\circ$C per hour; 5, 10, 15, 20, 25, 30 mM MgCl$_2$ mixtures are folded with a 17 hour temperature ramp from 60-44$^\circ$C. This initial screen narrows down which MgCl$_2$ concentration range and starting temperature for the annealing will produce the highest folding yield. Subsequent detailed screenings with annealing protocols ranging from 4-17 hour ramps and over a smaller range of MgCl$_2$ concentrations provide an optimal folding condition. The results of the folding screens and optimal conditions for each DNA origami particle are shown in SI Section~\ref{sec:folding}.

\subsection{Agarose gel electrophoresis}\label{subsec:electrophoresis}
To assess the outcome of folding, we separate the folding mixture using agarose gel electrophoresis. Gel electrophoresis requires the preparation of the gel and the buffer. The gel is prepared by heating a solution of 1.5\% w/w agarose, 0.5x TBE to boiling in a microwave. The solution is cooled to 60 $^{\circ}$C. At this point, we add MgCl$_2$ solution and SYBR-safe (Invitrogen) to adjust the concentration of the gel to 5.5 mM MgCl$_2$ and 0.5x SYBR-safe. The solution is then quickly cast into an Owl B2 gel cast, and further cooled to room temperature. The buffer solution contains 0.5x TBE and 5.5 mM MgCl$_2$, and is chilled to 4$^{\circ}$C before use. Agarose gel electrophoresis is performed at 110 V for 1.5 to 2 hours in a cold room kept at 4$^{\circ}$C. The gel is then scanned with a Typhoon FLA 9500 laser scanner (GE Healthcare) at 100 $\upmu$m resolution.

\subsection{Gel purification and resuspension}\label{subsec:purification}
After folding, DNA-origami particles are purified to remove all excess staples and misfolded aggregates using gel purification. The folded particles are run through an agarose gel (now at a 1xSYBR-safe concentration for visualization) using a custom gel comb, which can hold around 2 ml of solution per gel. We use a blue fluorescent table to identify the gel band containing the monomers. The monomer band is then extracted using a razor blade, which is further sliced into smaller pieces. We place the gel slices into a Freeze 'N Squeeze spin column (Bio-Rad), freeze it in a -20$^\circ$C freezer for 5 minutes, and then spin the solution down for 3 minutes at 12 krcf. The concentration of the DNA-origami particles is measured using a Nanodrop (Thermofisher), assuming that the solution consists only of monomers, where each monomer has 8064 base pairs.

Since the concentration of particles obtained after gel purification is typically not high enough for assembly, we concentrate the solution through ultrafiltration \cite{wagenbauer_how_2017}. First, a 0.5 ml Amicon 100kDA ultrafiltration spin column is equilibrated by centrifuging down 0.5 ml of FoB5 buffer at 5 krcf for 7 minutes. Then, the DNA origami solution is added up to 0.5 ml and centrifuged at 14 krcf for 15 minutes. We remove the flow-through and repeat the process until all of the DNA origami solution is filtered. Finally, we flip the filter upside down into a new Amicon tube and spin down the solution at 1 krcf for 2 minutes. The concentration of the final DNA origami solution is then measured using a Nanodrop (Thermo Scientific).

\subsection{Estimating the binding free energies between subunits}\label{SIsubsec:dimergel}
In this work, we also use gel electrophoresis to analyze the binding strengths between the sides of the triangles (see SI Section~\ref{sec:bindingenergy}). Here, the conditions for gel electrophoresis are slightly different. In order to maintain the same binding strengths as the assembly conditions, the MgCl$_2$ concentration is matched with that from the assembly experiment and the bias voltage is decreased to 80 V. We also exchange the buffer solution every 45 minutes to minimize MgCl$_2$ precipitation and excess heat generation.

\subsection{Tubule assembly}\label{subsec:assembly}
All assembly experiments are conducted at a DNA-origami particle concentration of 10 nM. Note that for the binary species assembly, both A-triangle and B-triangle are at 10 nM. By mixing the concentrated DNA origami solution after ultrafiltration with FoB of suitable MgCl$_2$ concentration, we make 50 $\upmu$l of 10 nM DNA origami at MgCl$_2$ concentrations ranging from 5--30 mM. The solution is carefully pipetted into 0.1 ml strip tubes (Rotor-Gene) and capped to prevent evaporation and condensation of water within the tubes. Each tube is then further sealed in a 0.2 ml strip tube and loaded into a rotating incubator (Roto-Therm, Benchmark Scientific) at 40$^\circ$C.

\subsection{Fluorescence microscopy}\label{subsec:fluorescenceMethod}
We incubate our DNA-origami tubules with YOYO-1 dye (Invitrogen) at room temperature for a minimum of half an hour in a solution of 5 nM monomers, and 500 nM YOYO-1, 1xFoB at the MgCl$_2$ concentration of assembly. This ratio of YOYO-1 to origami is chosen so that there are 100 dye particles per structure, a limit in which the dye's impact on the structural integrity of the origami should be negligible~\cite{gunther2010mechanical}. 1.6 $\upmu$L of the solution is pipetted onto a microscope slide that has been cleaned with Alconox, ethanol (90\%), acetone, deionized water, and subsequently plasma-cleaned. After deposition, a plasma-cleaned coverslip is placed on the droplet at an angle and carefully lowered so that the liquid film is as thin as possible. We find that this reduces the sample thickness to about the width of a tubule without damaging the tubules, allowing them to lie flat on the surface. Samples are imaged on a TE2000 Nikon inverted microscope with a Blackfly USB3 (FLIR) camera.

\subsection{Negative stain TEM}\label{subsec:TEM}
We first prepare a solution of uranyl formate (UFo). ddH$_2$O is boiled to deoxygenate it and then mixed with uranyl formate powder to create a 2\% w/w UFo solution. The solution is covered with aluminum foil to avoid light exposure, then vortexed vigorously for 20 minutes. The solution is filtered using a 0.2 $\upmu$m filter. The solution is divided into 0.2 ml aliquots, which are stored in a -80$^\circ$C freezer until further use.

Prior to each negative-stain TEM experiment, a 0.2 ml aliquot is taken out from the freezer to thaw at room temperature. We add 4 $\upmu$L of 1 M NaOH and vortex the solution vigorously for 15 seconds. The solution is centrifuged at 4$^\circ$C and 16 krcf for 8 minutes. We extract 170 $\upmu$l of the supernatant for staining and discard the rest. 

The EM samples are prepared using FCF400-Cu grids (Electron Microscopy Sciences). We glow discharge the grid prior to use at 45 mA for 90 seconds at 0.1 mbar, using a Quorum Emitech K100X glow discharger. We place 3 $\upmu$l of the sample that initially contained 10 nM of monomers on the carbon side of the grid for 1 minute to allow adsorption of the sample to the grid. During this time 5 $\upmu$l and 18 $\upmu$l droplets of UFo solution are placed on a piece of parafilm. After the adsorption period, the remaining sample solution is blotted on 11 $\upmu$ Whatman filter paper. We then touch the carbon side of the grid to the 5 $\upmu$l drop and blot it away immediately to wash away any buffer solution from the grid. This step is followed by picking up the 18 $\upmu$l UFo drop onto the carbon side of the grid and letting it rest for 30 seconds to deposit the stain. The UFo solution is then blotted and we vacuum away any excess fluid. Grids are allowed to dry for a minimum of 15 minutes before insertion into the TEM.

We image the grids using an FEI Morgagni TEM operated at 80 kV with a Nanosprint5 CMOS camera (AMT). The microscope is operated at 80 kV and images are acquired between x8,000 to x20,000. The images are high-pass filtered and the contrast is adjusted using Adobe Photoshop. 

\subsection{TEM tomography}\label{subsec:tomography}
To obtain a tilt series, we use an FEI F20 equipped with a Gatan Ultrascan 4kx4k CCD camera, operated at 200 kV. The grid is observed at x18000, x29000, or x31000 magnification from -50 degrees to 50 degrees in 2 degree increments. The data is analyzed and the z stack is reconstructed using IMOD \cite{kremer_computer_1996}.

\subsection{Cryo-electron microscopy}\label{subsec:cryo}
Higher concentrations of DNA origami are used for cryo-EM grids than for assembly experiments, as listed in SI Table~\ref{STable:cryoconditions}. When using triangles that could assemble into tubules, we typically found excess aggregation and binding. We avoid these issues by using passive monomers, which we prepare by adding a 5-nucleotide-long poly-T overhang at the end of the blunt ends for all lock-and-key structures. To prepare samples we fold between 1-2 ml of the folding mixture (50 nM scaffold concentration), gel purify it, and concentrate the sample by ultrafiltration, as described above. EM samples are prepared on glow-discharged C-flat 1.2/1.3 400 mesh grids (Protochip). Plunge-freezing of grids in liquid ethane is performed with an FEI Vitrobot with sample volumes of 3 $\upmu$L, blot times of 5-8 s, a blot force of -1, and a drain time of 0 s at 20$^\circ$C and 95\% humidity. All cryo-EM images are acquired with a Tecnai F30 TEM with the field emission gun electron source operated at 300 kV and equipped with an FEI Falcon II direct electron detector at a magnification of x39000. Single particle acquisition is performed with SerialEM. The defocus is set to -2 $\upmu$m for all acquisitions with a pixel size of 2.87 Angstrom. 

\subsection{Single-particle reconstruction}\label{subsec:recontruction}
Image processing is performed using RELION-3~\cite{zivanov2018RELION3}. Contrast-transfer-function (CTF) estimation is performed using CTFFIND4.1~\cite{ctffind}. After picking single particles we performed a reference-free 2D classification from which the best 2D class averages are selected for processing, estimated by visual inspection. The particles in these 2D class averages are used to calculate an initial 3D model. A single round of 3D classification is used to remove heterogeneous monomers and the remaining particles are used for 3D auto-refinement and post-processing.  A summary of the cryo-EM reconstructions is shown in SI Table~\ref{STable:cryoconditions}. All post-processed maps are deposited in the Electron Microscopy Data Bank.

%%%%%%%%%%%%%%%%%%%%%%%%%%%%%%%%%%%%%%%%%%%%%%%%%%%%%%%%%%%%

\section{Design criteria for triangular subunits}\label{sec:tubeGeom}

%mention yoshimura tiling?

\begin{figure*}[ht]
 \centering
 \includegraphics[width=0.9\textwidth]{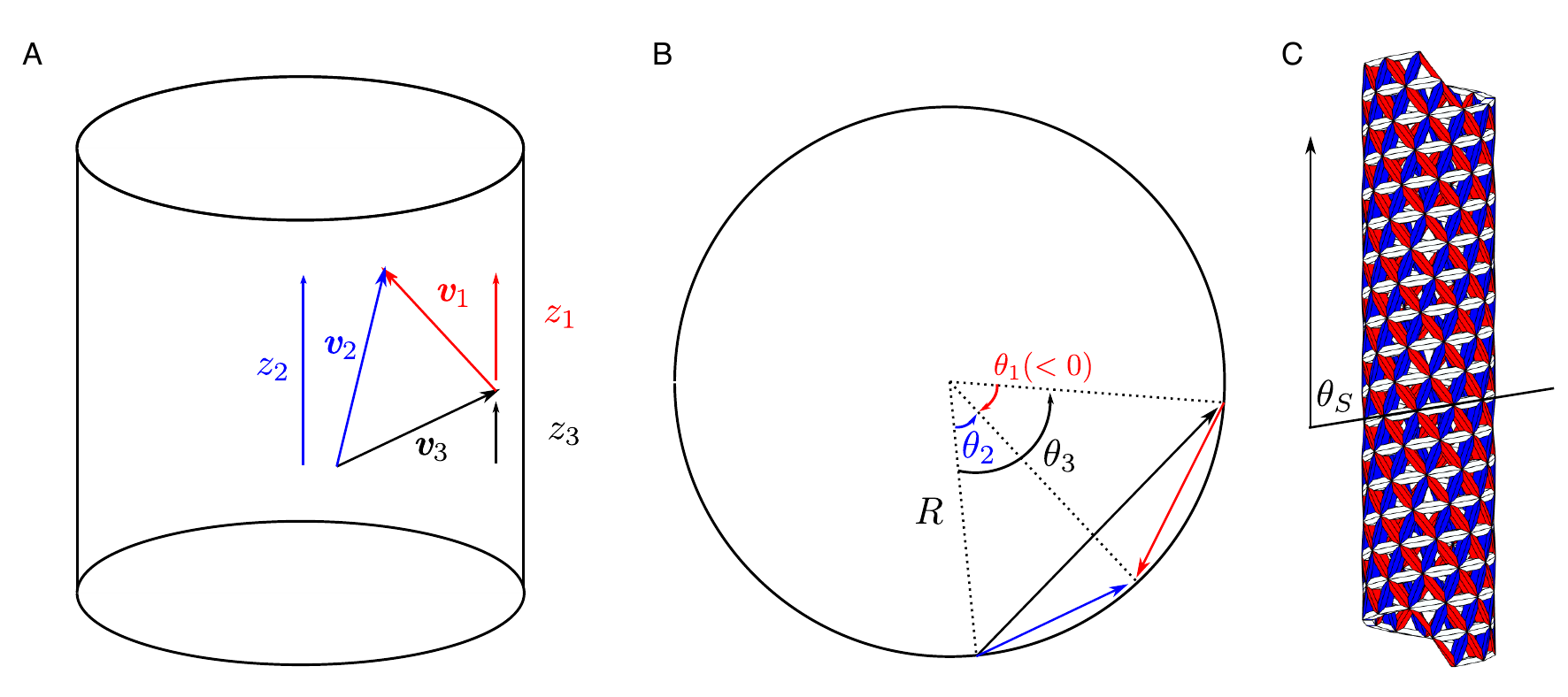}
 \caption{\textbf{Calculations of the tubule lattice coordinates.} The location of the vertices of a triangle on the surface of a cylinder is shown from the side, to see the relative differences in $z$ (A) , and along the cylinder axis, to see relative differences in the angular coordinate (B). (C) A tubule, viewed from the side, shows the the maximum seam angle, $\theta_S$.}
 \label{Sfig:tubeGeometry}
\end{figure*}

\begin{figure*}[bth]
 \centering
 \includegraphics[width=0.99\textwidth]{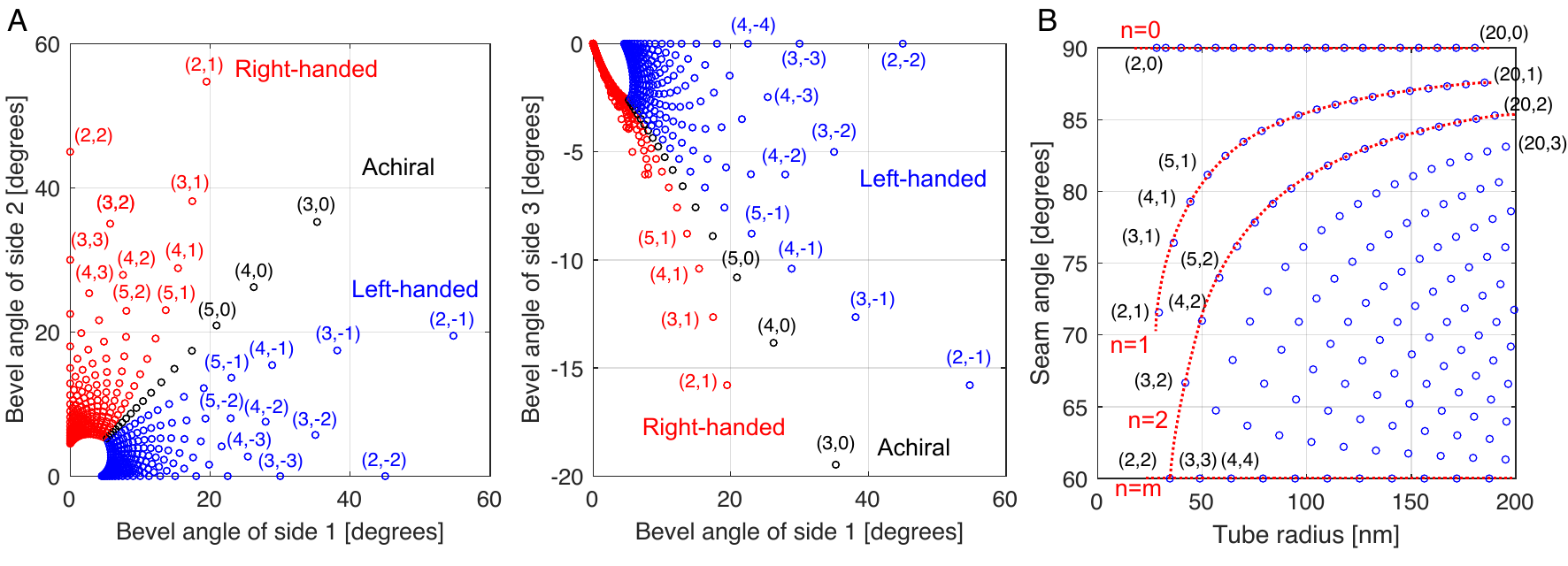}
 \caption{\textbf{The geometry of tubules with different lattice numbers.}
(A) A plot of the bevel angles of sides 1, 2, and 3 for different lattice numbers. (B) A plot of the maximum seam angle versus the tubule radius for different tubule types. Red dotted lines are guides for the eyes for different $n$ values and are not fitted lines. To calculate the radius, we assume that the length of a side of a triangle is 56.44 nm. 
 }
 \label{Sfig:anglePlot}
\end{figure*}

We derive the coordinates and the dihedral angles of the triangular subunits by assuming that they are rigid triangles whose vertices lie on the surface of a cylinder. This scheme corresponds to the polyhedral model proposed in \cite{Lee2009Jan}. Let us choose one triangle on a tubule and label the three sides side 1, 2, and 3, as in Fig.~\ref{Sfig:tubeGeometry}A. For convenience, we choose side 3 to be the side that is closest to being perpendicular to the axis of the tubule. We choose the 3 vectors of the tubule $\textbf{v}_1$, $\textbf{v}_2$, and $\textbf{v}_3$ as in Fig.~\ref{Sfig:tubeGeometry}A, whose length are all 1. The corresponding height and angular change along the vector in cylindrical coordinate is $z_1$, $z_2$, $z_3$, and $\theta_1 (\leq 0)$, $\theta_2$, $\theta_3$, respectively, and the radius of the tubule is $R$ (Fig.~\ref{Sfig:tubeGeometry}B). Since the length of each side is 1,
\begin{equation}
    4R^2 \sin^2{\frac{\theta_i}{2}} + z_i^2 = 1,  \: (i\in[1,3]).
    \label{Seq:lengthCondition}
\end{equation}
Additionally, for the three vectors to form a closed triangle,
\begin{eqnarray}
    z_1 + z_3 &= z_2
    \\
    \theta_1 + \theta_3 &= \theta_2.
    \label{Seq:triangleCondition}
\end{eqnarray}
Finally, for a right-handed, ($m,n$) tubule, we require closed loops after m steps along edge 3 and n steps along edge 1, as illustrated in Fig.~1D of the main text, yielding
\begin{eqnarray}
    mz_3 - nz_1 &= 0
    \\
    m\theta_3 - n\theta_1 &= 2\pi.
  \label{Seq:latticeCondition}
\end{eqnarray}
In the case of left-handed tubules, where $n$ and $z_3$ are both negative, the loop constraint is given by
\begin{eqnarray}
    mz_3 - nz_2 &= 0
    \\
    m\theta_3 - n\theta_2 &= 2\pi.
  \label{Seq:latticeConditionLeft}
\end{eqnarray}
Solving simultaneous the equations~\ref{Seq:lengthCondition},\ref{Seq:triangleCondition},\ref{Seq:latticeCondition}, for variables $z_i$, $\theta_i$, and $R$, one obtains the coordinates for all the triangle vertices on the tubule. From the vertex coordinates, we derive bevel angles of triangles $\theta_{Bi}$ for side $i$ as listed on Table~\ref{STable:bevel} and plotted in Fig.~\ref{Sfig:anglePlot}A. Here, we used the relationship
\begin{equation}
    \theta_{Bi} = \frac{180 - \phi_i}{2}
    \label{Seq:dihedral2bevel}
\end{equation}
to calculate the bevel angle $\theta_{Bi}$ from the dihedral angle $\phi_i$. In the case of left-handed tubules, equation~\ref{Seq:latticeConditionLeft} results in flipped bevel angles between sides 1 and 2. Additionally, we also obtain the seam angles, which we later use to classify the tubules into different lattice-number types (Fig.~\ref{Sfig:anglePlot}B). Here, we define the seam angle $\theta_S$ as the angle of side 3 with respect to the tubule axis (Fig.~\ref{Sfig:tubeGeometry}C). Coming back to the premise that side 3 is the side that is closest to being perpendicular to the axis of a tubule, the seam angle takes values between 60--90 degrees. Specifically, it is calculated as  
\begin{equation}
    \theta_S = \arctan{\frac{2R\sin{\frac{\theta_3}{2}}}{z_3}}.
\end{equation}

%%%%%%%%%%%%%%%%%%%%%%%%%%%%%%%%%%%%%%%%%%%%%%%%%%%%%%%%%%%%

\section{Designing triangular subunits using DNA origami}\label{sec:design}
\begin{figure*}[ht]
 \centering
 \includegraphics[width=0.9\textwidth]{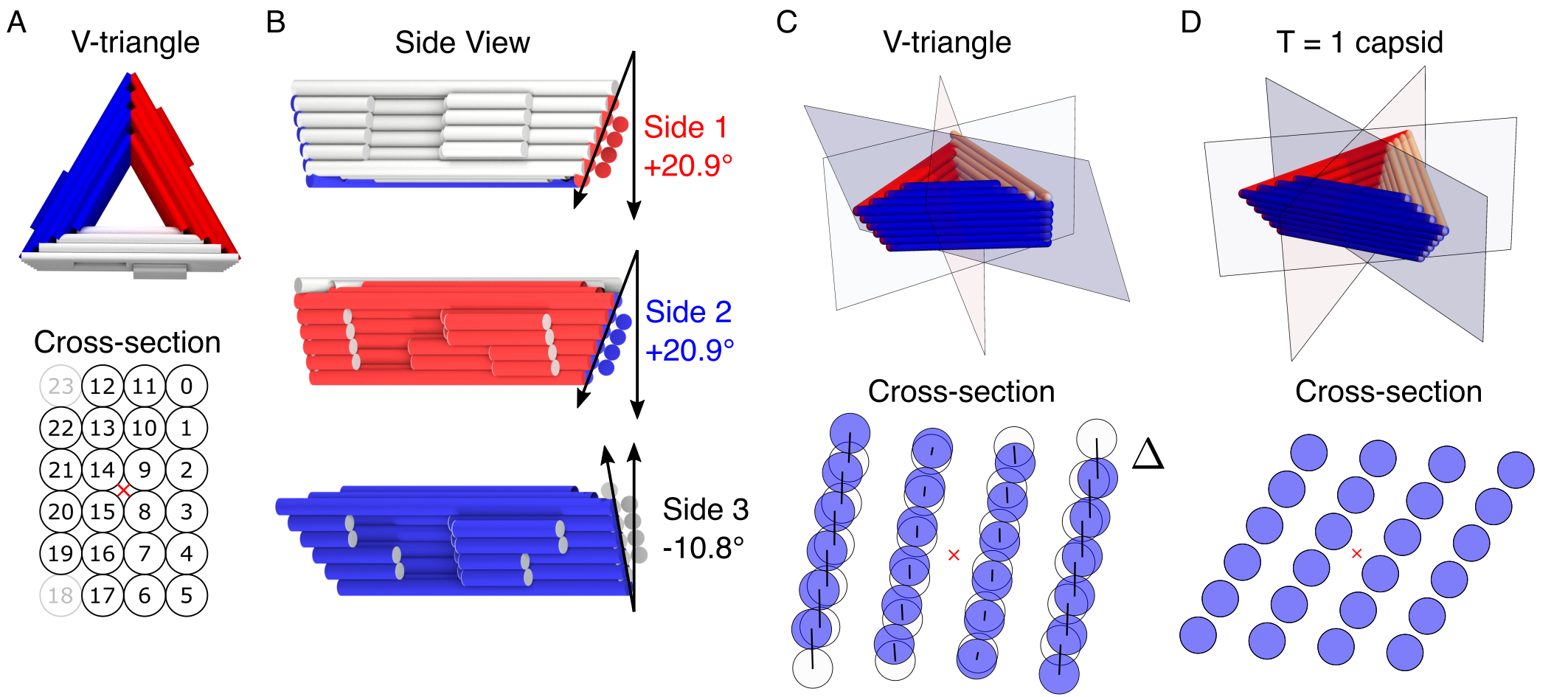}
 \caption{\textbf{Triangle design.}
 (A) Schematic of the V-triangle and the cross-section of each brick composing a side. The numbers are labeled as in the caDNAno design. The red cross indicates the axis about which we rotate the brick. Helix 18 and helix 23 are excluded in the actual design. (B) Side view of the V-triangle with the bevel angles for each side. (C) (Top) Three planes that cut through the vertices of the V-triangle and (bottom) the cross-section of the vertex between sides 2 and 3. Again, the red cross shows the center of the bricks, which matches between sides 2 and 3. $\Delta$ shows the mismatching distance for each helix, illustrated by the straight black lines. (D) shows the same schematic as in (C) but for the T=1 capsid monomer design, where all bevel angles are 21.6 degrees~\cite{sigl_programmable_2021}. %\dha{why are vertical helices not touching for (d)?}
 }
 \label{Sfig:triangleDesign}
\end{figure*}

We adapt the triangle design scheme initially developed by C. Sigl \textit{et al.} \cite{sigl_programmable_2021} for our monomer design. Our triangle is created from three bricks that are 6-by-4 DNA double helices arranged on a square lattice (Fig.~\ref{Sfig:triangleDesign}A). For generating the geometry, we assume 2.6 nm for the interhelical spacing, 0.34 nm for the dsDNA helical rise, and 0.5 nm for the length of a single unpaired nucleotide. To compensate for the 10.667 bp/turn helical pitch imposed by the square lattice scheme in the caDNAno software \cite{douglas_rapid_2009}, we place a `skip' every 32 bp, except in regions near the locks and keys. In essence, the triangles are simply three bricks that are rotated about the brick axes by the preassigned bevel angles (Fig.~\ref{Sfig:triangleDesign}B). The three bricks are then connected at the vertices by single stranded domains of the scaffold and staple strands to make a closed object. In the design steps, we exclude helices 18 and 23 in Fig.~\ref{Sfig:triangleDesign}A from our design to increase the size of the triangle. Also, on the faces of the bricks, we place locks and keys that are self-complementary, each spanning 4 helices in length and 1 helix deep.

Since the bevel angles are not the same on all sides, one needs to take care in how the corners are connected. In Fig.~\ref{Sfig:triangleDesign}C, the white helices from side 3 and blue helices from side 2 do not match up at the vertex, as can be seen from the cross-sectional view. We refer to this mismatching distance between the helices as $\Delta$. Such staggering at the corners is a result of the bevel angle mismatch between the neighboring bricks. When there is no bevel angle mismatch, such as the T=1 capsid monomer from ref.~\cite{sigl_programmable_2021}, the helices meet perfectly at the corners (Fig.~\ref{Sfig:triangleDesign}D). 

Our goal is to come up with a design method for minimizing the effect of having mismatched helices at the corners; poorly designed corners may compromise the programmed bevel angles of the sides. First, we make an equilateral triangle with three bricks.  Initially, the three bricks are placed so that the central axes of the bricks, marked by the red cross in Fig.~\ref{Sfig:triangleDesign}A, forms an equilateral triangle. We call this the base equilateral triangle. The bricks are then rotated by the prescribed bevel angles around the brick axes. This rotation leaves the center of the brick fixed, but the helices around it are displaced. Then, we optimize the orientation of the plane used to cut through the corners, as shown at the top of Fig.~\ref{Sfig:triangleDesign}C. At the corners, there are infinitely many ways one can come up with to terminate a helix from one brick and transition it to the other. Here, we search for a flat plane that simultaneously goes through the corner of the base equilateral triangle and minimizes the sum of all mismatching distances $\Delta$. The helices from both sides terminate where they meet this plane. Finally, the remaining mismatch of the helices is compensated by leaving a portion of the scaffold single stranded at the corners. Similarly, at the corners, we insert 5T to the staples to remove any potential base-stacking interactions that may add stress to the corners. Finally, the triangles are scaled so that the required scaffold length is 8064 bases in total, which is the length of the p8064 scaffold (Tilibit nanosystems). The final caDNAno designs for the triangles are shown in Fig.~\ref{Sfig:XVcadnano} and~\ref{Sfig:ABcadnano}, and the corresponding sequences can be found in Tables~\ref{vseq}-\ref{bseq}.

\subsection{Designing the lock-and-key interactions}

\begin{figure*}[thb]
 \centering
 \includegraphics[width=0.99\textwidth]{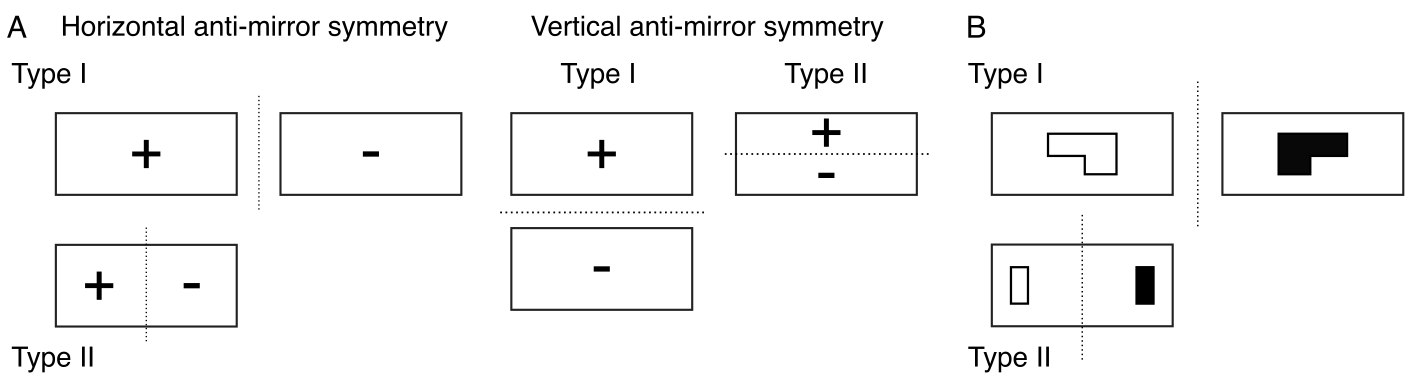}
 \caption{\textbf{Symmetries of protrusions and recesses.} (A) Two symmetry patterns of protrusions and recesses are allowed on the faces of the triangles. For both the horizontal and vertical anti-mirror symmetries, the mirror plane can be located either between the two faces (Type I) or at the center of the face of a triangle (Type II). Dotted lines indicate the mirror plane. $+$ and $-$ signs indicate inversion of the protrusion and the recess. (B) Example protrusion and recess patterns for the triangles. A non-self-complementary interaction (Type I) and a self-complementary interaction (Type II) are shown. Protrusions are shown as outlines, whereas recesses are shown as filled shapes.
 }
 \label{Sfig:binaryInteraction}
\end{figure*}

The interactions between our DNA-origami subunits are based on lock-and-key shape complementarity. In other words, we design the shapes of protrusions and recesses that fit together. In order for one face of our monomer to bind to another in the right orientation, the shapes of the protrusions and recesses need to be anti-mirror symmetric (see Fig.~\ref{Sfig:binaryInteraction}A). Here, the term `anti' refers to the inversion between the protrusion and the recess, expressed as '$+$' and '$-$'. Namely, for the two faces to bind to each other, the protrusions and recesses on one face have to be mirror symmetric in shape and inverted compared to those on the other face. In this paper, we only consider mirror symmetry that is horizontal, though the use of vertical mirror symmetry may be useful for other design applications. Regardless of the orientation of the symmetry, there are two locations where the mirror plane can be placed: in between two distinct faces, which we name Type I, or in the middle of the face, which we call Type II. By nature, Type I symmetries impose non-self-complementary interactions, whereas Type II symmetry imposes self-complementary interactions. Figure~\ref{Sfig:binaryInteraction}B shows examples of protrusions and recesses for binding, where open shapes denote recesses and filled shapes denote protrusions.

%%%%%%%%%%%%%%%%%%%%%%%%%%%%%%%%%%%%%%%%%%%%%%%%%%%%%%%%%%%%

\section{Design criteria for binary tilings of tubules}\label{sec:binaryDesign}

\begin{figure*}[ht]
 \centering
 \includegraphics[width=0.99\textwidth]{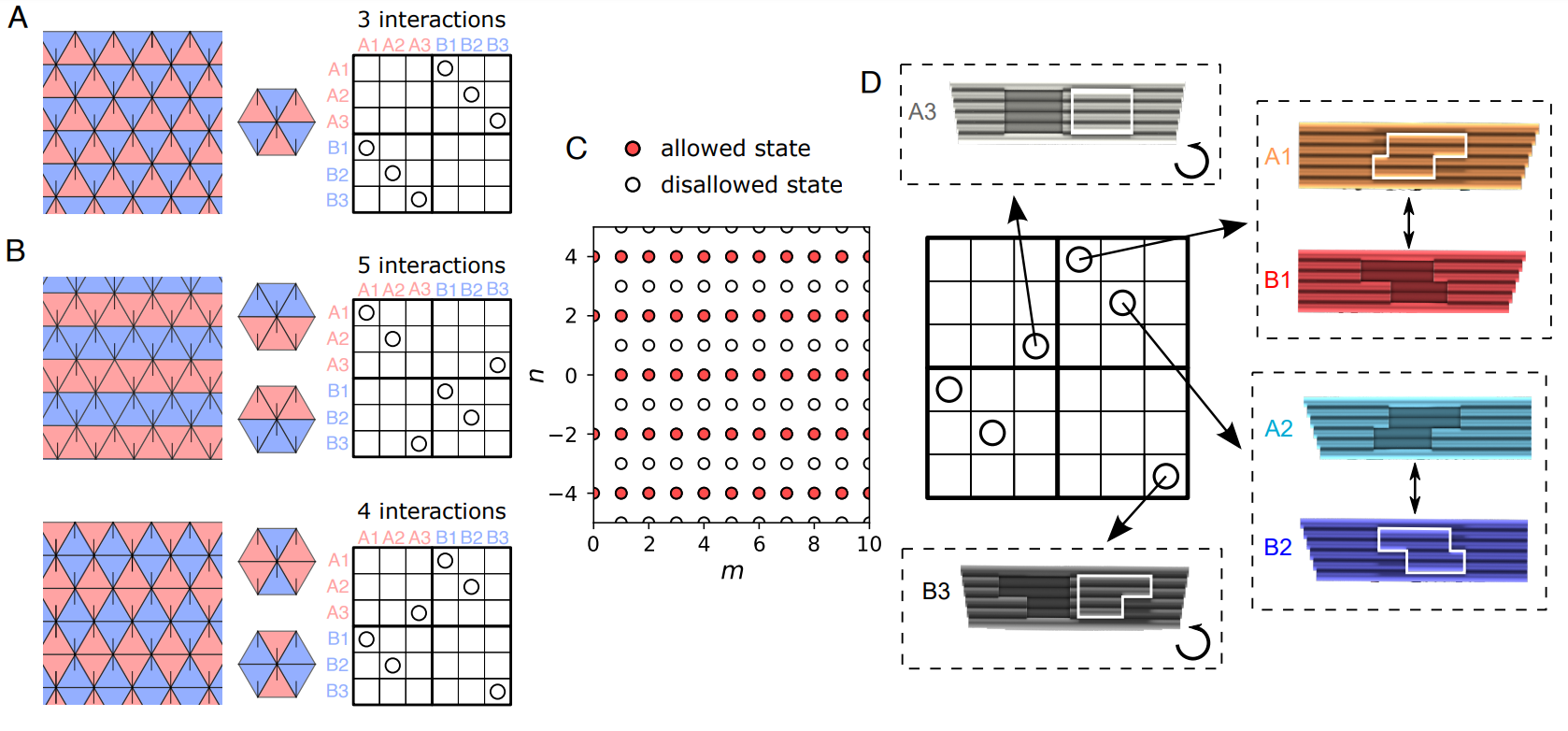}
 \caption{\textbf{Two species design considerations.} (A) and (B) show the three possible tilings using two species that can fold into tubules. Red and blue triangles represent A and B triangles, respectively. A line inside the triangle points to the vertex in between sides 1 and 2. Each pattern has next to it the unique vertices in the pattern and the interaction matrix needed to create the tiling. Circles in the interaction matrix represent favorable interactions. The pattern in (A) contains only a single unique vertex, and thus, does not limit any tubule state. (C) Space of allowed ($m,n$) tubule types results from the tilings in (B). (D) Design for lock-and-key interactions for a given interaction matrix. Note that when you have self-complementary sides both a protrusion and a recess are needed on the same face.
 }
 \label{Sfig:TwoSpeciesDesgin}
\end{figure*}

Previously, we found that there are only three possible tiling patterns that could assemble into a tubule using two species of triangles  \cite{videbaek2021tiling}. These patterns are shown in Fig.~\ref{Sfig:TwoSpeciesDesgin}A and B. For each pattern, there is also an associated interaction matrix that describes which sides of the particles need to have favorable interactions to produce the given tiling. The basic premise is that by increasing the number of species of triangles in the system, the number of distinct vertices in the tiling increases, leading to a reduction in the number of accessible tubule states since some combinations of vertices cannot close in a periodic way. Since the pattern in Fig.~\ref{Sfig:TwoSpeciesDesgin}A has only a single vertex type, it does not remove any accessible tubules, while both patterns in Fig.~\ref{Sfig:TwoSpeciesDesgin}B have two distinct vertices, thereby removing half of the possible tubule types. The available tubule types for these patterns are shown in Fig.~\ref{Sfig:TwoSpeciesDesgin}C.

An important consideration when designing the monomers for a multiple-species ensemble is determining how many unique interactions need to be encoded into the system. From the interaction matrices shown in Fig.~\ref{Sfig:TwoSpeciesDesgin}A and B, we can see how many interactions are needed by counting the number of points in the upper triangular region of the matrix. We can see that the patterns in Fig.~\ref{Sfig:TwoSpeciesDesgin}B require different numbers of unique interactions even though they provide the same state restrictions. The fewer interactions that one needs to use, the easier it is to design interactions that have minimal cross-talk.

In the main text we use the interactions shown in Fig.~\ref{Sfig:TwoSpeciesDesgin}D. This illustrates another aspect of how the lock-and-key interactions correspond to where the interaction appears in the interaction matrix. If the interaction is on the main diagonal, then the interaction needs to be self-complementary, or type II, illustrated in Fig.~\ref{Sfig:binaryInteraction}A. For such interactions, both a protrusion and a recess are necessary since they involve interactions with a single species of particle. Off-diagonal interactions are inherently between two species of particles and only require a protrusion or a recess, as in type I. 

%%%%%%%%%%%%%%%%%%%%%%%%%%%%%%%%%%%%%%%%%%%%%%%%%%%%%%%%%%%%

\section{DNA origami folding conditions} \label{sec:folding}
\begin{figure*}[htb]
 \centering
 \includegraphics{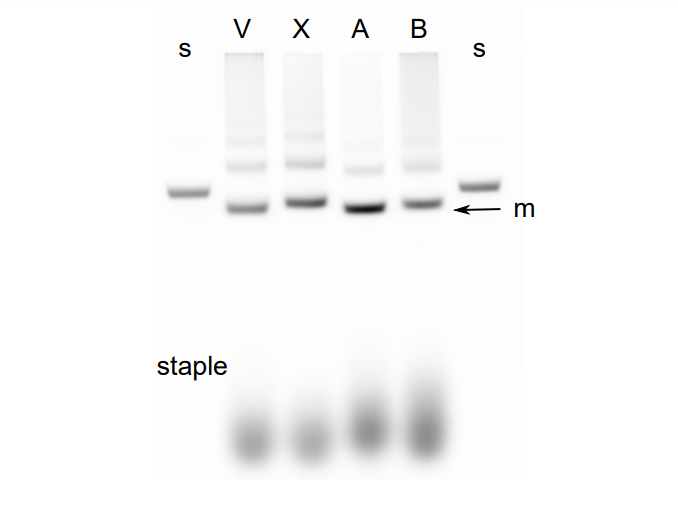}
  \caption{\textbf{Gel electrophoresis of folded solutions.}
 Gel electrophoresis of folding solutions for each triangle type. The ``s'' band at both ends of the gel contains only the scaffold solution for reference. The monomer bands for the triangles are all lower than the scaffold band, denoted as ``m''.}
 \label{Sfig:foldingCondition}
\end{figure*}

The DNA-origami particles have folding conditions that are quite sensitive to the temperature ramp range and the salt concentration \cite{sobczak_rapid_2012}. For each particle, we conduct a folding screen at different Mg$^{2+}$ concentrations and temperature ramp ranges to determine the optimal conditions. The folding conditions are summarized in Table~\ref{STable:condition}. In the end, the folded solution contains a relatively high yield of monomers for all of the designs that we created, as can be seen from Fig.~\ref{Sfig:foldingCondition}. Typically, the gel purification yield of the triangles is around 20-30\% of the initial scaffold amount.

\begin{table}[tbh]
\centering
\begin{tabular}{c c c}
Object & Temperature ramp (1$^\circ$C/1hr) & Folding Mg$^{2+}$ concentration (mM)\\ 

\hline 
V-triangle & 60-44$^\circ$C & 15\\
X-triangle & 58-50$^\circ$C & 15\\
A-triangle & 58-50$^\circ$C & 20\\
B-triangle & 58-50$^\circ$C & 20\\

\end{tabular}
\caption{\textbf{Summary of folding and assembly conditions.} }
\label{STable:condition}
\end{table}

%%%%%%%%%%%%%%%%%%%%%%%%%%%%%%%%%%%%%%%%%%%%%%%%%%%%%%%%%%%%

\section{Binding energies and assembly conditions for triangles} \label{sec:bindingenergy}

\subsection{Binding energies of V- and X-triangles} \label{subsec:bindingVX}

Our DNA-origami subunits interact through blunt-end stacking at the ends of the lock-and-key shapes that are located on each side of the triangle. The blunt-end interactions of dsDNA originate from the $\pi-\pi$ stacking between base pairs that are adjacent along the helical direction of DNA and are sequence dependent \cite{SantaLucia_1996Jan,kilchherr_single-molecule_2016}. In designing our DNA origami, we attempt to maximize the free energy of all intended blunt-end interactions, as predicted from ref.~\cite{kilchherr_single-molecule_2016}, while also keeping the differences between the free energy of each side's blunt-end interactions below 4 kcal/mol, or 6.75 $k_\mathrm{B}T$ at 25$^\circ$C (see Table~\ref{STable:bindingEnergy}). Since the scaffold DNA is a loop, this task is accomplished by permuting the starting point of the loop through the 8064 possible positions until our free-energy criteria are satisfied as best as possible. Note that for sides 1 and 2 of A- and B-triangles, there are either protrusions or recesses, but not both on a single side. Having only a protrusion or a recess reduces the blunt-end stacking sites by half compared to the other cases, thus, the interaction is considerably weaker compared to that of side 3. For these triangles, we permute the starting point of the scaffold loop for the two triangles to maximize the total binding energy of all interactions, while keeping the maximum difference between the sides to below 10 kcal/mol, or 16.9 $k_\mathrm{B}T$.

\begin{table}[tbh]
\centering
\begin{tabular}{c | c c c | c c c}
\multicolumn{1}{c}{}&
\multicolumn{3}{c}{Predicted from \cite{kilchherr_single-molecule_2016} ($k_\mathrm{B}T$)} &
\multicolumn{3}{c}{Measured ($k_\mathrm{B}T$)}  \\

Object & Side 1 & Side 2 & Side 3 & Side 1 & Side 2 & Side 3\\ 

\hline 
V-triangle & -53.95 & -54.18 & -59.25 & -20.55 & -22.24 & -24.27 \\
X-triangle & -56.28 & -51.48 & -57.09 & -18.44 & -18.64 & -18.93 \\
A-triangle & -33.37 & -33.00 & -47.97 & Not measurable & Not measurable & -19.59\\
B-triangle & -33.37 & -33.00 & -47.13 & Not measurable & Not measurable & -19.13\\

\end{tabular}
\caption{\textbf{Summary of predicted and measured free energies of binding at 20 mM MgCl$_2$.} }
\label{STable:bindingEnergy}
\end{table}

\begin{figure*}[tbh]
 \centering
 \includegraphics[width=0.8\textwidth]{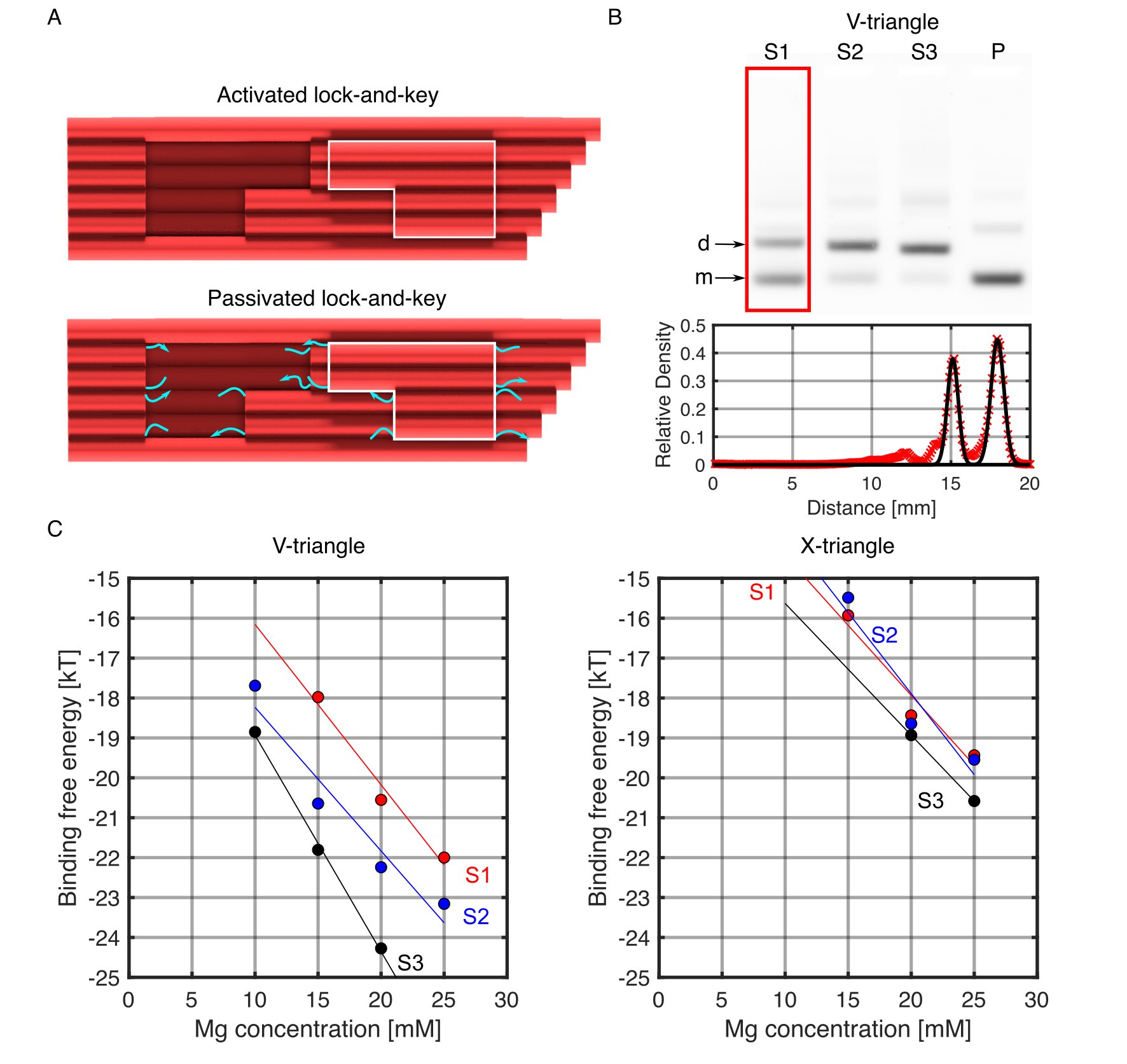}
 \caption{\textbf{Bindings strength for V-triangle and X-triangle.}
 (A) An illustration of our method for passivating triangle interactions. Passivated triangles have staples with an extra 5T domain hanging from the blunt ends, illustrated in cyan. The arrows indicate the direction from the 5' to 3' ends of the DNA. The length of the poly-T sequence is not drawn to scale. (B) Gel electrophoresis of dimer samples at 15 mM MgCl$_2$ in a 1.5\% agarose gel. The number after S indicates the side that is activated, while P indicates a fully passivated sample. A profile of the gel in the red rectangle is shown below. The monomer band, located at around 15 mm, and the dimer band, located at around 18 mm, are both fitted with a Gaussian, shown in black. (C) shows the measured binding free energies for V-triangle (left) and X-triangle (right). Red, blue, and black dots indicate the free energies for sides 1, 2, and 3, respectively. Lines are linear fits to the data.}
 \label{Sfig:unaryDimer}
\end{figure*}

We experimentally measure the binding free energy of each side using gel electrophoresis. First, triangles are folded and purified, with only one target side activated and the other two sides are passivated. We passivate the sides by adding 5T to the ends of the staple strands at the locations of the blunt-end interactions of the locks and keys, effectively prohibiting them from binding (Fig.~\ref{Sfig:unaryDimer}A). Samples at 10 nM monomers are incubated at 40$^\circ$C in different MgCl$_2$ concentration buffers, then run in a 1.5\% agarose gel with the same MgCl$_2$ concentration. As can be seen from the gel scan in Fig.~\ref{Sfig:unaryDimer}B, we observe distinct monomer and dimer bands. We individually fit the monomer and dimer band with Gaussian distributions to find their relative concentrations. 

To calculate the binding free energy we consider the rate equation for dimerization 
\begin{equation}
    M + M \rightleftharpoons D,
\end{equation}
where $M$ is a monomer and $D$ is a dimer. The chemical equilibrium constant, $K$, for this reaction equation is related to the free energy difference, $\Delta G$, between the initial and the end state,
\begin{equation}
    K = \frac{[D]u_0}{[M]^2} = \exp{\left(-\frac{\Delta G}{k_\mathrm{B}T}\right)},
\end{equation}
where $u_0$ is a reference concentration, set to 1 M in this case. Then the free energy is simply
\begin{equation}
    \Delta G = -k_\mathrm{B}T \log{\left(\frac{[D]u_0}{[M]^2}\right)}.
    \label{Seq:freeEnergy}
\end{equation}
Additionally, since the initial monomer concentration is 10 nM, we have 
\begin{equation}
    \left[M\right] + 2\left[D\right] = 10\ \mathrm{nM}
    \label{Seq:totMono}
\end{equation}
as a constraint. Therefore, using the measured concentration ratio between the monomer and the dimer band with Eqn.~\ref{Seq:freeEnergy} and~\ref{Seq:totMono}, we obtain the free energy of binding for each side, as in Fig.~\ref{Sfig:unaryDimer}C.

The binding free energies are relatively similar between the three sides of the triangles, as initially designed. However, the magnitude of the binding energies are less than half the value predicted from reference \cite{kilchherr_single-molecule_2016} (see Table~\ref{STable:bindingEnergy}). This can be attributed to a few different factors. First and foremost, the assumption that the total free energy is the sum of the individual blunt-end stacking interactions may oversimplify the problem. Though the same assumption is used in the reference, the goal is entirely different. Kilchherr \textit{et al.} \cite{kilchherr_single-molecule_2016} used a rigid origami rod that interacts with another origami rod in the direction of the DNA double helices, while our blunt ends are placed perpendicular to the direction of the binding between triangles. Geometry adds further complications. At the molecular scale, our designed lock-and-key configuration may be floppy, which could lower the binding strength if not all blunt ends in the lock-and-key interactions are in contact. Finally, gel electrophoresis is a non-equilibrium assay, and thus, the result might not accurately reflect the assembly conditions. During gel electrophoresis, the buffer is lowered to 4$^\circ$ C, an electric field is applied, and monomers become separated from the dimers, taking local concentrations far from their equilibrium values. For accurate measurement, we propose \textit{in situ} measurements such as light-scattering.

\subsection{Assembly condition for V- and X- triangles}\label{subsec:assemblyVX}

\begin{figure*}[htb]
 \centering
 \includegraphics[width= 0.7\textwidth]{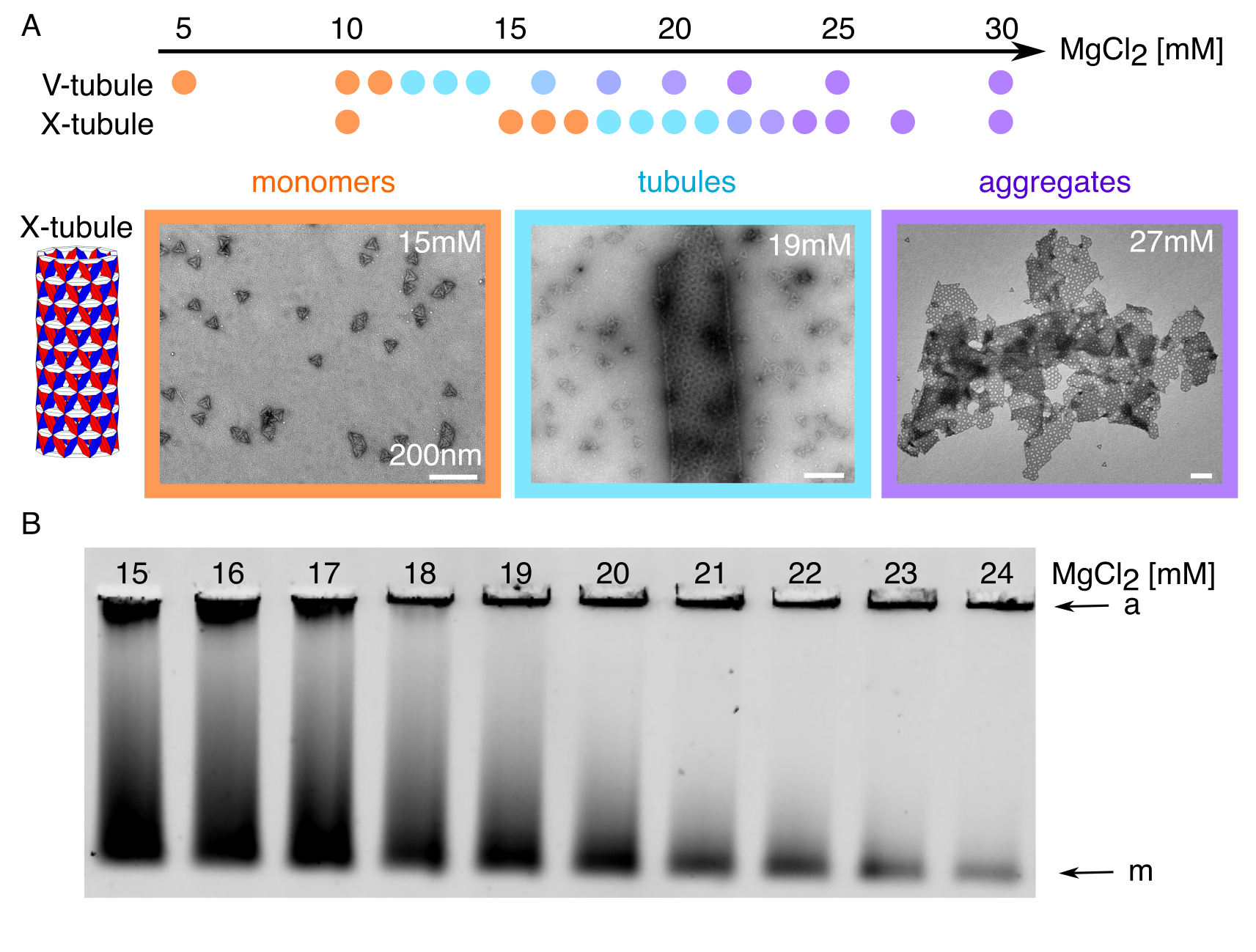}
  \caption{\textbf{Assembly conditions for the V-tubule and X-tubule.}
(A) A scan of the Mg$^{2+}$ concentration for assembly of the V-tubule and X-tubule. Representative TEM images for the X-tubules are shown at the bottom (for V-tubule images, see Fig.~2 in the main text). The gradient in color from tubule assembly to aggregation is meant to convey a qualitative change in the aggregate fraction. (B) Gel electrophoresis of assembled X-triangles in 0.5\% agarose and 20 mM MgCl$_2$ buffer. Different lanes show different Mg$^{2+}$ concentrations of assembly. ``m'' denotes the location of the monomer band and ``a'' is the location of the assembly or aggregate band. The contrast of the image has been enhanced to highlight the bands.}
 \label{Sfig:assemblyCondition}
\end{figure*}

After gel purification of folded monomers, we adjust the monomer concentration and Mg$^{2+}$ concentration, then incubate the solution in a rotating incubator at 40$^\circ$C for one week. At low salt concentrations, monomers remain unbound or form small clusters. Depending on the triangle types, we observe the spontaneous assembly of tubules between 10 and 20 mM MgCl$_2$ (Fig.~\ref{Sfig:assemblyCondition}A). The Mg$^{2+}$ concentration for the transition is usually consistent within 1 mM MgCl$_2$ over repeated trials. As the Mg$^{2+}$ concentration is increased even higher, fewer and fewer monomers are present (Fig.~\ref{Sfig:assemblyCondition}B), since the free energy of binding is lowered. However, this also enhances the chance for cluster-to-cluster binding to occur, resulting in aggregated structures, as can be seen in Fig.~\ref{Sfig:assemblyCondition}A. As a result, at high Mg$^{2+}$ concentrations, we observe mostly aggregated structures in TEM, rather than tubules or free monomers. In essence, there is a Goldilocks principle for assembly. The Mg$^{2+}$ concentration should not be too high nor too low to obtain tubules. It should be noted that the transition from tubule assembly to aggregate formation is gradual, as represented qualitatively by the color gradient of circles at the top of Fig.~\ref{Sfig:assemblyCondition}A.

The Mg$^{2+}$ concentration at which assembly begins can be compared against the binding energy. Tubule assembly starts around 12 mM MgCl$_2$ for the V-triangle and 18 mM for the X-triangle. By interpolating the free energy of binding using a linear fit, as in Fig.~\ref{Sfig:unaryDimer}C, we obtain $-18.6\ k_\mathrm{B}T$ and $-17.5\ k_\mathrm{B}T$ as the average binding energy of assembly of the three sides for V- and X-tubule, respectively. Not surprisingly, the free energy for tubule assembly of the two systems is similar.

\subsection{Binding free energies and assembly conditions for binary species triangles}\label{subsec:assemblybinary}

Unlike the unary species triangles, we expect asymmetry in the binding strengths for the binary species triangles. To be specific, side 1 and side 2 of both the A- and B-triangles have only half the base-stacking sites as side 3. Using the same gel electrophoresis assay, we could not detect any dimer formation for side 1 or side 2 between triangles A and B, even at 20 mM MgCl$_2$ (Table~\ref{STable:bindingEnergy}). 

\begin{figure*}[tbh]
 \centering
 \includegraphics[width=0.8\textwidth]{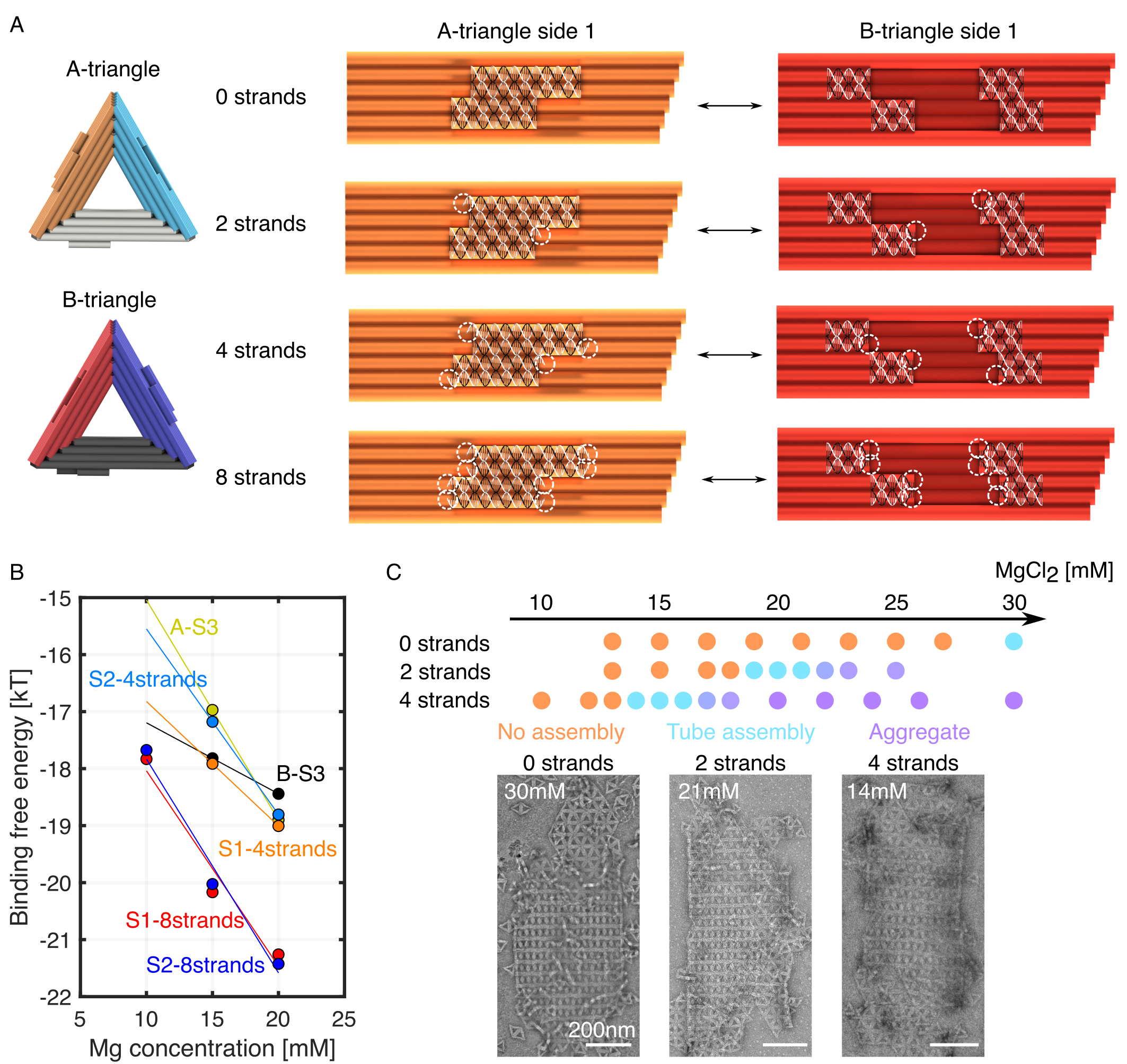}
  \caption{\textbf{Binary species interaction strengths and assembly conditions.}
(A) Schematics of the binary species triangles and the modification of lock-and-key shapes to incorporate hybridization domains for binding side 1 of the A-triangle to side 1 of the B-triangle. All of the DNA is explicitly illustrated for the protrusion on the A-triangle, whereas only DNA close to the hole is illustrated on B-triangle. The phosphate backbone and nucleotides colored in black indicate scaffold DNA, which is unchanged for all cases. Strands colored in white are staples, which are elongated by 3 nucleotides for the A-triangle and shortened by 3 nucleotides for the B-triangle. Locations where staples are elongated or shortened are circled in white. (B) A plot of the binding free energy between different sides for both  triangle species. The number following ``S'' indicates the interacting side. When sides 1 and 2 have less than 2 strands with added hybridization domains, the binding energy is too weak to measure and is there not reported. (C) Assembly conditions for the binary tubules with different numbers of hybridization domains for sides 1 and 2. Example TEM micrographs of tubules for 0, 2, and 4 hybridization domains are shown.} 
 \label{Sfig:binaryIllustration}
\end{figure*}

To overcome the discrepancy in the binding energy between different sides, we enhance the binding strength of sides 1 and 2 by adding hybridization interactions (Fig.~\ref{Sfig:binaryIllustration}A). We intentionally elongate or shorten the staple sequences that are at the end of the protrusions and recesses, so that base pairs can form in addition to the base stacking interactions. Here, we shortened either 2, 4, or 8 strands at the end of the recess of side 1 of the B-triangle, each by 3 nucleotides. At the same time, the corresponding staples on the protrusion of side 1 of the A-triangle were elongated by 3 nucleotides. The sequences of the elongated nucleotides were designed so that they were complementary to the exposed scaffold sequence on B-triangles. We also implemented similar designs for side 2 of the A- and B-triangle to enhance the interaction strength.

We measured the binding strength for the binary species particles, including different variants of sides 1 and 2, using gel electrophoresis (Fig.~\ref{Sfig:binaryIllustration}B). We find that the dimer bands for sides 1 and 2 are not detectable when we use less than 2 hybridization domains. The binding strength of sides 1 and 2 are almost equivalent to side 3 when 4 hybridization domains are used, and much stronger when all 8 domains are used. Note that for the analysis of binding between two different triangles, we use different equations from those discussed in the previous section. Here, the reaction equation is
\begin{equation}
    A + B \rightleftharpoons AB,
\end{equation}
where $AB$ is a dimer complex of the A-triangle and the B-triangle. In the solution, we have total of 10 nM of both the A-triangle and the B-triangle, thus
\begin{eqnarray}
    \left[A\right] + \left[AB\right] &= 10\ \mathrm{nM},   \\ \nonumber
    \left[B\right] + \left[AB\right] &= 10\ \mathrm{nM}.
  \label{Seq:totMonoBinary}
\end{eqnarray}
The free energy is then obtained from the equilibrium concentration of the A and B triangles as
\begin{equation}
    \Delta G = -k_\mathrm{B}T \log{\left(\frac{[AB]u_0}{[A][B]}\right)}.
    \label{Seq:freeEnergyBinary}
\end{equation}

As a result of the different interaction strengths between the variants, they assemble at different Mg$^{2+}$ concentrations. For a system with 4 hybridization domains for both sides 1 and 2, assembly occurs at around 14 mM MgCl$_2$. The average binding strength of the three sides was $-18.06\ k_\mathrm{B}T$, which is very similar to the threshold free energy that we found for the V- and X-tubules. As the number of hybridizing domains was reduced, the Mg$^{2+}$ concentration necessary to assemble tubules increased. For 2 domains, assembly occurred at 20 mM MgCl$_2$, while for 0 domains, it was 30 mM MgCl$_2$.

%%%%%%%%%%%%%%%%%%%%%%%%%%%%%%%%%%%%%%%%%%%%%%%%%%%%%%%%%%%%

\section{Characterization of tubule assemblies}\label{sec:tubeExp}

\subsection{Measuring tubule lengths from fluorescence microscopy}\label{subsec:fluorescence}

\begin{figure*}[tb]
 \centering
 \includegraphics[width=0.99\textwidth]{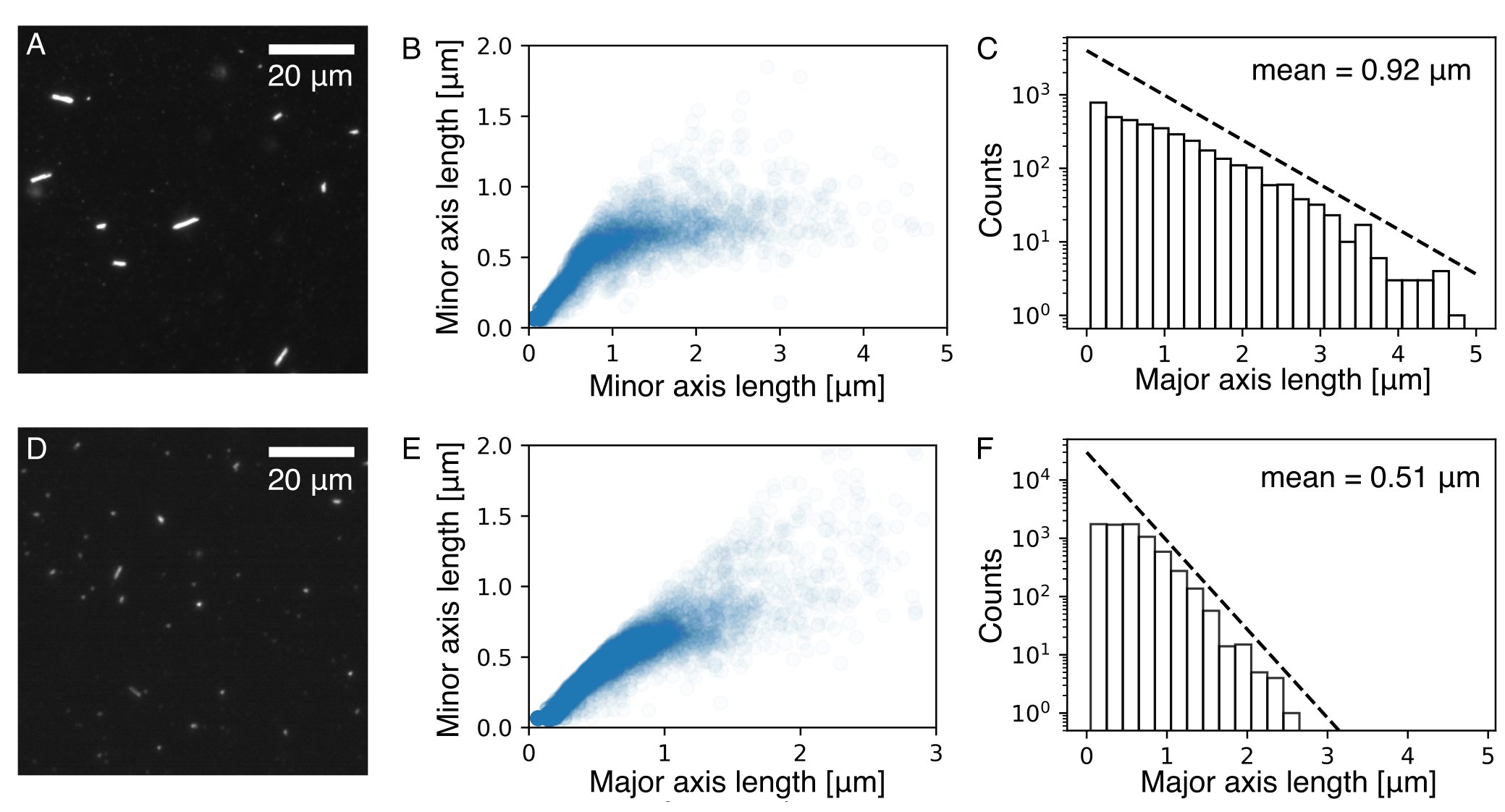}
 \caption{\textbf{X-tubule and V-tubule length distributions.}
 (A) Epi-fluorescence image of X-tubules dyed with YOYO-1 and imaged with a 100x objective. (B) Plot of the major and minor axes lengths of particles found in epi-fluorescence. (C) Histogram of major axis lengths found for 3710 particles. Particles with a minor axis length greater than 1 $\upmu$m are excluded.  (D) Epi-fluorescence image of V-tubules, along with their (E) measured lengths and (F) a histogram for 7362 particles. Dashed lines are guides for the eye to highlight the exponential decay. }
 \label{Sfig:fluorescence}
\end{figure*}

In addition to TEM imaging, we also use fluorescence microscopy to assess the lengths of tubules that form (Fig.~\ref{Sfig:fluorescence}A). To characterize the tubules we perform some simple image analysis. Epi-fluorescence images of tubules are denoised by convolution with a Gaussian whose width is 2 pixels, after which we binarize the image with a single threshold. We use FIJI to extract particles and fit them with ellipses. Looking at the values of the major and minor axes lengths of these fits (Fig.~\ref{Sfig:fluorescence}B) we see a saturation of the minor axis length at about 0.6 $\upmu$m, which is roughly the width of a tubule. Using the value of the major axis length as the length of an assembly, we construct a histogram for the distribution of the lengths, including only particles with a minor axis length less than 1 micrometer to exclude aggregates. Figure~\ref{Sfig:fluorescence}C shows such a histogram. We perform this imaging and analysis for X-tubules (Fig.~\ref{Sfig:fluorescence}A-C) and V-tubules (Fig.~\ref{Sfig:fluorescence}D-F). 

\subsection{Determining tubule handedness from EM tomography}\label{subsec:handedness}
Through tomography reconstruction, we obtain a z-stack of the tubules, which yields their handedness. Ideally, if one has complete information about the electron microscope, the orientation of the grid, the direction of the tilting, and the reconstruction algorithms, the handedness of a tubule can be inferred correctly. However, this task is unfeasible in reality. To overcome this challenge, we measure the chirality of an object with known handedness and compare it with tubules using the same measurement and analysis methods \cite{briegel_challenge_2013}. 

\begin{figure*}[t]
 \centering
 \includegraphics[width=0.8\textwidth]{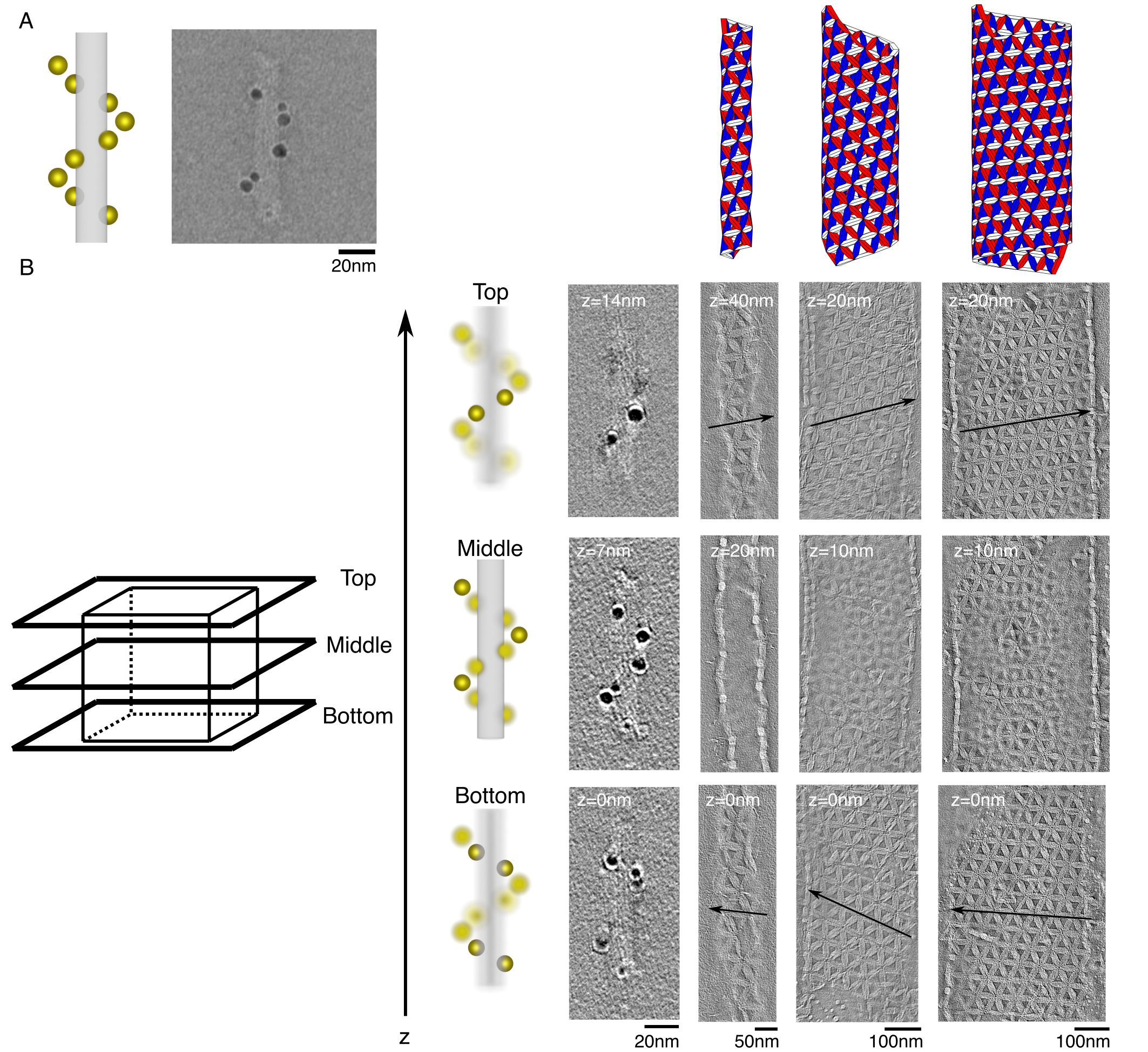}
 \caption{\textbf{Determining the handedness of the tubules from tomography.}
(A) Illustration of a right-handed gold nanohelix and its TEM image. (B) Tomography reconstruction of a right-handed gold nanohelix, a V-tubule, an X-tubule, and a binary species tubule. Z-slices from the top, middle, and bottom of each structure are shown. The order of sequences are all adjusted in the same way, so that the right-handed gold nanohelix appears right-handed. The black arrows indicate the direction of the seam along side 3. Above the TEM images of tubules, illustrations of the assembled tubules are also shown, which corresponds to (4,1), (9,4), and (14,2) tubules, for V-, X-, and binary tubules, respectively. The height of each image with respect to the position of the bottom image is labeled.}
 \label{Sfig:handedness}
\end{figure*}

Here, we use a right-handed gold nanohelix to determine the handedness of the tubules \cite{kuzyk_dna-based_2012,briegel_challenge_2013}. The gold nanohelix (Gattaquant DNA Nanotechnologies) is a 24 helix bundle DNA-origami rod with 9 gold nanoparticles arranged helically (Fig.~\ref{Sfig:handedness}(A)). Though the TEM image we obtain only contains 7 gold nanoparticles, it is still sufficient to determine the handedness. Through tomography reconstruction, we obtain z-stacks of a right-handed gold nanohelix and the tubules. In Fig.~\ref{Sfig:handedness}(B), the z-stack of the gold nanohelix is ordered so that it appears right-handed, as prescribed. The V-, X-, and binary species tubules in Fig.~\ref{Sfig:handedness}(B) are ordered in the same way as the gold nanohelix, to preserve the correct handedness. In all of the tubules shown, the seam of side 3 is right-handed. In total, we observe 26 V-tubules of which 8 are achiral and 14 are right-handed, 15 X-tubules which are all right-handed, and 9 binary species tubules of which 2 are achiral and 7 are right-handed.

\subsection{Determining tubule types from negative-stain TEM}\label{subsec:classification}

For all of the tubules we discussed, we find that there are distributions of assembled tubule types. The fundamental idea behind classification is illustrated in Fig.~\ref{Sfig:anglePlot}B. Different tubule types can be distinguished from the radius and the seam angle of the tubule. However, we use slightly different classification methods for V-, X-, and binary tubules. This choice is because they have very different widths and seam angles compared to each other. For example, V-tubules have small lattice numbers, which makes it easy to distinguish the tubule types. It is rather difficult to measure the width and seam angles to characterize these tubules since they vary greatly along the tubule axis. On the other hand, X-tubules are larger, and they lay close to flat on the TEM grid surface. This makes them suitable for measurement simply using the width and the seam angles as in Fig.~\ref{Sfig:anglePlot}B.

\subsubsection{TEM image simulation}
\begin{figure*}[htb]
\centering
\includegraphics[width=0.9\textwidth]{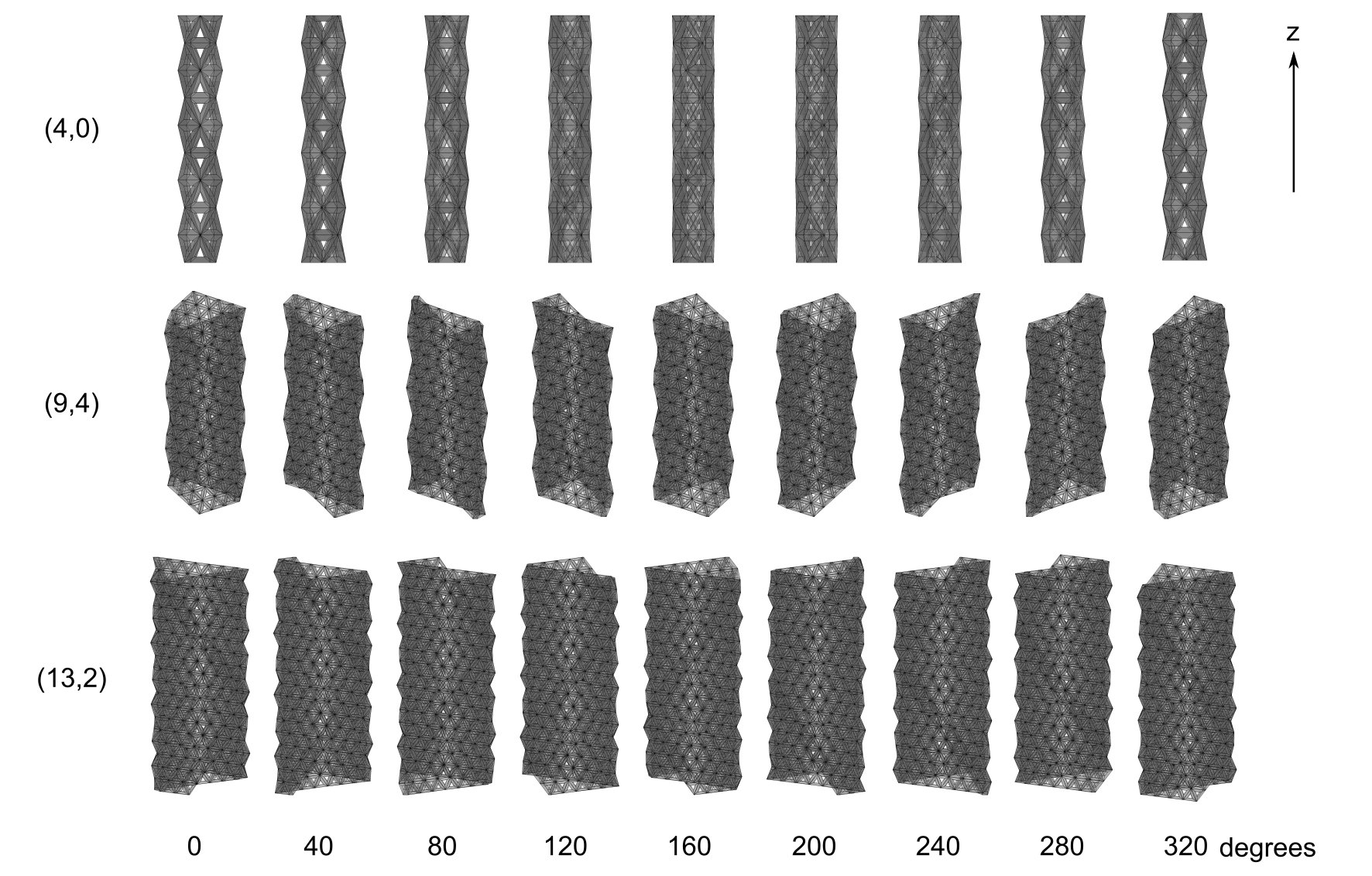}
\caption{\textbf{Simulations of TEM micrographs for representative tubule types.} From top to bottom, (4,0), (9,4), and (13,2) tubules are illustrated. The angles denote the rotation of the tubule about the z-axis. The method we used to generate these images is described in the SI text.
}
\label{Sfig:TEMsimulation}
\end{figure*}

We first generate semi-transparent representation of tubules to generate Moir\'e patterns for comparison against TEM micrographs. Using the method described in Supplementary Section 2, coordinates of triangles can be calculated, given the lattice numbers $m$ and $n$ of a tubule. Triangles are modeled using transparent patches whose dimensions are in proportion to the original design. We assume the length of a side of a triangle is 56 nm, whereas the height and width are 6 by 4 helices or 15 nm by 10 nm. The transparent triangles allow us to see both the top and the bottom of the triangles overlapping with each other, just as in TEM. The simulation results for representative tubules are shown in Fig.~\ref{Sfig:TEMsimulation}. In Fig.~2C, we observe that the cross-section of V-tubules are ellipse on TEM grid, whereas X-tubules are flattened. We reflect this feature in our simulation. In Fig.~\ref{Sfig:TEMsimulation}, the (4,0) tubule is round, but the (9,4) and (13,2) tubules are flattened before image generation. For the (9,4) and (13,2) tubules, we observe that the edges of the tubules are staggered, unlike those from experimental images. Although this simulation may not be accurate near the edges, we argue that the characteristics of the Moir\'e patterns, which are generated by overlapping two triangular lattices that are mirror-symmetric to each other, are well-represented in the simulations. We also find that the Moir\'e patterns do not change depending on the orientation that the tubule lands on the grid when the tubules are chiral. Unfortunately, the handedness of the tubules cannot be determined solely from the Moir\'e pattern. Since the top and bottom layers of the tubules are mirror symmetric, the same pattern appears for tubules with opposite handedness.

\subsubsection{Classification of V-tubules}
The width of V-tubules in the experiment are relatively small, with circumference ranging from 3 to 5 triangles. Since the dihedral angles between two triangles are small, tubules appear bumpy and inconsistent. This makes the determination of the tubule type by measurement of the width and the seam angle less reliable. Thus, we characterize the V-tubules by pattern matching by eye. The distinction between achiral and chiral tubules are not difficult; the angle along side 3, or seam angle, is perpendicular against tubule axis for achiral case, whereas for (4,1) tubule, it is further tilted by about 10 degrees. Within achiral tubules, (3,0), (4,0), and (5,0) can be distinguished by the width. (4,0) tubules typically exhibit the trademark hexagonal pattern, while (3,0) is narrower and (5,0) would be wider. Similarly, for achiral case, when the tubule width is around one hexagonal pattern wide, the tubule has an $m$ number of 4. Different $n$ numbers can further be categorized by the seam angle, which varies by around 10 degrees for a difference of 1. %To help the intuition, the FIGURE?? shows the TEM simulation for different tubule types. %The examples of pattern matching using TEM micrographs are shown in FIGURE??.

\subsubsection{Classification of X-tubules and binary tubules}
On the contrary, X-tubules and binary species tubules can be classified by measuring the width and the seam angle. Because the tubule is flattened on the grid, as confirmed from the tomography reconstruction (Fig.~2C), the circumference of the tubule can be inferred approximately as twice the width of the tubule on grid. Combined with the measurement of seam angle, we obtain the lattice number of the tubule, which we define as the tubule state with the closest circumference and seam angle to the measured values. %(FIGURE). 
To find the closest tubule state, we define 'distance' as
\begin{equation}
    d = \sqrt{\frac{(\theta_{(m,n)} - \theta)^2}{\theta^2} + \frac{(c_{(m,n)} - c)^2}{c^2}},
\end{equation}
where $\theta_{(m,n)}$ and $c_{(m,n)}$ is the seam angle and circumference of an ideal tubule state ${(m,n)}$, respectively, and $\theta$ and $c$ is the measured seam angle and circumference, respectively. The observed tubule is classified into a lattice number $(m,n)$ whose distance is the shortest.

Additionally, in the case of binary tubules, Moir\'e pattern of the tubules provide more accurate and convenient estimation of $n$ numbers. %FIGURE?? shows TEM simulations of different tubule types in the range that binary tubes can assemble. 
Surprisingly, $n=$0 and 2 tubes within this range has unique Moir\'e patterns which allows for a straightforward classification. $n=0$ tubules have parallel seam on the two sides of the tubules, whereas $n=2$ tubules shows distinct hexagonal pattern as shown in Fig. \ref{Sfig:TEMsimulation}. On the other hand, $n=1$ or $n$ higher than 3 states do not exhibit Moir\'e pattern that stands out. In the case we are able to identify the Moir\'e pattern, we decouple the lattice number classification by obtaining $n$ from Moir\'e pattern and $m$ from the circumference. It should be noted that not all tubules benefit from this Moir\'e pattern classification due to staining inconsistencies; when the layer of staining is thin, we sometimes only observe one layer of the triangle, and thus no Moir\'e pattern. Nevertheless, we emphasize that we only observe tubules with even $n$ numbers.

%%%%%%%%%%%%%%%%%%%%%%%%%%%%%%%%%%%%%%%%%%%%%%%%%%%%%%%%%%%%

\section{Determination of bevel angles from cryo-EM maps}\label{sec:PAM}

A key aspect of the construction of tubules is the ability to have different bevel angles for each side of the triangular monomers. To supplement the TEM measurements of the tubule types that form, we also perform quantitative measurements on the cryo-EM reconstructions we have taken for each monomer type to see how close they are to our designed structures. To measure the bevel angles we fit pseudo-atomic models (PAM) to our reconstructions using the methods in ref.~\cite{kube2020revealing} implemented in https://github.com/elija-feigl/dnaFit. In brief, this method (i) takes the design file from caDNAno and simulates an atomic model of the structure using mrdna~\cite{maffeo2020mrdna}, it then (ii) uses the electron density of the cryo-EM reconstruction as a potential to relax the atomic model into a position that fits the cryo-EM map. This allows one to choose helices and base pairs from the design and find their coordinates in the cryo-EM map.

\begin{figure*}[t]
 \centering
 \includegraphics[width=0.9\textwidth]{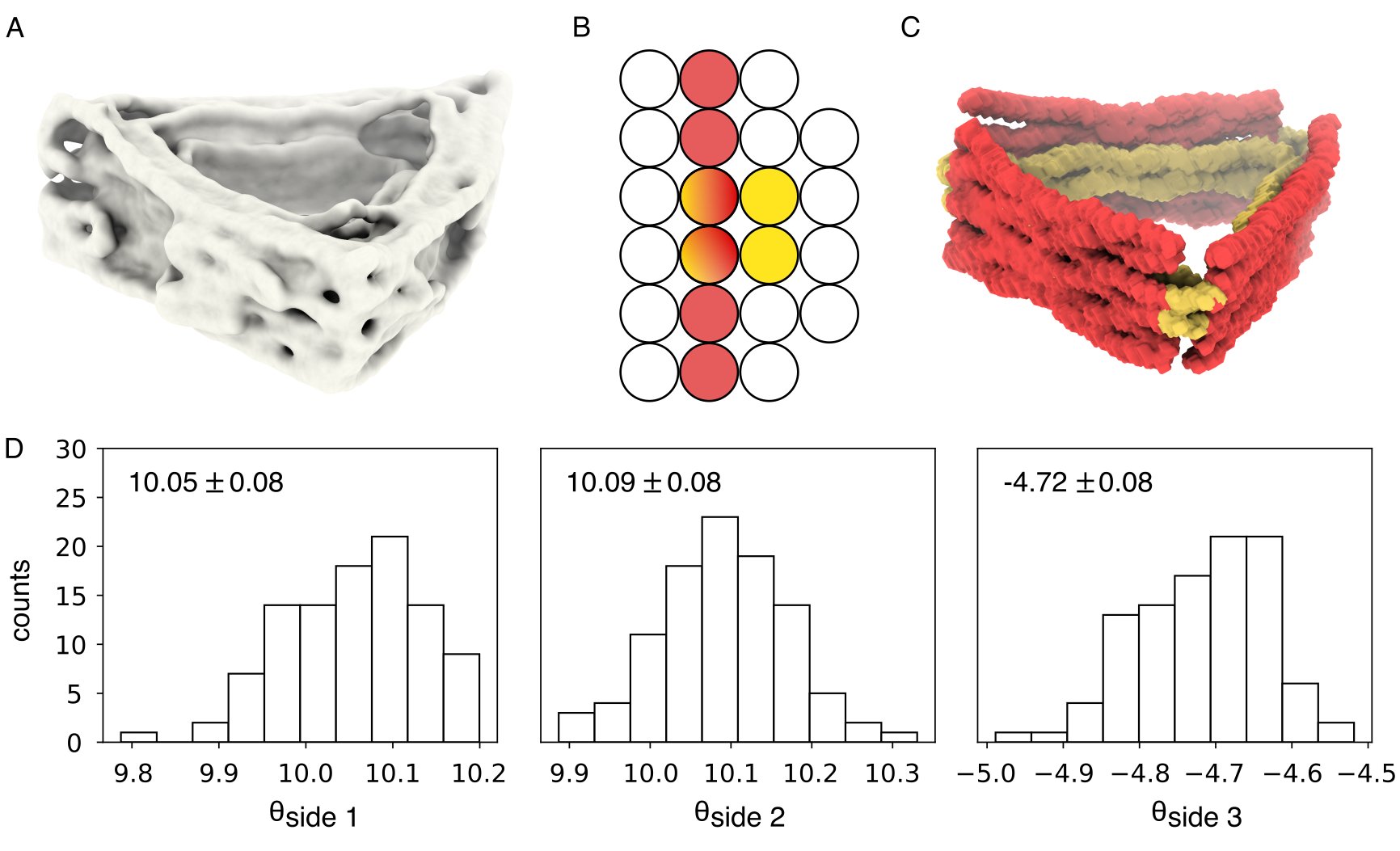}
 \caption{\textbf{Fitting cryo-EM maps to determine the bevel angles.}
(A) Cryo-EM reconstruction of the B-triangle. (B) Schematic of the cross-section of the triangle design. The four inner helices (colored yellow) are used to define the inner core, while the second layer of helices (colored red) is used to define the plane of each side. (C) Extracted portions of the cryo-EM map that are used to fit the planes. Red sections are the second layer of the sides and the yellow section is the inner core, to define the orientation of the monomer.  (D) Histograms of the fit angles for each side from the bootstrap method for 100 iterations. Average and standard deviations are shown for each histogram.}
 \label{Sfig:bevelangle}
\end{figure*}

The process of fitting the bevel angles is shown in Fig.~\ref{Sfig:bevelangle}. Given a cryo-EM map (Fig.~\ref{Sfig:bevelangle}A), we can generate a PAM from which we extract coordinates. To get the angle of all three sides we will need to find four distinct planes: a plane to define the orientation of the monomer and a plane for the orientation of each side. Figure~\ref{Sfig:bevelangle}B shows our choice of helices to define the monomer and side orientations: the inner four helices (yellow) are used for the monomer and the six helices of the second layer (red) are used for each side. The extracted coordinates from the PAM are shown in Fig.~\ref{Sfig:bevelangle}C.

After extracting the coordinates we can fit a plane to each selection. Each plane can be used to generate a normal vector, $\textbf{n}$. Taking the normal for the monomer and a side we can get the bevel angle from the following equation
\begin{equation}
    \cos(\pi/2 - \theta_\mathrm{side}) = \textbf{n}_{\mathrm{side}} \cdot \textbf{n}_\mathrm{monomer}
\end{equation}
where $\theta_\mathrm{side}$ is the deviation of the side normal being perpendicular to the normal of the monomer. To estimate the uncertainty of these measurements we use a bootstrap method. We perform measurements of the bevel angles from 100 random subsets of half of the positions in each selection from the PAM. Histograms of these values are shown in Fig.~\ref{Sfig:bevelangle}D. In general, our uncertainty is of the order 0.1 degrees. Measured bevel angles for all monomers are shown in Table~\ref{STable:bevelCryo}. We note that the uncertainty written here does not reflect any thermal fluctuations that may occur.

%%%%%%%%%%%%%%%%%%%%%%%%%%%%%%%%%%%%%%%%%%%%%%%%%%%%%%%%%%%%

\section{Estimate of bending rigidity}\label{sec:helfrich}

Using the widths distribution of Fig.~3B we make an estimate of the intersubunit bending energy. The Helfrich energy goes as $E=\frac{1}{2}BA(\Delta\kappa_\perp)^2$ where $A$ is the area of the assembly, $B$ is the bending rigidity, and $\Delta\kappa_\perp$ is the fluctuation of the curvature in the circumference direction. Based on prior research~\cite{videbaek2021tiling} we assume that the tubule type is frozen in at the point that the growing triangular sheet first closes on itself, forming a tubule. If the binding energies are similar, we can assume that the assembly will grow isotropically to form a patch-like disk. At the point of closure, we assume this disk just closes and will have an area of $A=\pi^3 R^2$, where $R$ is the radius of curvature of the tubule. We estimate the fluctuation of the curvature as
\begin{eqnarray}
    (\Delta \kappa_\perp)^2 &=& \left( \frac{1}{R}- \frac{1}{R+\Delta R} \right)^2 \approx \frac{1}{R^2} \left( \frac{\Delta R}{R}\right)^2.
\end{eqnarray}
In our experiment we measure the circumference, $C$, and its fluctuations; we note that $(\Delta C/C) = (\Delta R/R)$. Combining these two we get that the Helfrich energy is
\begin{equation}
E_\mathrm{H} = \frac{1}{2}\pi^3 B \left( \frac{\Delta C}{C} \right)^2
\label{eqn:EHelfrich}
\end{equation}
We then assume that the widths are distributed with a Boltzmann probability $P=\exp(-E_\mathrm{H}/k_\mathrm{B}T)/Z$, where $Z$ is the partition function. By measuring the variance of the circumference distribution and assuming Gaussian fluctuations we can get an estimate for $B$ as
\begin{equation}
    B = \frac{1}{\pi^3 }\frac{\langle C \rangle^2}{\langle C^2 \rangle}.
\end{equation}
Using this method we find the following bending rigidities: $1.4\ k_\mathrm{B}T$ for the V-tubule, $2.0\ k_\mathrm{B}T$ for the X-tubule, and $3.8\ k_\mathrm{B}T$ for the binary tubule.

To get a sense of the ($m,n$) distribution that would be expected from Gaussian fluctuations, as well as get a second estimate of the bending rigidity, we consider a simple elastic model for the energy of our assembly. Each side of our triangle has a preferred binding angle with its neighbors and deviations from this angle will result in some bending energy. For an assembly consisting of $N$ subunits, the elastic energy is
\begin{equation}
    E = \frac{1}{4}NB \sum_{i\in 1,2,3} (\theta_i - \theta_{0,i})^2
\end{equation}
where $B$ is the bending rigidity, $\theta_{0,i}$ is the preferred binding angle, and $\theta_i$ is the assembly binding angle. Using the same estimate of the assembly area at closure we can estimate $N$ for any given tubule type. Dividing the disk area by the size of a triangle gives $N=(4\pi/\sqrt{3})(C/l_0)^2$, where $l_0$ is the edge length of a subunit and $C$ is the circumference of the tubule. We also assume that the tubule distribution follows a Boltzmann distribution of the elastic energy, $P(m,n)=\exp(-E(m,n)/k_\mathrm{B}T)/Z$, where both the angles and $N$ depend on the tubule type and $Z$ is the partition function.

Using this elastic model we can fit our experimental distributions to a target ($m,n$) and bending rigidity $B$. The fits give targets of (4,0) and $B=1.1\ k_\mathrm{B}T$ for V-tubules, (9,4) and $B=3.0\ k_\mathrm{B}T$ for X-tubules, and (13,2) and $B=3.1\ k_\mathrm{B}T$ for binary tubules. These bending rigidity values are consistent with the estimate from the Helfrich model. In the main text, we showed contours corresponding to the model distributions overlaid on the experiment distributions in Fig. 3A and 4E; these come from the best fit target ($m,n$) given here and the average value of $B=1.8\ k_\mathrm{B}T$ for the V- and X-tubules and $B=3.8\ k_\mathrm{B}T$ for the binary tubule.

\begin{figure*}[!ht]
 \centering
 \includegraphics[width=0.7\textwidth]{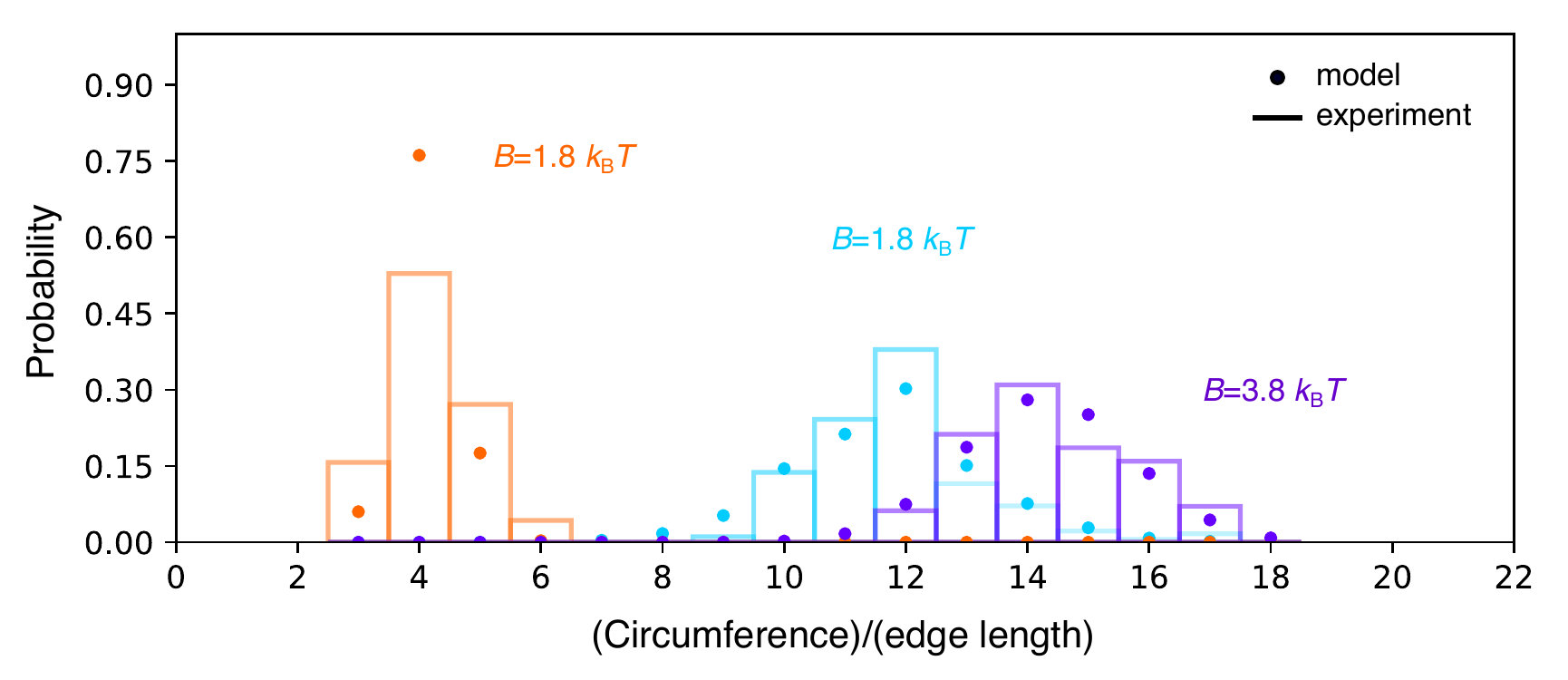}
 \caption{\textbf{Estimation of the bending rigidity using the Helfrich energy.}
Circumference distributions, shown as histograms, for the experimental V-, X-, and binary tubules are shown in orange, cyan, and purple respectively. Each distribution has a corresponding elastic energy model comparison, shown by the circles. The bending rigidity used to generate the model distributions comes from the Helfrich model estimate. For the V-tubule and X-tubule the $B$ estimate of 1.8 $k_\mathrm{B}T$ comes from the average value of their respective Helfrich model estimates. The binary tubule value of $3.8\ k_\mathrm{B}T$ comes from the experimental distribution of the binary tubules.}
 \label{Sfig:modelcomp}
\end{figure*}

%%%%%%%%%%%%%%%%%%%%%%%%%%%%%%%%%%%%%%%%%%%%%%%%%%%%%%%%%%%%

\setlength{\tabcolsep}{9pt}

\begin{table}[tbh]
\centering
\begin{tabular}{c c c c c}
Object & Concentration [nM] & \# of refined particles & Dose (e/A$^2$) & Resolution of final 3D map [A] \\ \hline \\
V-triangle & 90 & 1228 & 25-35 & 23.48 \\
X-triangle & 170 & 3075 & 25-35 & 23.48 \\
A-triangle & 300 & 3627 & 25-35 & 19.97 \\
B-triangle & 150 & 1636 & 25-35 & 24.03 \\
\end{tabular}
\caption{\textbf{Summary of cryo-EM reconstructions.} For each DNA origami structure used we have 3D reconstructions from cryo-EM. Here are listed the parameters summarizing each structure.}
\label{STable:cryoconditions}
\end{table}

\begin{table}[tbh]
\centering
\begin{tabular}{c c c c c c c}
Object & measurement & $\theta_{\mathrm{side}\ 1}$ & $\theta_{\mathrm{side}\ 2}$ & $\theta_{\mathrm{side}\ 3}$ & Estimated  & Experimental \\ 
 & & & & & (\textit{m}, \textit{n}) & (\textit{m}, \textit{n}) \\

\hline \\
V-triangle & design & $20.9$ & $20.9$ & $-10.8$ & (5,0) & (4,0) \\
% & surface (cryo-EM) & $19.9\pm1.8$ & $15.9\pm1.3$ & $-4.5\pm1.5$ & (6,1) &  \\
 & 2$^{\mathrm{nd}}$ layer (PAM) & $21.3\pm0.1$ & $21.3\pm0.1$ & $-5.5\pm0.1$ & (5,0) &  \\
\hline \\

X-triangle & design & $10.4$ & $10.4$ & $-5.3$ & (10,0) & (9,4) \\
% & surface (cryo-EM) & $9.4\pm1.7$ & $3.6\pm1.3$ & $0.0\pm1.4$ & (12,6) &  \\
 & 2$^{\mathrm{nd}}$ layer (PAM) & $11.4\pm0.1$ & $9.0\pm0.1$ & $-6.4\pm0.1$ & (9,-1) &  \\
\hline \\

A-triangle & design & $10.4$ & $10.4$ & $-5.3$ & (10,0) &  \\
% & surface (cryo-EM) & $6.7\pm1.0$ & $4.8\pm1.2$ & $-2.3\pm0.9$ &  &  \\
 & 2$^{\mathrm{nd}}$ layer (PAM) & $7.2\pm0.1$ & $4.2\pm0.1$ & $-0.4\pm0.1$ &  &  \\
\hline \\

B-triangle & design & $10.4$ & $10.4$ & $-5.3$ & (10,0) & \\
% & surface (cryo-EM) & $6.6\pm1.4$ & $6.9\pm0.9$ & $-6.4\pm1.0$ & &  \\
 & 2$^{\mathrm{nd}}$ layer (PAM) & $9.9\pm0.1$ & $9.7\pm0.1$ & $-5.0\pm0.1$ &  &  \\

%2-species combined & surface (cryo-EM) & $6.7\pm1.2$ & $5.9\pm1.1$ & $-4.4\pm1.0$ & (15,1) & (14,1) \\
2-species combined & 2$^{\mathrm{nd}}$ layer (PAM) & $8.6\pm0.1$ & $7.0\pm0.1$ & $-2.7\pm0.1$ & (13,-1) & (13,2) \\

\end{tabular}
\caption{\textbf{Summary of designed and measured bevel angles.} This table shows the bevel angles for the different triangle designs used in this research. The "measurement" column denotes if the following quantities come from the design, from surface estimates of the cryo-EM maps, or from fits to pseudo-atomic model (PAM). $\theta$ denotes the bevel angle of a given side. Estimated (\textit{m,n}) shows which tubule type has the minimum deviation of angles from the given bevel angles. Experimental (\textit{m,n}) is the tubule type that most closely matches the observed tubule distributions.}
\label{STable:bevelCryo}
\end{table}

%%%%%%%%%%%% BEGIN CRYO-EM FIGURES %%%%%%%%%%%%%%%%%%%%%%

\begin{figure*}[htb]
 \centering
 \includegraphics[width=0.99\textwidth]{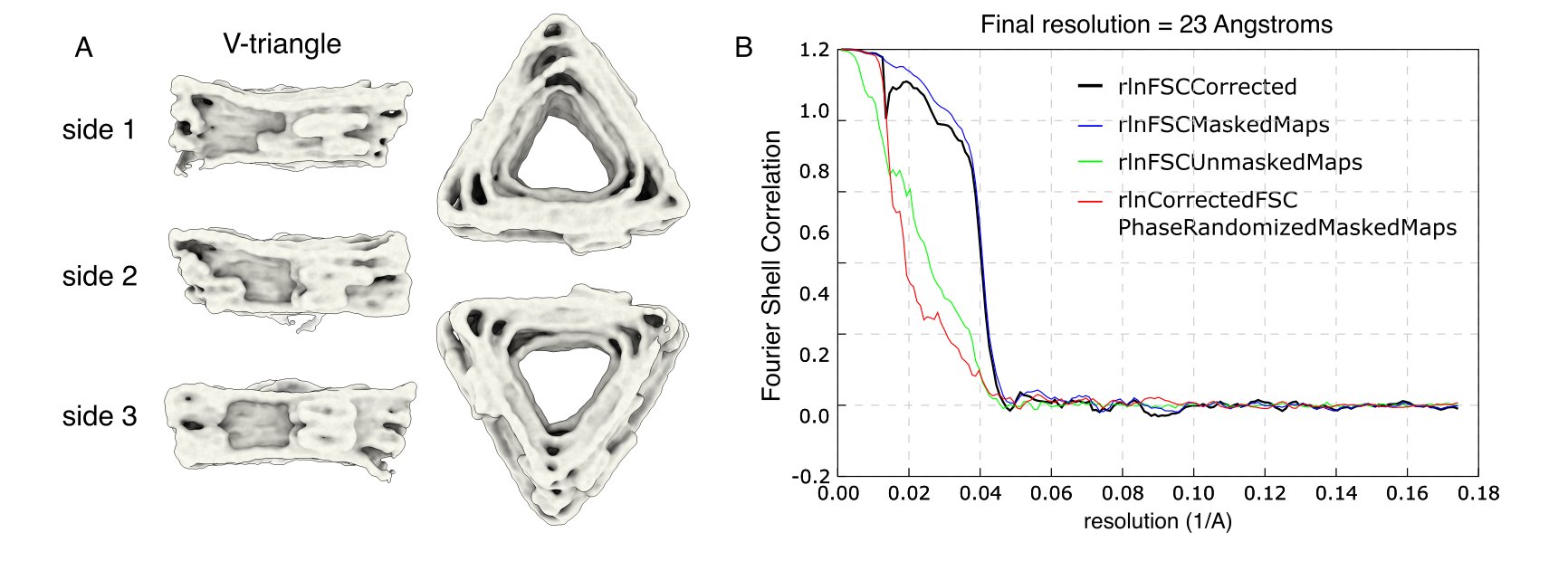}
 \caption{\textbf{Cryo-EM reconstruction of the V-triangle.}
(A) Electron density maps of the V-triangle shown from different viewing angles. (B) Plot of the FSC curves used to estimate the resolution for the V-triangle.}
 \label{Sfig:cryo-EM-V}
\end{figure*}

\begin{figure*}[htb]
 \centering
 \includegraphics[width=0.99\textwidth]{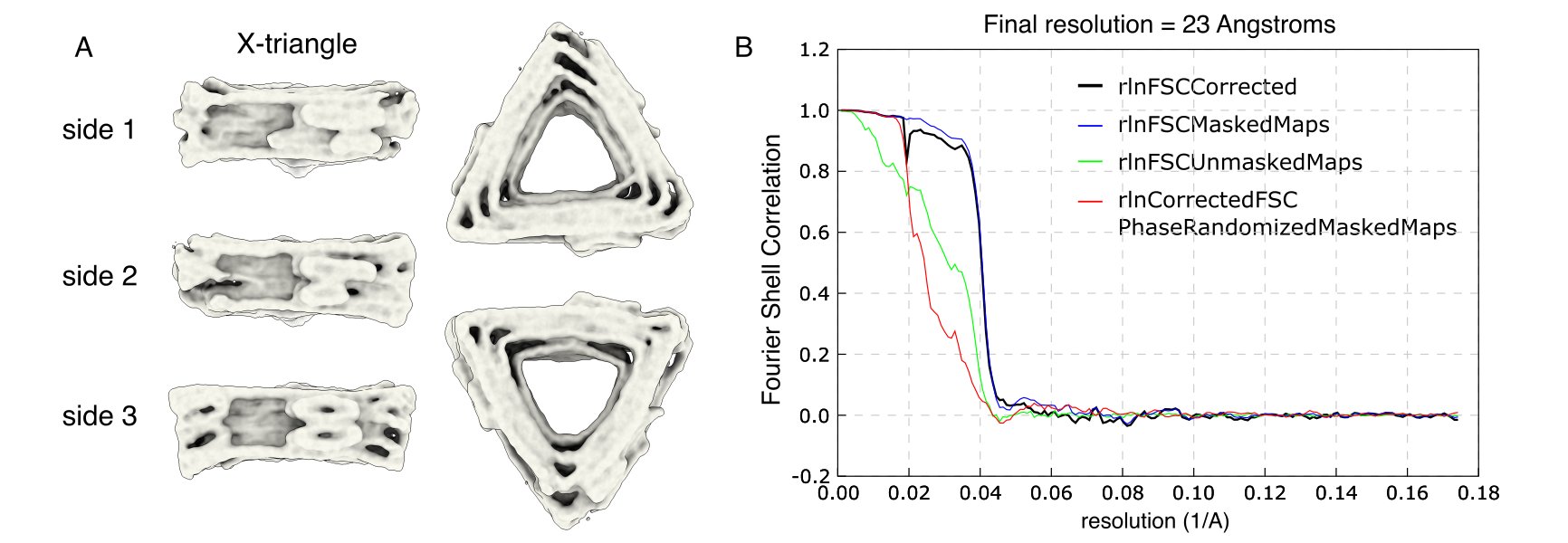}
 \caption{\textbf{Cryo-EM reconstruction of the X-triangle.}
(A) Electron density maps of the X-triangle shown from different viewing angles. (B) Plot of the FSC curves used to estimate the resolution of the X-triangle.}
 \label{Sfig:cryo-EM-X}
\end{figure*}

\begin{figure*}[htb]
 \centering
 \includegraphics[width=0.99\textwidth]{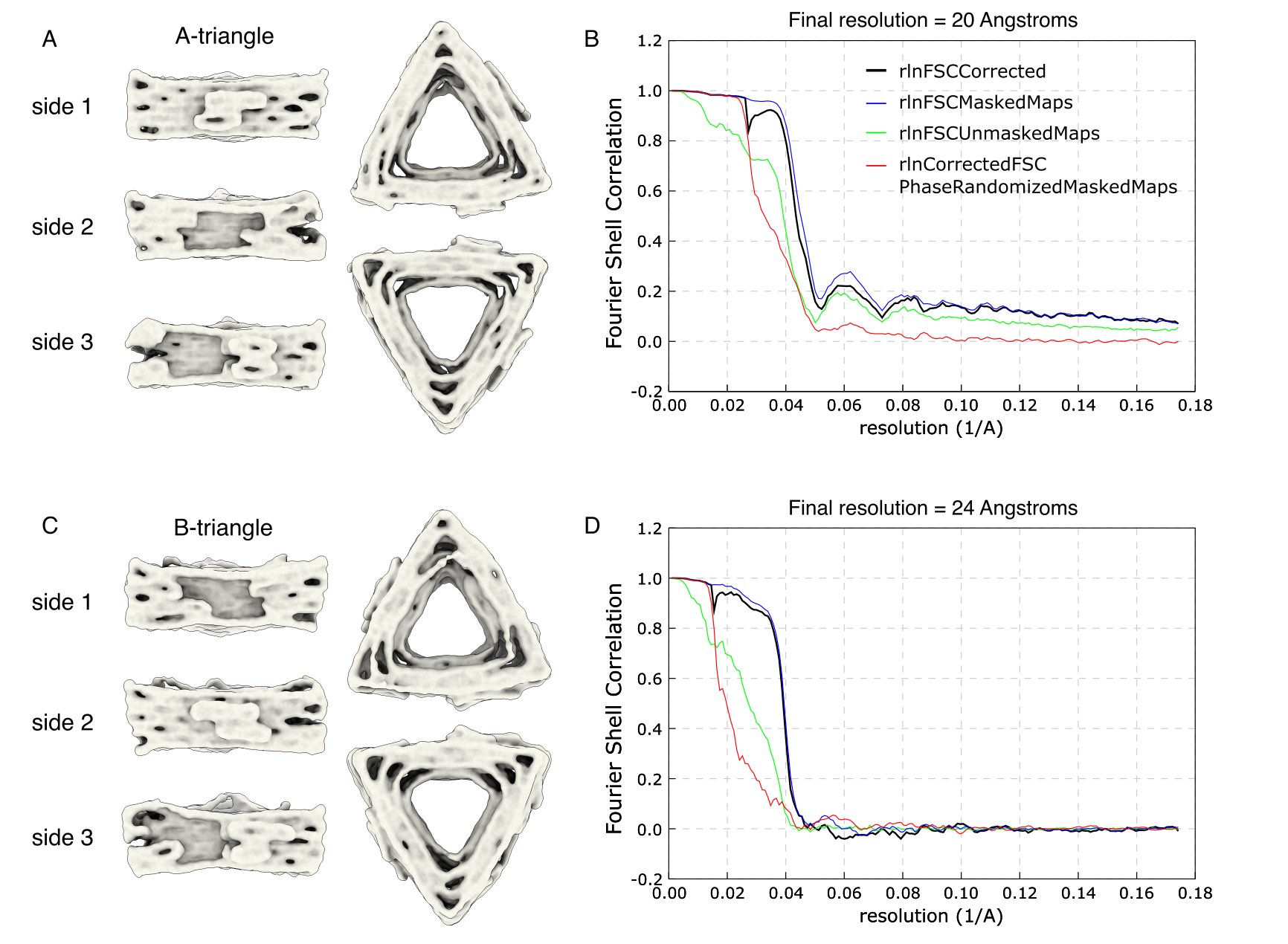}
 \caption{\textbf{Cryo-EM reconstruction of the 2-species X-triangles.}
(A) Electron density maps of the A-triangle shown from different viewing angles. (B) Plot of the FSC curves used to estimate the resolution of the A-triangle. (C) Electron density maps of the B-triangle shown from different viewing angles. (D) Plot of the FSC curves used to estimate the resolution of the B-triangle.}
 \label{Sfig:cryo-EM-2J}
\end{figure*}

%\bibliography{supp.bib}

%\bibliography{main.bib}

\clearpage

%%%%%%%%%%% tubule type table %%%%%%%%%%%%%%
\begin{longtable}{c c c c c c c c}
m &  n &  type &  Side1 &  Side2 &  Side3 &  Tube Radius & Seam Angle \\ 
\hline 
\endfirsthead

m &  n &  type &  Side1 &  Side2 &  Side3 &  Tube Radius & Seam Angle\\ 
\hline 
\endhead

\endfoot

\endlastfoot
3 &  0 &  armchair &  70.5288 &  70.5288 &  -38.9424 &  0.57735 &  90 \\ 
3 &  1 &  chiral &  34.8794 &  76.289 &  -25.2939 &  0.64526 &  76.4267 \\ 
3 &  2 &  chiral &  11.4318 &  70.0346 &  -10.0106 &  0.74313 &  66.6665 \\ 
3 &  3 &  zigzag &  0 &  60 &  0 &  0.86603 &  60 \\ 
4 &  0 &  armchair &  52.4484 &  52.4484 &  -27.6723 &  0.70711 &  90 \\ 
4 &  1 &  chiral &  30.8016 &  57.7027 &  -20.8118 &  0.78561 &  79.2816 \\ 
4 &  2 &  chiral &  15.0912 &  55.8337 &  -12.0981 &  0.88462 &  70.9921 \\ 
4 &  3 &  chiral &  5.4309 &  50.7492 &  -4.9378 &  1.0019 &  64.7355 \\ 
4 &  4 &  zigzag &  0 &  45 &  0 &  1.1315 &  60 \\ 
5 &  0 &  armchair &  41.8103 &  41.8103 &  -21.6246 &  0.85065 &  90 \\ 
5 &  1 &  chiral &  27.3046 &  46.0569 &  -17.5731 &  0.93259 &  81.1551 \\ 
5 &  2 &  chiral &  16.0536 &  45.862 &  -12.0747 &  1.0312 &  73.9826 \\ 
5 &  3 &  chiral &  8.2454 &  43.2505 &  -6.9803 &  1.1444 &  68.2498 \\ 
5 &  4 &  chiral &  3.1803 &  39.6875 &  -2.9532 &  1.2689 &  63.6781 \\ 
5 &  5 &  zigzag &  0 &  36 &  0 &  1.4013 &  60 \\ 
6 &  0 &  armchair &  34.7781 &  34.7781 &  -17.7989 &  1 &  90 \\ 
6 &  1 &  chiral &  24.4024 &  38.1678 &  -15.159 &  1.0832 &  82.4765 \\ 
6 &  2 &  chiral &  15.9669 &  38.6429 &  -11.4411 &  1.1808 &  76.1687 \\ 
6 &  3 &  chiral &  9.6541 &  37.3291 &  -7.7299 &  1.2909 &  70.9337 \\ 
6 &  4 &  chiral &  5.175 &  35.1086 &  -4.5289 &  1.4112 &  66.6031 \\ 
6 &  5 &  chiral &  2.0912 &  32.558 &  -1.9682 &  1.5392 &  63.0079 \\ 
6 &  6 &  zigzag &  0 &  30 &  0 &  1.673 &  60 \\ 
7 &  0 &  armchair &  29.7777 &  29.7777 &  -15.1445 &  1.1524 &  90 \\ 
7 &  1 &  chiral &  21.9995 &  32.5095 &  -13.3042 &  1.236 &  83.4572 \\ 
7 &  2 &  chiral &  15.4541 &  33.2418 &  -10.6588 &  1.3324 &  77.8349 \\ 
7 &  3 &  chiral &  10.2954 &  32.6383 &  -7.8818 &  1.44 &  73.0418 \\ 
7 &  4 &  chiral &  6.4037 &  31.2534 &  -5.3381 &  1.5568 &  68.9691 \\ 
7 &  5 &  chiral &  3.5464 &  29.4859 &  -3.1733 &  1.6811 &  65.5047 \\ 
7 &  6 &  chiral &  1.4808 &  27.5905 &  -1.4067 &  1.8112 &  62.5447 \\ 
7 &  7 &  zigzag &  0 &  25.7143 &  0 &  1.9459 &  60 \\ 
8 &  0 &  armchair &  26.0376 &  26.0376 &  -13.189 &  1.3066 &  90 \\ 
8 &  1 &  chiral &  19.996 &  28.2713 &  -11.8409 &  1.3903 &  84.2133 \\ 
8 &  2 &  chiral &  14.7784 &  29.081 &  -9.8827 &  1.4856 &  79.1461 \\ 
8 &  3 &  chiral &  10.508 &  28.8779 &  -7.7527 &  1.5909 &  74.738 \\ 
8 &  4 &  chiral &  7.1397 &  28.0234 &  -5.7144 &  1.7049 &  70.9154 \\ 
8 &  5 &  chiral &  4.5469 &  26.7963 &  -3.8979 &  1.8259 &  67.6008 \\ 
8 &  6 &  chiral &  2.5811 &  25.3912 &  -2.3465 &  1.9528 &  64.7196 \\ 
8 &  7 &  chiral &  1.104 &  23.934 &  -1.056 &  2.0843 &  62.2052 \\ 
8 &  8 &  zigzag &  0 &  22.5 &  0 &  2.2195 &  60 \\ 
9 &  0 &  armchair &  23.1338 &  23.1338 &  -11.6859 &  1.4619 &  90 \\ 
9 &  1 &  chiral &  18.3088 &  24.9873 &  -10.6601 &  1.5455 &  84.8137 \\ 
9 &  2 &  chiral &  14.0578 &  25.7944 &  -9.1641 &  1.6398 &  80.2044 \\ 
9 &  3 &  chiral &  10.4774 &  25.8206 &  -7.4941 &  1.7433 &  76.1307 \\ 
9 &  4 &  chiral &  7.5564 &  25.3093 &  -5.8424 &  1.8548 &  72.5412 \\ 
9 &  5 &  chiral &  5.2252 &  24.4584 &  -4.3174 &  1.973 &  69.3804 \\ 
9 &  6 &  chiral &  3.3914 &  23.4156 &  -2.9677 &  2.0969 &  66.5938 \\ 
9 &  7 &  chiral &  1.9623 &  22.2839 &  -1.8054 &  2.2255 &  64.1306 \\ 
9 &  8 &  chiral &  0.85495 &  21.1312 &  -0.82207 &  2.3579 &  61.9457 \\ 
9 &  9 &  zigzag &  0 &  20 &  0 &  2.4936 &  60 \\ 
10 &  0 &  armchair &  20.8135 &  20.8135 &  -10.4933 &  1.618 &  90 \\ 
10 &  1 &  chiral &  16.8731 &  22.3729 &  -9.6889 &  1.7015 &  85.3018 \\ 
10 &  2 &  chiral &  13.3463 &  23.1424 &  -8.516 &  1.7948 &  81.076 \\ 
10 &  3 &  chiral &  10.3088 &  23.3001 &  -7.1813 &  1.8967 &  77.2937 \\ 
10 &  4 &  chiral &  7.7644 &  23.0145 &  -5.8272 &  2.006 &  73.9176 \\ 
10 &  5 &  chiral &  5.6752 &  22.4284 &  -4.5415 &  2.1218 &  70.9073 \\ 
10 &  6 &  chiral &  3.983 &  21.6541 &  -3.3707 &  2.2431 &  68.2218 \\ 
10 &  7 &  chiral &  2.6252 &  20.7746 &  -2.3336 &  2.369 &  65.822 \\ 
10 &  8 &  chiral &  1.5422 &  19.8484 &  -1.432 &  2.4989 &  63.6723 \\ 
10 &  9 &  chiral &  0.68173 &  18.915 &  -0.65822 &  2.632 &  61.7409 \\ 
10 &  10 &  zigzag &  0 &  18 &  0 &  2.768 &  60 \\ 
11 &  0 &  armchair &  18.9167 &  18.9167 &  -9.5233 &  1.7747 &  90 \\ 
11 &  1 &  chiral &  15.6393 &  20.2449 &  -8.8771 &  1.8581 &  85.7062 \\ 
11 &  2 &  chiral &  12.6684 &  20.9633 &  -7.9374 &  1.9505 &  81.8062 \\ 
11 &  3 &  chiral &  10.0635 &  21.195 &  -6.8526 &  2.0509 &  78.2789 \\ 
11 &  4 &  chiral &  7.835 &  21.0594 &  -5.7298 &  2.1584 &  75.097 \\ 
11 &  5 &  chiral &  5.9628 &  20.6617 &  -4.6394 &  2.272 &  72.23 \\ 
11 &  6 &  chiral &  4.41 &  20.0877 &  -3.6231 &  2.3909 &  69.6467 \\ 
11 &  7 &  chiral &  3.1338 &  19.4038 &  -2.7018 &  2.5144 &  67.3168 \\ 
11 &  8 &  chiral &  2.0915 &  18.6593 &  -1.8824 &  2.6419 &  65.2117 \\ 
11 &  9 &  chiral &  1.2438 &  17.8896 &  -1.1636 &  2.7727 &  63.3055 \\ 
11 &  10 &  chiral &  0.55636 &  17.1189 &  -0.53898 &  2.9064 &  61.5751 \\ 
11 &  11 &  zigzag &  0 &  16.3636 &  0 &  3.0426 &  60 \\ 
12 &  0 &  armchair &  17.3372 &  17.3372 &  -8.7185 &  1.9319 &  90 \\ 
12 &  1 &  chiral &  14.5689 &  18.4809 &  -8.1889 &  2.015 &  86.0468 \\ 
12 &  2 &  chiral &  12.0335 &  19.1446 &  -7.4222 &  2.1066 &  82.4266 \\ 
12 &  3 &  chiral &  9.778 &  19.4156 &  -6.5274 &  2.2058 &  79.1239 \\ 
12 &  4 &  chiral &  7.8149 &  19.3807 &  -5.5863 &  2.3116 &  76.118 \\ 
12 &  5 &  chiral &  6.1343 &  19.1183 &  -4.6555 &  2.4233 &  73.3858 \\ 
12 &  6 &  chiral &  4.7129 &  18.6946 &  -3.7711 &  2.5401 &  70.903 \\ 
12 &  7 &  chiral &  3.5213 &  18.1629 &  -2.9536 &  2.6615 &  68.6456 \\ 
12 &  8 &  chiral &  2.5286 &  17.5643 &  -2.2126 &  2.7867 &  66.5907 \\ 
12 &  9 &  chiral &  1.7052 &  16.9294 &  -1.5502 &  2.9153 &  64.717 \\ 
12 &  10 &  chiral &  1.0244 &  16.2807 &  -0.96416 &  3.0469 &  63.0053 \\ 
12 &  11 &  chiral &  0.4627 &  15.634 &  -0.44948 &  3.181 &  61.4381 \\ 
12 &  12 &  zigzag &  0 &  15 &  0 &  3.3174 &  60 \\ 
13 &  0 &  armchair &  16.0012 &  16.0012 &  -8.0398 &  2.0893 &  90 \\ 
13 &  1 &  chiral &  13.6325 &  16.9958 &  -7.5984 &  2.1723 &  86.3375 \\ 
13 &  2 &  chiral &  11.4444 &  17.606 &  -6.963 &  2.2632 &  82.9602 \\ 
13 &  3 &  chiral &  9.4743 &  17.8954 &  -6.2154 &  2.3612 &  79.8563 \\ 
13 &  4 &  chiral &  7.735 &  17.9281 &  -5.4187 &  2.4656 &  77.0101 \\ 
13 &  5 &  chiral &  6.2226 &  17.7638 &  -4.6188 &  2.5755 &  74.4037 \\ 
13 &  6 &  chiral &  4.9222 &  17.4538 &  -3.8464 &  2.6905 &  72.0178 \\ 
13 &  7 &  chiral &  3.8136 &  17.041 &  -3.1206 &  2.8098 &  69.8333 \\ 
13 &  8 &  chiral &  2.8745 &  16.5595 &  -2.4519 &  2.933 &  67.8317 \\ 
13 &  9 &  chiral &  2.0825 &  16.0357 &  -1.8445 &  3.0596 &  65.9953 \\ 
13 &  10 &  chiral &  1.4167 &  15.4897 &  -1.2987 &  3.1891 &  64.3081 \\ 
13 &  11 &  chiral &  0.85835 &  14.9362 &  -0.81195 &  3.3213 &  62.7551 \\ 
13 &  12 &  chiral &  0.39086 &  14.3859 &  -0.38058 &  3.4558 &  61.3231 \\ 
13 &  13 &  zigzag &  0 &  13.8462 &  0 &  3.5924 &  60 \\ 
14 &  0 &  armchair &  14.8566 &  14.8566 &  -7.4596 &  2.247 &  90 \\ 
14 &  1 &  chiral &  12.807 &  15.7289 &  -7.0865 &  2.3298 &  86.5885 \\ 
14 &  2 &  chiral &  10.9001 &  16.289 &  -6.5527 &  2.4201 &  83.4239 \\ 
14 &  3 &  chiral &  9.1657 &  16.5839 &  -5.9208 &  2.5171 &  80.4971 \\ 
14 &  4 &  chiral &  7.6162 &  16.662 &  -5.2402 &  2.6201 &  77.796 \\ 
14 &  5 &  chiral &  6.2509 &  16.5691 &  -4.5482 &  2.7285 &  75.3064 \\ 
14 &  6 &  chiral &  5.0605 &  16.3457 &  -3.8709 &  2.8418 &  73.013 \\ 
14 &  7 &  chiral &  4.0312 &  16.0262 &  -3.2255 &  2.9593 &  70.9004 \\ 
14 &  8 &  chiral &  3.1466 &  15.639 &  -2.6224 &  3.0806 &  68.9533 \\ 
14 &  9 &  chiral &  2.3898 &  15.2068 &  -2.0669 &  3.2053 &  67.1572 \\ 
14 &  10 &  chiral &  1.7445 &  14.7471 &  -1.5609 &  3.3329 &  65.4984 \\ 
14 &  11 &  chiral &  1.1956 &  14.2734 &  -1.1036 &  3.4633 &  63.9643 \\ 
14 &  12 &  chiral &  0.72962 &  13.796 &  -0.69314 &  3.5959 &  62.5434 \\ 
14 &  13 &  chiral &  0.33456 &  13.3221 &  -0.3264 &  3.7307 &  61.2251 \\ 
14 &  14 &  zigzag &  0 &  12.8571 &  0 &  3.8674 &  60 \\ 
15 &  0 &  armchair &  13.8649 &  13.8649 &  -6.9579 &  2.4049 &  90 \\ 
15 &  1 &  chiral &  12.0742 &  14.636 &  -6.6385 &  2.4876 &  86.8074 \\ 
15 &  2 &  chiral &  10.3979 &  15.15 &  -6.1848 &  2.5773 &  83.8305 \\ 
15 &  3 &  chiral &  8.8604 &  15.4425 &  -5.6451 &  2.6734 &  81.0624 \\ 
15 &  4 &  chiral &  7.4727 &  15.5507 &  -5.0589 &  2.7752 &  78.4934 \\ 
15 &  5 &  chiral &  6.2361 &  15.5102 &  -4.4565 &  2.8822 &  76.1122 \\ 
15 &  6 &  chiral &  5.145 &  15.353 &  -3.8599 &  2.9938 &  73.9065 \\ 
15 &  7 &  chiral &  4.1899 &  15.1072 &  -3.2846 &  3.1097 &  71.8638 \\ 
15 &  8 &  chiral &  3.3588 &  14.7962 &  -2.7403 &  3.2293 &  69.9713 \\ 
15 &  9 &  chiral &  2.6389 &  14.4395 &  -2.233 &  3.3522 &  68.217 \\ 
15 &  10 &  chiral &  2.0174 &  14.0523 &  -1.7653 &  3.4781 &  66.5893 \\ 
15 &  11 &  chiral &  1.4823 &  13.647 &  -1.3377 &  3.6066 &  65.0774 \\ 
15 &  12 &  chiral &  1.0224 &  13.2328 &  -0.9494 &  3.7376 &  63.6713 \\ 
15 &  13 &  chiral &  0.62783 &  12.817 &  -0.59863 &  3.8707 &  62.3618 \\ 
15 &  14 &  chiral &  0.28961 &  12.4047 &  -0.28303 &  4.0058 &  61.1406 \\ 
15 &  15 &  zigzag &  0 &  12 &  0 &  4.1425 &  60 \\ 
16 &  0 &  armchair &  12.9974 &  12.9974 &  -6.5197 &  2.5629 &  90 \\ 
16 &  1 &  chiral &  11.4196 &  13.6836 &  -6.2434 &  2.6455 &  87 \\ 
16 &  2 &  chiral &  9.9348 &  14.1559 &  -5.8536 &  2.7347 &  84.19 \\ 
16 &  3 &  chiral &  8.563 &  14.4414 &  -5.3886 &  2.8299 &  81.5647 \\ 
16 &  4 &  chiral &  7.3141 &  14.5691 &  -4.8798 &  2.9306 &  79.1162 \\ 
16 &  5 &  chiral &  6.1903 &  14.567 &  -4.3521 &  3.0364 &  76.8356 \\ 
16 &  6 &  chiral &  5.1886 &  14.4607 &  -3.8244 &  3.1466 &  74.7128 \\ 
16 &  7 &  chiral &  4.3022 &  14.2733 &  -3.3101 &  3.2609 &  72.7373 \\ 
16 &  8 &  chiral &  3.5225 &  14.0242 &  -2.8184 &  3.3789 &  70.8988 \\ 
16 &  9 &  chiral &  2.8398 &  13.7299 &  -2.3551 &  3.5002 &  69.1869 \\ 
16 &  10 &  chiral &  2.2439 &  13.4038 &  -1.9234 &  3.6244 &  67.5919 \\ 
16 &  11 &  chiral &  1.7253 &  13.0568 &  -1.5248 &  3.7513 &  66.1046 \\ 
16 &  12 &  chiral &  1.2749 &  12.6976 &  -1.159 &  3.8807 &  64.7161 \\ 
16 &  13 &  chiral &  0.8843 &  12.3327 &  -0.82535 &  4.0122 &  63.4186 \\ 
16 &  14 &  chiral &  0.54595 &  11.9674 &  -0.52222 &  4.1456 &  62.2045 \\ 
16 &  15 &  chiral &  0.25316 &  11.6055 &  -0.24777 &  4.2809 &  61.067 \\ 
16 &  16 &  zigzag &  0 &  11.25 &  0 &  4.4177 &  60 \\ 
17 &  0 &  armchair &  12.2321 &  12.2321 &  -6.1335 &  2.7211 &  90 \\ 
17 &  1 &  chiral &  10.8314 &  12.8467 &  -5.8923 &  2.8035 &  87.1706 \\ 
17 &  2 &  chiral &  9.5074 &  13.2812 &  -5.5544 &  2.8923 &  84.51 \\ 
17 &  3 &  chiral &  8.2764 &  13.557 &  -5.1503 &  2.9867 &  82.0138 \\ 
17 &  4 &  chiral &  7.1472 &  13.6968 &  -4.7055 &  3.0865 &  79.6758 \\ 
17 &  5 &  chiral &  6.1225 &  13.7228 &  -4.2408 &  3.191 &  77.4885 \\ 
17 &  6 &  chiral &  5.2009 &  13.656 &  -3.7719 &  3.3 &  75.4437 \\ 
17 &  7 &  chiral &  4.3779 &  13.515 &  -3.3108 &  3.4129 &  73.5327 \\ 
17 &  8 &  chiral &  3.6469 &  13.3163 &  -2.8658 &  3.5293 &  71.7469 \\ 
17 &  9 &  chiral &  3.0006 &  13.0737 &  -2.4426 &  3.6491 &  70.0775 \\ 
17 &  10 &  chiral &  2.4312 &  12.7991 &  -2.0445 &  3.7718 &  68.5161 \\ 
17 &  11 &  chiral &  1.9309 &  12.5021 &  -1.6735 &  3.8971 &  67.0549 \\ 
17 &  12 &  chiral &  1.4922 &  12.1903 &  -1.33 &  4.0249 &  65.6861 \\ 
17 &  13 &  chiral &  1.1081 &  11.8702 &  -1.0138 &  4.1549 &  64.4028 \\ 
17 &  14 &  chiral &  0.77237 &  11.5465 &  -0.7241 &  4.2868 &  63.1984 \\ 
17 &  15 &  chiral &  0.47911 &  11.2231 &  -0.45955 &  4.4206 &  62.0668 \\ 
17 &  16 &  chiral &  0.22318 &  10.903 &  -0.21872 &  4.556 &  61.0024 \\ 
17 &  17 &  zigzag &  0 &  10.5882 &  0 &  4.693 &  60 \\ 
18 &  0 &  armchair &  11.5519 &  11.5519 &  -5.7907 &  2.8794 &  90 \\ 
18 &  1 &  chiral &  10.3002 &  12.1054 &  -5.5783 &  2.9617 &  87.323 \\ 
18 &  2 &  chiral &  9.1123 &  12.506 &  -5.283 &  3.05 &  84.7968 \\ 
18 &  3 &  chiral &  8.0018 &  12.7707 &  -4.9292 &  3.1438 &  82.4178 \\ 
18 &  4 &  chiral &  6.9765 &  12.9173 &  -4.5379 &  3.2426 &  80.1811 \\ 
18 &  5 &  chiral &  6.0392 &  12.964 &  -4.1264 &  3.3461 &  78.0805 \\ 
18 &  6 &  chiral &  5.1896 &  12.9277 &  -3.7081 &  3.4538 &  76.1091 \\ 
18 &  7 &  chiral &  4.4245 &  12.8238 &  -3.2934 &  3.5654 &  74.2598 \\ 
18 &  8 &  chiral &  3.7393 &  12.6662 &  -2.8898 &  3.6805 &  72.525 \\ 
18 &  9 &  chiral &  3.1283 &  12.4667 &  -2.5028 &  3.7988 &  70.8976 \\ 
18 &  10 &  chiral &  2.5853 &  12.2356 &  -2.1358 &  3.9201 &  69.3703 \\ 
18 &  11 &  chiral &  2.1042 &  11.9812 &  -1.7908 &  4.0439 &  67.9361 \\ 
18 &  12 &  chiral &  1.6787 &  11.7107 &  -1.4689 &  4.1702 &  66.5885 \\ 
18 &  13 &  chiral &  1.3031 &  11.4298 &  -1.1701 &  4.2988 &  65.3212 \\ 
18 &  14 &  chiral &  0.972 &  11.143 &  -0.89424 &  4.4293 &  64.1285 \\ 
18 &  15 &  chiral &  0.6804 &  10.854 &  -0.64038 &  4.5616 &  63.0049 \\ 
18 &  16 &  chiral &  0.42383 &  10.5658 &  -0.40753 &  4.6957 &  61.9453 \\ 
18 &  17 &  chiral &  0.19824 &  10.2806 &  -0.19449 &  4.8313 &  60.9451 \\ 
18 &  18 &  zigzag &  0 &  10 &  0 &  4.9683 &  60 \\ 
19 &  0 &  armchair &  10.9434 &  10.9434 &  -5.4842 &  3.0378 &  90 \\ 
19 &  1 &  chiral &  9.8182 &  11.4445 &  -5.2959 &  3.12 &  87.4598 \\ 
19 &  2 &  chiral &  8.7466 &  11.8146 &  -5.0358 &  3.2079 &  85.0551 \\ 
19 &  3 &  chiral &  7.74 &  12.0674 &  -4.724 &  3.301 &  82.7832 \\ 
19 &  4 &  chiral &  6.8052 &  12.2173 &  -4.3778 &  3.3991 &  80.6397 \\ 
19 &  5 &  chiral &  5.9452 &  12.279 &  -4.0115 &  3.5016 &  78.6197 \\ 
19 &  6 &  chiral &  5.1602 &  12.2663 &  -3.6368 &  3.6082 &  76.7174 \\ 
19 &  7 &  chiral &  4.4481 &  12.1922 &  -3.2627 &  3.7185 &  74.9267 \\ 
19 &  8 &  chiral &  3.8056 &  12.0683 &  -2.896 &  3.8323 &  73.2414 \\ 
19 &  9 &  chiral &  3.2283 &  11.9047 &  -2.5417 &  3.9493 &  71.6551 \\ 
19 &  10 &  chiral &  2.7113 &  11.7103 &  -2.2031 &  4.0692 &  70.1617 \\ 
19 &  11 &  chiral &  2.2497 &  11.4926 &  -1.8825 &  4.1917 &  68.7551 \\ 
19 &  12 &  chiral &  1.8384 &  11.2578 &  -1.5811 &  4.3166 &  67.4297 \\ 
19 &  13 &  chiral &  1.4726 &  11.0112 &  -1.2994 &  4.4437 &  66.1798 \\ 
19 &  14 &  chiral &  1.1477 &  10.7571 &  -1.0374 &  4.5728 &  65.0003 \\ 
19 &  15 &  chiral &  0.85945 &  10.499 &  -0.79459 &  4.7038 &  63.8863 \\ 
19 &  16 &  chiral &  0.60393 &  10.2396 &  -0.57038 &  4.8366 &  62.8334 \\ 
19 &  17 &  chiral &  0.3776 &  9.9811 &  -0.36387 &  4.9708 &  61.8373 \\ 
19 &  18 &  chiral &  0.17725 &  9.7254 &  -0.17409 &  5.1065 &  60.894 \\ 
19 &  19 &  zigzag &  0 &  9.4737 &  0 &  5.2436 &  60 \\ 
20 &  0 &  armchair &  10.3959 &  10.3959 &  -5.2087 &  3.1962 &  90 \\ 
20 &  1 &  chiral &  9.3789 &  10.8515 &  -5.0406 &  3.2783 &  87.5833 \\ 
20 &  2 &  chiral &  8.4074 &  11.1943 &  -4.81 &  3.3659 &  85.2891 \\ 
20 &  3 &  chiral &  7.491 &  11.435 &  -4.5335 &  3.4585 &  83.1151 \\ 
20 &  4 &  chiral &  6.6356 &  11.5856 &  -4.2255 &  3.5557 &  81.0577 \\ 
20 &  5 &  chiral &  5.8441 &  11.658 &  -3.898 &  3.6573 &  79.1128 \\ 
20 &  6 &  chiral &  5.1172 &  11.6638 &  -3.561 &  3.7629 &  77.2754 \\ 
20 &  7 &  chiral &  4.4535 &  11.6136 &  -3.2225 &  3.8721 &  75.5405 \\ 
20 &  8 &  chiral &  3.8506 &  11.5174 &  -2.8885 &  3.9847 &  73.9027 \\ 
20 &  9 &  chiral &  3.3053 &  11.3838 &  -2.5636 &  4.1004 &  72.3566 \\ 
20 &  10 &  chiral &  2.8136 &  11.2205 &  -2.251 &  4.219 &  70.8968 \\ 
20 &  11 &  chiral &  2.3715 &  11.0342 &  -1.9531 &  4.3402 &  69.5181 \\ 
20 &  12 &  chiral &  1.9749 &  10.8305 &  -1.6711 &  4.4638 &  68.2153 \\ 
20 &  13 &  chiral &  1.6197 &  10.6141 &  -1.4058 &  4.5896 &  66.9837 \\ 
20 &  14 &  chiral &  1.3021 &  10.3889 &  -1.1575 &  4.7174 &  65.8186 \\ 
20 &  15 &  chiral &  1.0185 &  10.1582 &  -0.92588 &  4.8471 &  64.7157 \\ 
20 &  16 &  chiral &  0.76535 &  9.9248 &  -0.71068 &  4.9785 &  63.6709 \\ 
20 &  17 &  chiral &  0.53965 &  9.6907 &  -0.51125 &  5.1115 &  62.6804 \\ 
20 &  18 &  chiral &  0.33854 &  9.4576 &  -0.32687 &  5.246 &  61.7406 \\ 
20 &  19 &  chiral &  0.15943 &  9.2271 &  -0.15673 &  5.3819 &  60.8482 \\ 
20 &  20 &  zigzag &  0 &  9 &  0 &  5.519 &  60 \\ 
\\
\caption{\textbf{Summary of bevel angles, tubule radii, and seam angles for different tubules.} Bevel angles and seam angles are in the unit of degrees. Tubule radius is in units of the monomer edge length. To get opposite chirality tubules, $n<0$, swap the bevel angles of sides 1 and 2.}
\label{STable:bevel}
\end{longtable}

%%%%%%%%%%%% origami design %%%%%%%%%%%%%%%

\begin{figure*}[htb]
 \centering
 \includegraphics[width=\textwidth]{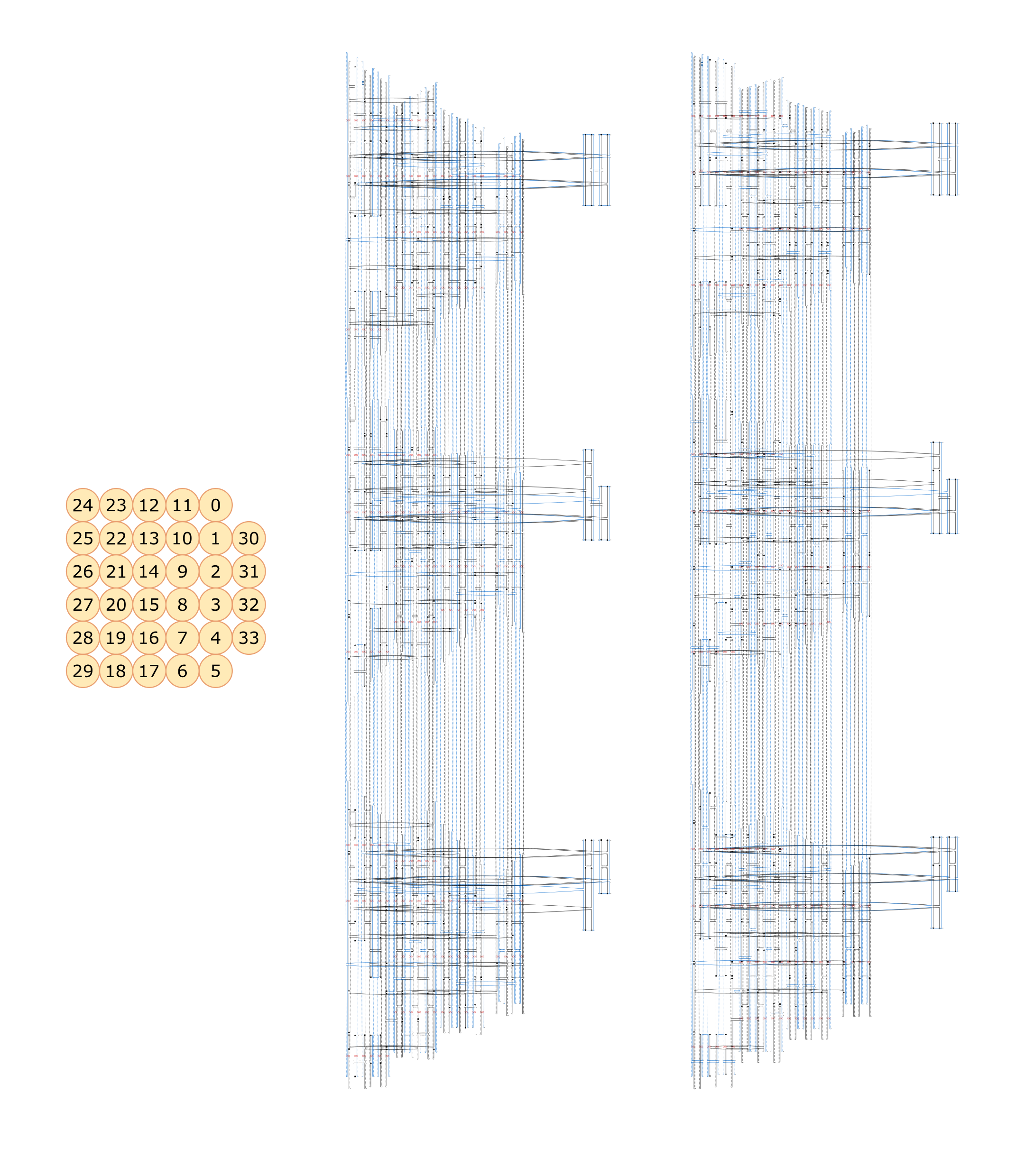}
 \caption{\textbf{Helical numbers and caDNAno designs for V-triangle (middle) and X-triangle (right).}}
 \label{Sfig:XVcadnano}
\end{figure*}

\begin{figure*}[htb]
 \centering
 \includegraphics[width=\textwidth]{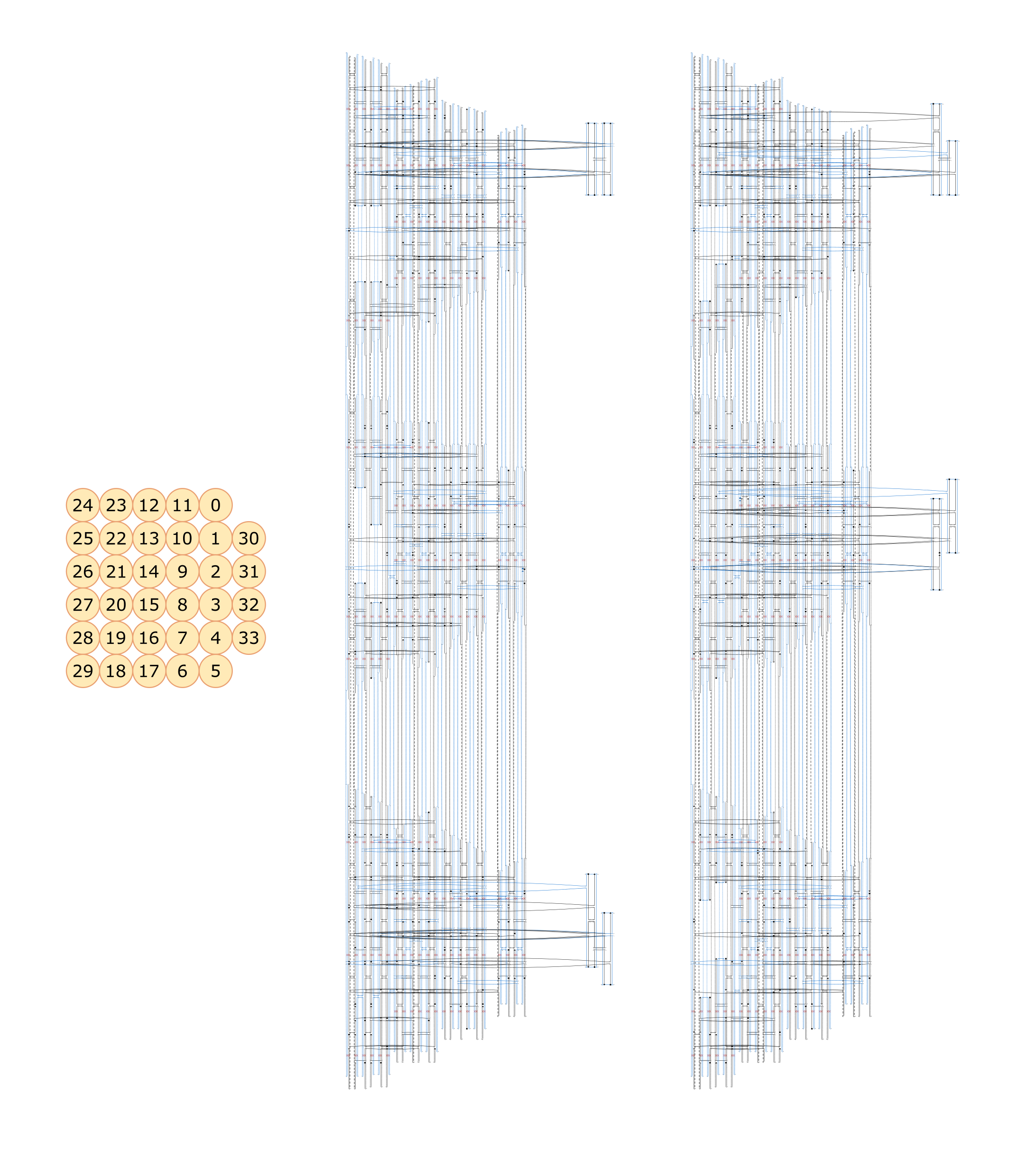}
 \caption{\textbf{Helical numbers and caDNAno designs for A-triangle (middle) and B-triangle (right).}}
 \label{Sfig:ABcadnano}
\end{figure*}

\clearpage

\fontsize{8}{10}\selectfont

\begin{longtable}{c c}
Name &  Sequence \\ 
\hline 
\endfirsthead

Name &  Sequence \\ 
\hline 
\endhead

\endfoot

\endlastfoot
Core 1 Side1 1 &  TAATCAGTTTAGAATCAGAGCGGGAGCTAAAGAAAGGAAGGGAAGACCGTAAAG \\ 
Core 1 Side1 2 &  CGCATCGTTTGCCGCCATCAACATAAGGTGGCAACATCAG \\ 
Core 1 Side1 3 &  GCTTGACGGGGAAAGAGCCCCCGCAATATAAGGAATTATCAAACAATGATTTAGA \\ 
Core 1 Side1 4 &  CCGCCGGGATTGCAGGAGGTGTCCAGCATCAGAGACAGGAACGGTACG \\ 
Core 1 Side1 5 &  CACTCAATCGGCAATGAAATAGCAATAGAAAATACATA \\ 
Core 1 Side1 6 &  TGCTCATAACCGTGTGACCGTGGAAGATCGCACTCCAGCCAGCTT \\ 
Core 1 Side1 7 &  AAAAAGAGAAAGCCGCCGCAAGAATGCCAAC \\ 
Core 1 Side1 8 &  CGTGGCGACAGGAGGCCCAGAATCCTGAGAAGTGTTTTTA \\ 
Core 1 Side1 9 &  CGATTAAAGGGATTTTAATACCAATTTTACATTAATGGAATTGTTTGG \\ 
Core 1 Side1 10 &  AGAAAATTCGCCTGATTGCTTTGCGGGGTC \\ 
Core 1 Side1 11 &  ATAAACAGTAAATTTCGTGTACATGTGCCATCCCACGCAACCAGCTT \\ 
Core 1 Side1 12 &  AATTCATATTTAGACGCGGAGCCAGCTTTC \\ 
Core 1 Side1 13 &  AACCACCATCAGCAAATCGGAACCCTAAAGGGCCGGCGAAACAGTACC \\ 
Core 1 Side1 14 &  CCTTATCTACCAGCGCCAAACCTGAAAATTC \\ 
Core 1 Side1 15 &  ACCTTGCTGAACCTCAAATATCATGCGGCTGGTATGAGC \\ 
Core 1 Side1 16 &  AGTATTAAATCAATACAATCAATTTCCTGATTGTTTGCG \\ 
Core 1 Side1 17 &  AAGTTTATTTTGTCACAGTAAGCAATGATGGCAACAATAA \\ 
Core 1 Side1 18 &  GATAGCCGAGCCCTTTTTAAGAAATTCTGAACGGGAGA \\ 
Core 1 Side1 19 &  ATTATACGACTTTACATCATAATCTGGTCAGTTGGC \\ 
Core 1 Side1 20 &  TACCTATCTTACCGAAACAAAGTTACCAGACCCACAC \\ 
Core 1 Side1 21 &  TGTAGGTAAAGATTCACCCAGTCACGACGTTG \\ 
Core 1 Side1 22 &  AAAGCGCCCAGGCTGCCGTCGGTGCGACATAAAAAAATC \\ 
Core 1 Side1 23 &  ACTAGAAAAAGCCTGTGTAAAGGTTCATAATTAAAGCATC \\ 
Core 1 Side1 24 &  TTAGTATCATATGCGTTATACAAAGATAATCACGGGTCAC \\ 
Core 1 Side1 25 &  GGAAGGTATGTTGCCCAACGACAAAAGGGCGA \\ 
Core 1 Side1 26 &  GAGGGAGAGTATAAAGCCAACGCTCAACAGAGGCTATCAAAGGTGA \\ 
Core 1 Side1 27 &  TTAAATACCGAACGCCTAAAACATCGCCAT \\ 
Core 1 Side1 28 &  CACCGGAATGGTGTGTGAAGGAGCTCCTAATTTTCCCTTAATACAGTA \\ 
Core 1 Side1 29 &  GGCCTTCCTATATAAAAGAAACGCAAAGAAT \\ 
Core 1 Side1 30 &  TAAATGTAAATAATTTAAAATATATGTACCCCGGTTTTCTTACC \\ 
Core 1 Side1 31 &  AACAACCCATCAAACTCCATATTAAAGAATTGAGTTAAGCCCAATAAGGCGGTT \\ 
Core 1 Side1 32 &  CACTAAATCGTTAACGGCATCAATAAGGCG \\ 
Core 1 Side2 1 &  CCAGAGCACATCCTCATAACGGAAGTTAAACGATGCTGATTGCCGT \\ 
Core 1 Side2 2 &  AGTAGCATTAACATCCAATAAATCAGGTAGAGGTCATTTATTACCTTAGCTTGCT \\ 
Core 1 Side2 3 &  GTGGCATCAATTCTAAAAATAGCAGAGTACCTTATACCACAAAAGGA \\ 
Core 1 Side2 4 &  AGTCCCGGAAAATAGCATTAACTGAACACCCTGAACAAA \\ 
Core 1 Side2 5 &  AGCGCCATGTTTACCATCGGCGATCAAGATTTTTTAGCGAAGCCGTTACCGCACT \\ 
Core 1 Side2 6 &  AGCTAAATAAGCAATACAAGGCAAAGAATTAGCGTCTCGTCGCTGGCA \\ 
Core 1 Side2 7 &  GAGCATAAAAAGACGTTAGTAAATGAATGTAGTTTATC \\ 
Core 1 Side2 8 &  ACGCGGTCCGTTTTTTACGGGAGAAGCCTTTA \\ 
Core 1 Side2 9 &  TTCCAAACAGGGAAGCGCATTAGCAAAATT \\ 
Core 1 Side2 10 &  TTAACGTCAAAAATGAATTTGTGTCCGGCAAGCCTCCGG \\ 
Core 1 Side2 11 &  CTCAGAACGTTAGCGTAACGATCTAAAGTTTGGCTTAG \\ 
Core 1 Side2 12 &  CCTTAAAAACGTACCGGTTAATCATTGTGA \\ 
Core 1 Side2 13 &  TTCGAAGGTTTCAGCGATTTTGCTAAACAACTTACAATTCAGAGAG \\ 
Core 1 Side2 14 &  CGAGCATTAAGAACATTTTTAGAACCTCACCGGAAACA \\ 
Core 1 Side2 15 &  ATCCCATCGCCTGTTTATCAACAATAGATAATTTTGCGGAAAAACAT \\ 
Core 1 Side2 16 &  ACCGATATATTCGCTGCCATCGCC \\ 
Core 1 Side2 17 &  GTCGTCGCTGAGGCTTGCAGGGAGTTAGAACCGCCACCCAGAATGGA \\ 
Core 1 Side2 18 &  CATCGAGGACAACAAAACAAGAACGACTTGCTTTGTACC \\ 
Core 1 Side2 19 &  ATTTACCGTTCCAGTAAGCGCAGGTTAATGCTAAATTGGTTTAAGAA \\ 
Core 1 Side2 20 &  CGAGAATGTTTTGCAAAAGAAGTTGGCGCGAGCTGAAAAG \\ 
Core 1 Side2 21 &  TAGTTGCGCCGACAATTTCAACAGTTTTGAAGAATAACAT \\ 
Core 1 Side2 22 &  GAGTGAGAATAGAAATAAGCTTGATACCGACACGCATA \\ 
Core 1 Side2 23 &  GGAACAACTTGTATCGCCGCCACCCTCCTTTTAGCTTAATTGCTGAAT \\ 
Core 1 Side2 24 &  TTATCCTGTATTTTGCACCCAGCAACAAGCAACCTCCAAATAAT \\ 
Core 1 Side2 25 &  TCATCAGTGAATAAGGGTTGGGAACTGGCTCATTTAATTG \\ 
Core 1 Side2 26 &  GACGTGACAATCATAAGGGAACCAGAATCATCCCTCAG \\ 
Core 1 Side2 27 &  TCAGACTGACGAGGCGCAGACGGTCCCCCAGC \\ 
Core 1 Side2 28 &  ATATAAAGAGAGCCAGCAAAATCAGGATAAAAGCGAGGCGAGTAAAGA \\ 
Core 1 Side2 29 &  GAGAATTACCATGTAATAAGAGACAGAACGCCTAATTTA \\ 
Core 1 Side2 30 &  GGCCGCTAGCAGCACCGTAATCAGTAGCGACGAACTGAACAGCATCACGAGTAG \\ 
Core 1 Side2 31 &  AGCCGCCACCCTCAAAAGCGAAAGCCAACTTT \\ 
Core 1 Side2 32 &  GGGAATTTACCGACAATTTAGGCAGAGGCATTTTCGAG \\ 
Core 1 Side2 33 &  TATGACCCATTTCAACGGGGGTGAATTTCTTAAACAACATCTCTGACCTCAGC \\ 
Core 1 Side2 34 &  TTTATTTCAACGCAACCAGTAGCACCAGCC \\ 
Core 1 Side2 35 &  CTTAGAAACACCAGAGGAACGAGTTAAAGCC \\ 
Core1 Side12 1 &  ACGTGCCGGACTTGTAGAACGTCAGCTTGTGGTGCTGGTCTGGTCA \\ 
Core1 Side12 2 &  GGCGATCGGTGCGGGCGCAGCAACACAGGCGGCCTTTATTGTGATGAAGGGTAA \\ 
Core1 Side12 3 &  CTCTTCGCTATTATTCGCCAGCTGGCGAGGGTAACGC \\ 
Core1 Side12 4 &  ACAGCGGGTCGGATTATGTTAGATAGAAGGCTTTTATCCGGTATTC \\ 
Core1 Side12 5 &  AGAGATAGACTTTCTGCCACGGGACTCCTTATTTTACGCAGACGGCCAG \\ 
Core1 Side12 6 &  TAACCCACTTTATCCCAATCCAAATAAGATTAACGATTTTTTGT \\ 
Core1 Side12 7 &  GTCAGAGGGTAATTGAGTTCGCTAATATCAGAGAGACCGTAAAAACGCAGAA \\ 
Core1 Side12 8 &  GTAGAAATTCCAATCAATAATCGGCTGTCTTTAGCAAGCAGCCATCAAGAGCGAGT \\ 
Core1 Side12 9 &  CATTCAACCGAATAAACAACATGTTTTCAGCTAATG \\ 
Core1 Side12 10 &  CAGGGTTTTAAAGTTGGTGAGAAAGGCCGGAGACAGTCAAATATATATTTAGCCATTT \\ 
Core1 Side12 11 &  CACGGAATATATGGTTATTCCAAGAACGGGTATTTTAAACCAAGT \\ 
Core1 Side12 12 &  TTTATTTTCATCGTATTGGAATCATTACCGCGAGGAAACCGAAAATAC \\ 
Core1 Side12 13 &  AATCTTACCAACGCTAACGAGCGTTTCTTTCCAGAGCCTAATTTGCCAGT \\ 
Core1 Side12 14 &  TAAATTAATGCCGGTTAGAGGGTAGCTATTT \\ 
Core1 Side12 15 &  AAAAGGTTTAAAGTAATTGAGAATCGCCATATTTTTAACAACGC \\ 
Core1 Side12 16 &  ATAGGAACAATCAGATCAAACGTAGAGGAAACGCAATTTAATAACGGAATACCCATGC \\ 
Core1 Side12 17 &  TAAATGCATTATGCCTGAGTAATTTTTTAACCAATTATCACCGTCACCGACTTTTG \\ 
Core 2 Side3 1 &  ATTCTGCGAACGAGTATAACAGTTGATTCCCAGTTTCGTCCAGGGATA \\ 
Core 2 Side3 2 &  AGAAGAACATTGCCCTCCCGCCGCCCAGCTGCAATAGTGAATTTAGGAATAAATCA \\ 
Core 2 Side3 3 &  GAAAGACTTCAAATAATTAAGAGTCAGTACCGCTGAGACTAAGTTTT \\ 
Core 2 Side3 4 &  TCACGCTGAAATTGCGAAAGAAGATGATGAAACAAACATCGCCGCTCACCACA \\ 
Core 2 Side3 5 &  TGACGAGTGGCAAGGAATAGCTGAACCACGGTTTGC \\ 
Core 2 Side3 6 &  CACGTATATTGTAGCAATACTTCTTTGATTAAATCAAAATGTAGCGG \\ 
Core 2 Side3 7 &  CAAAACATGTACCGTAACACTGATATCCAG \\ 
Core 2 Side3 8 &  CAATAGGAAACAGGAAAACAATATTCGGCCTTGCTGGTAA \\ 
Core 2 Side3 9 &  TACATTTAACAACAGAAGAGTCATTAATGGCTCACT \\ 
Core 2 Side3 10 &  GAGAAGGAGGACTCATTACGTAATGCCACTA \\ 
Core 2 Side3 11 &  TTATCAAAATCATAGGTCTGTAATAACAACAGTTAATGAAAACT \\ 
Core 2 Side3 12 &  TGCAAGGATATATAACTTTAACCTCCGGCTTAAACAGTACTTAGCGGG \\ 
Core 2 Side3 13 &  AGGAGCACTTTTAAAATGCGTTGCAATCGGCC \\ 
Core 2 Side3 14 &  AGAACCACGCCTTGATATTCACAAATGAAAGTCCAACCTAGGATTGCA \\ 
Core 2 Side3 15 &  CAGCATTGAATCGTCTGGAAATACCTACATTTTCCAGTCG \\ 
Core 2 Side3 16 &  ACAAAGAACGGCAACAGTGCCACATATCTTT \\ 
Core 2 Side3 17 &  TGATACAGGAGTGTAGGTTGGGTGGTTGAGG \\ 
Core 2 Side3 18 &  TAACCTTGACGCTGAGAAATAAAGCGCGTAACACAGGGCGATCACTTGCCTGAGT \\ 
Core 2 Side3 19 &  ATATAAGTATAAGTGCCGTCGAGCTGGTAATCCTCAACAGGTCAGGTAAAA \\ 
Core 2 Side3 20 &  TACCTTTTTTAATGGAAGGGTTGGCAAGCCGTTTTGC \\ 
Core 2 Side3 21 &  ATTTGAATATATATGTGAGTGAATATAGAGACTACCTT \\ 
Core 2 Side3 22 &  GGTGCCTAATGAGTGACCGGGGGTTTCTGC \\ 
Core 2 Side3 23 &  ATTAGCGTTTGCCATCGATTCACCAGTCACACATTATTTACATTGGCA \\ 
Core 2 Side3 24 &  GAGATAGAACCCTTCTGACCTGACAGCACGCCTGACCTAACATTAAT \\ 
Core 2 Side3 25 &  AAGCGTAACGGTCAGTATTAACACCGCCTCGATTCATCTT \\ 
Core 2 Side3 26 &  AAGTTTCCCGAAGGCAATTAAGAGAGGTTTCACGTTGAAACCACCACCTAATTCGA \\ 
Core 2 Side3 27 &  ATTAAACGGACGATTGCACCAGAGCCGCCGC \\ 
Core 2 Side3 28 &  TCAAAAAGTCGCGTTTCTCATTTTACCAGTACAAACTACAACGCCTG \\ 
Core 2 Side3 29 &  GAAGCCCAAACGAACCTGTCGTGGCCATTGCACCTTAAT \\ 
Core 2 Side3 30 &  TCACCCAGCGGTGAGCTAACTCAATTTAATGGTTTGAGGTGAGG \\ 
Core 2 Side23 1 &  ATAACCTGTTTAGCTTTTTTTTTTCATTTGGTTGCCAGAGTTTTTTTAATAGTAAAATG \\ 
Core 2 Side23 2 &  AGCAAATTTTCCAACAGGTCAAACATGTTTTATTTTTATGCAACTATGGTCA \\ 
Core 2 Side23 3 &  GTGTCTGGTAGCATTCCACTTTGACAGCCCTCATA \\ 
Core 2 Side23 4 &  AAGTTTCGCTTCAAAGCGAACCAGTCAGAAAAAAAATTGGCTCATCAAAA \\ 
Core 2 Side23 5 &  CGCCACCCTCAGATTTCCGCCACCCTCAGAGATCTCCAGCAAAGC \\ 
Core 2 Side23 6 &  CCCCTGCCGCCTTTAATAAAGGATTTTTTCGAATAATAATTTCGG \\ 
Core 2 Side23 7 &  TATTTCGGAACCTTTTTTATTCTGAAAC \\ 
Core 2 Side23 8 &  ACAAATAATTTTCCTCAGGTAGCAAACCGCCAGTTTGCCTTTAGCG \\ 
Core 2 Side23 9 &  TCAGAGCCACCACTTTTTCAGAGCCGCCACCTTTTCATATCATGAGG \\ 
Core 2 Side23 10 &  AAGCGTTTTTTTTATGGCTTTTGAAACGGGGTCAGTGCTTTTTTTGTAACAGTGCCCGTA \\ 
Core 2 Side23 11 &  TCCCCCTCAATTTTTTTTTAAACAGTTCAGAAAA \\ 
Core 2 Side23 12 &  ATAGCCCGGAATAGGTGTTTTTTTCCGTACTCAGGAGGTTTATTTCTGTATGGG \\ 
Core 2 Side23 13 &  GATTATACCAAGCGCGATTTTTTAAGTACACCATGTTTTTTTTAGCCGGA \\ 
Core 2 Side23 14 &  ACCTGCTACGGAGATTTGTAATACACTTTAAAACACTCAT \\ 
Core 2 Side23 15 &  TAGCGCGTTTTTTTTTCGGCATTTTCGGTCATAAATCCGCG \\ 
Core 2 Side23 16 &  ACGGCTACAGATTTGCTTTGAGGACTAAAGACTTTTATCAAAA \\ 
Core 2 Side23 17 &  ATCAGGTCTTTTTTTCCTGCATTGAATTTAGACTGGATAGCTTTCGCAAAAAGTACG \\ 
Core 2 Side31 1 &  GAGGCCACCGAGTAAATTTTTTTCTGTCCATCACGCAATGTTGTTCCAGT \\ 
Core 2 Side31 2 &  ACAAGAGTCAAGGGCGAATTTTTTTGTCTATCAGGGCGATGGC \\ 
Core 2 Side31 3 &  CCTGAGCATAGATTTTCAGGTTTATTTCGTCAGATGAAT \\ 
Core 2 Side31 4 &  GTTACAAAATCGCGCAGAGGCGAATTATTTTCATTTCAATTA \\ 
Core 2 Side31 5 &  GAATCCTTGAAATTTCATAGCGATAGCTTAACATTATCATT \\ 
Core 2 Side31 6 &  GTTTGAGTAGATTAAGCTTCTGTATTTTTTGTCGCTATTAATTGA \\ 
Core 2 Side31 7 &  GCGGAACAAAGAGGAATTGAGGAATTTTTTATCTAAA \\ 
Core 2 Side31 8 &  AAATCAACCATTTGAGTCGACAACTCGTATTTTTTTTTCTTTGCCCGAACGT \\ 
Core 2 Side31 9 &  GCTGAGATTTTCAGCAGCAAATGAAAAATCTTAAAAATAAGAATAAA \\ 
Core 2 Side31 10 &  TATATGTATAATAGATTAGAGCCGTCTTTTTTTATAATAAGTTGAAA \\ 
Core 2 Side31 11 &  CCACTACGCCGAGATAGGGTTGAGATTAACCGACGTGCTTTTTTTTCG \\ 
Core 2 Side31 12 &  GGGTTAGAACCTACCTTTTTTCAAAATTATTTGCACGTAAAATTTC \\ 
Core 2 Side31 13 &  CCGGTGCCCCCTGCATCAGACGATTTTTTGCGCAGTGT \\ 
Core 2 Side31 14 &  CGGCCAGAATTTTTTGCGGGCCGTTTTCACGGTCAT \\ 
Core 2 Side31 15 &  TGCGCGAACTGATAGCAACCACCAGCAGAAGATAATTTACAGA \\ 
Core 2 Side31 16 &  GAATACGTGGCACAGACAATATTTTTTTTTTTGGCTATTAGTCTTTAA \\ 
Active Hubble1 1 &  CGGCGGATCATCCATTATTCGCTGCCAGTTTGAGGGGACGACGAC \\ 
Active Hubble1 2 &  AGTATCGGCCTCAAATGGGATATATTTTGTCGCGTCTAATATTGACGGAAAT \\ 
Active Hubble1 3 &  TCCAGACGACGACATTTATTCATTTGTCAATCTCGCATTAAATCGTAA \\ 
Active Hubble1 4 &  AACTAGCAAGGTCATTTGAACGGTAATTTTTGGGAACAAA \\ 
Active Hubble1 5 &  GTGCCGGAAACCAGGCTCCGGCACTCTCCGTG \\ 
Active Hubble1 6 &  GGCAGCACGCAACTGTTGGGAAGTAAAACGCGCTTCTG \\ 
Active Hubble1 7 &  CAAACAAGAGAATCGAGCCTGAGAATCTACAATAGGGCTTAATTCTG \\ 
Active Hubble1 8 &  TTAAATCAGCTCATGTTGAGAGGTCTGGAG \\ 
Active Hole1 1 &  GTGTAGATGGGACGGCTGGCGCTTTCGAGCAGTTGTAAGAGCA \\ 
Active Hole1 2 &  GAAAAGCCTAAATTGTAAACGTTAAGGTCACGTTG \\ 
Active Hole1 3 &  TGTATAAGCCTTACACTTCTTTGCTCGTCATGTAATGG \\ 
Active Hole1 4 &  AAACATCCAAATATTCCAAAAACAGGAAGAT \\ 
Active Hole1 5 &  AATCAAGTTTTTTTTTGGGAACGTCACACTATTAAAGAACG \\ 
Active Hole1 6 &  TGGACTCCTCGAGGTGAAGCGAAAGGATTTTGGGCGCTAGGGCGC \\ 
Active Hole1 7 &  AATACCGTTTCCGGGTGCTGCACTGCGCGCCTGTG \\ 
Active Hole1 8 &  CACTCTGTTGTGATAAGATGCCGGGTTTTTCCTGCAGCCA \\ 
Active Hubble2 1 &  GAAAAATCGACGACGAACTATCATAACCCTCGAAGATTCATCAGT \\ 
Active Hubble2 2 &  TGAGATTTAGGAAATAAAACGAACTAACGGAA \\ 
Active Hubble2 3 &  GATAAGATTTACCATACGTTATACCACATAGAGTAATCTTGACAAGAACC \\ 
Active Hubble2 4 &  GGATATTCATTACGACAGATGCCATCGATTTTGCGGGTGTAATAC \\ 
Active Hubble2 5 &  ATTACGAGGCATAGTACATAACGCGTCAGGACCTTGCCCT \\ 
Active Hubble2 6 &  ATAATGCTGTAGCTCGGATTAGGAGAGGCACCATAACAAAAGGA \\ 
Active Hubble2 7 &  CTGCTCATTCATCAACTAATGCAGATAAGAGCAACTAAAAACCCTAATAGT \\ 
Active Hubble2 8 &  ATGCGATGCTTGAGCTGACCTGAAAGAGCCAAATCAACGTAACAAAG \\ 
Active Hole2 1 &  CAACATTATTACATACAGGAAGCCTCATTGCGGATTGTCGTCT \\ 
Active Hole2 3 &  GACCAGGCGCATAGGCTGGATGGTTTAATCGTCAC \\ 
Active Hole2 4 &  CCATAGCAAGGCCGGAAACGTCACCAATGAAAAACGGTGTACA \\ 
Active Hole2 5 &  TGCCAAGCTTTCTTAGAGATTAAGTTAAGGGGGATGTGCTG \\ 
Active Hole2 6 &  CAAGGCGGTGGAGCCCCGTGGTGAAGGGATAGTTCTCTCACGG \\ 
Active Hole2 7 &  ACTGGCATGATTAAGAACGGATAACCCTCCACCATCAATATG \\ 
Active Hole2 8 &  ATATTCAACCGTTCTAGCTGACAACATGT \\ 
Active Hubble3 1 &  CGCCAGGGGGCAAAATTGTTTGATGGTGGTTCTCACCGCCTGGCC \\ 
Active Hubble3 2 &  CTGAGAGAGTTGCTCTTTTCACCAGTGAGACG \\ 
Active Hubble3 3 &  CGAAATCTGGTTTTAGCAAGCGGTTGAGGATCCCCGGGTACCG \\ 
Active Hubble3 4 &  AGCTCGAATTCGTTTGTTATCGGCCAACATTTTCAAATGACGCTCACAATCCA \\ 
Active Hubble3 5 &  CAGGCGAAAATCCCCCTTATAGTAATAACCGTACTATGGTTGCTT \\ 
Active Hubble3 6 &  GTTTCCTGAAAGTGTATCCTCACAGTCCACGCTGGTTTGCCCCAG \\ 
Active Hubble3 7 &  GCGTCCGTGAGCCAAGCCTGGGTATTGGGAACGCGCGGGGAGAGG \\ 
Active Hubble3 8 &  GAAGCATTGTGAAAAATCATGGTCATAGCTGTGCCTGTTCTTC \\ 
Active Hole3 1 &  GGCAACAGCTGTCAAACTATACCGCCAGCTTAATGCAAGAAAA \\ 
Active Hole3 3 &  CCACACAACATACGAGCCGGCCCGCTTTATATTTTAATGCTGAATCGCAAG \\ 
Active Hole3 4 &  GAAATGGGACCAGTAATAAAAGGGACATTCTCGCTCACAATT \\ 
Active Hole3 5 &  ATTCCATAGATTTAGTTTGACCATTAGATACAGTCCAATACTG \\ 
Active Hole3 6 &  CGGAATCGTCATAAATATTACTATTATAGACCGGA \\ 
Active Hole3 7 &  AAACGAAAGAGGCAAAAGATCATCGCCTGATA \\ 
Active Hole3 8 &  AATTGTGTCGAGCCCCCTTTCACCGGAACCAGTTTGCCACCACCGG \\ 
Passive Hubble1 1 &  CGGCGGATCATCCATTATTCGCTGCCAGTTTGAGGGGACGACGACTTTTT \\ 
Passive Hubble1 2 &  TTTTTAGTATCGGCCTCAAATGGGATATATTTTGTCGCGTCTAATATTGACGGAAAT \\ 
Passive Hubble1 3 &  TCCAGACGACGACATTTATTCATTTGTCAATCTCGCATTAAATCGTAATTTTT \\ 
Passive Hubble1 4 &  TTTTTAACTAGCAAGGTCATTTGAACGGTAATTTTTGGGAACAAA \\ 
Passive Hubble1 5 &  TTTTTGTGCCGGAAACCAGGCTCCGGCACTCTCCGTG \\ 
Passive Hubble1 6 &  GGCAGCACGCAACTGTTGGGAAGTAAAACGCGCTTCTGTTTTT \\ 
Passive Hubble1 7 &  TTTTTCAAACAAGAGAATCGAGCCTGAGAATCTACAATAGGGCTTAATTCTG \\ 
Passive Hubble1 8 &  TTAAATCAGCTCATGTTGAGAGGTCTGGAGTTTTT \\ 
Passive Hole1 1 &  TTTTTGTGTAGATGGGACGGCTGGCGCTTTCGAGCAGTTGTAAGAGCA \\ 
Passive Hole1 2 &  GAAAAGCCTAAATTGTAAACGTTAAGGTCACGTTGTTTTT \\ 
Passive Hole1 3 &  TTTTTTGTATAAGCCTTACACTTCTTTGCTCGTCATGTAATGG \\ 
Passive Hole1 4 &  AAACATCCAAATATTCCAAAAACAGGAAGATTTTTT \\ 
Passive Hole1 5 &  AATCAAGTTTTTTTTTGGGAACGTCACACTATTAAAGAACGTTTTT \\ 
Passive Hole1 6 &  TTTTTTGGACTCCTCGAGGTGAAGCGAAAGGATTTTGGGCGCTAGGGCGC \\ 
Passive Hole1 7 &  AATACCGTTTCCGGGTGCTGCACTGCGCGCCTGTGTTTTT \\ 
Passive Hole1 8 &  TTTTTCACTCTGTTGTGATAAGATGCCGGGTTTTTCCTGCAGCCA \\ 
Passive Hubble2 1 &  GAAAAATCGACGACGAACTATCATAACCCTCGAAGATTCATCAGTTTTTT \\ 
Passive Hubble2 2 &  TTTTTTGAGATTTAGGAAATAAAACGAACTAACGGAATTTTT \\ 
Passive Hubble2 3 &  GATAAGATTTACCATACGTTATACCACATAGAGTAATCTTGACAAGAACCTTTTT \\ 
Passive Hubble2 4 &  TTTTTGGATATTCATTACGACAGATGCCATCGATTTTGCGGGTGTAATAC \\ 
Passive Hubble2 5 &  TTTTTATTACGAGGCATAGTACATAACGCGTCAGGACCTTGCCCT \\ 
Passive Hubble2 6 &  ATAATGCTGTAGCTCGGATTAGGAGAGGCACCATAACAAAAGGATTTTT \\ 
Passive Hubble2 7 &  TTTTTCTGCTCATTCATCAACTAATGCAGATAAGAGCAACTAAAAACCCTAATAGT \\ 
Passive Hubble2 8 &  ATGCGATGCTTGAGCTGACCTGAAAGAGCCAAATCAACGTAACAAAGTTTTT \\ 
Passive Hole2 1 &  TTTTTCAACATTATTACATACAGGAAGCCTCATTGCGGATTGTCGTCT \\ 
Passive Hole2 3 &  TTTTTGACCAGGCGCATAGGCTGGATGGTTTAATCGTCAC \\ 
Passive Hole2 4 &  CCATAGCAAGGCCGGAAACGTCACCAATGAAAAACGGTGTACATTTTT \\ 
Passive Hole2 5 &  TGCCAAGCTTTCTTAGAGATTAAGTTAAGGGGGATGTGCTGTTTTT \\ 
Passive Hole2 6 &  TTTTTCAAGGCGGTGGAGCCCCGTGGTGAAGGGATAGTTCTCTCACGG \\ 
Passive Hole2 7 &  ACTGGCATGATTAAGAACGGATAACCCTCCACCATCAATATGTTTTT \\ 
Passive Hole2 8 &  TTTTTATATTCAACCGTTCTAGCTGACAACATGT \\ 
Passive Hubble3 1 &  CGCCAGGGGGCAAAATTGTTTGATGGTGGTTCTCACCGCCTGGCCTTTTT \\ 
Passive Hubble3 2 &  TTTTTCTGAGAGAGTTGCTCTTTTCACCAGTGAGACGTTTTT \\ 
Passive Hubble3 3 &  CGAAATCTGGTTTTAGCAAGCGGTTGAGGATCCCCGGGTACCGTTTTT \\ 
Passive Hubble3 4 &  TTTTTAGCTCGAATTCGTTTGTTATCGGCCAACATTTTCAAATGACGCTCACAATCCA \\ 
Passive Hubble3 5 &  TTTTTCAGGCGAAAATCCCCCTTATAGTAATAACCGTACTATGGTTGCTT \\ 
Passive Hubble3 6 &  GTTTCCTGAAAGTGTATCCTCACAGTCCACGCTGGTTTGCCCCAGTTTTT \\ 
Passive Hubble3 7 &  TTTTTGCGTCCGTGAGCCAAGCCTGGGTATTGGGAACGCGCGGGGAGAGG \\ 
Passive Hubble3 8 &  GAAGCATTGTGAAAAATCATGGTCATAGCTGTGCCTGTTCTTCTTTTT \\ 
Passive Hole3 1 &  TTTTTGGCAACAGCTGTCAAACTATACCGCCAGCTTAATGCAAGAAAA \\ 
Passive Hole3 3 &  TTTTTCCACACAACATACGAGCCGGCCCGCTTTATATTTTAATGCTGAATCGCAAG \\ 
Passive Hole3 4 &  GAAATGGGACCAGTAATAAAAGGGACATTCTCGCTCACAATTTTTTT \\ 
Passive Hole3 5 &  ATTCCATAGATTTAGTTTGACCATTAGATACAGTCCAATACTGTTTTT \\ 
Passive Hole3 6 &  TTTTTCGGAATCGTCATAAATATTACTATTATAGACCGGA \\ 
Passive Hole3 7 &  AAACGAAAGAGGCAAAAGATCATCGCCTGATATTTTT \\ 
Passive Hole3 8 &  TTTTTAATTGTGTCGAGCCCCCTTTCACCGGAACCAGTTTGCCACCACCGG \\ 
\\
\caption{\textbf{Staple sequences for V-triangle.}}
\label{vseq}
\end{longtable}

\begin{longtable}{c c}
Name &  Sequence \\ 
\hline 
\endfirsthead

Name &  Sequence \\ 
\hline 
\endhead

\endfoot

\endlastfoot
Core 1 Side1 1 &  TATAATCAACTATGGGTAAAGGTATGTCAA \\ 
Core 1 Side1 2 &  AGCCTTTATTTCAACAAAAGGGTGAGAATC \\ 
Core 1 Side1 3 &  CTGATAGCCCTGAGAAGTGTTTTTCCTTTG \\ 
Core 1 Side1 4 &  CCGATTGGCGTTTTCATCGATTTCTGCTCA \\ 
Core 1 Side1 5 &  AGTCACGACGTTGTAACCAGGCAGTGTAGGT \\ 
Core 1 Side1 6 &  ACCAGTCCCAGAGCCAGACGATTGGCCTTGA \\ 
Core 1 Side1 7 &  ACACTGGTGTGTTTCCACCATCATCACCGAC \\ 
Core 1 Side1 8 &  AGGAAGATATATTTTGTTAAAATTCGCATTAA \\ 
Core 1 Side1 9 &  AGCATCAGCGGGGTCATTGCAGGGTGCCGGGAAATTA \\ 
Core 1 Side1 10 &  GAACCAGAATCACCTAATCAGTAGCGACAGAATCAAGT \\ 
Core 1 Side1 11 &  CTCCGGCTCATATGTACCCCGGAAACTAGCTTCTTTGC \\ 
Core 1 Side1 12 &  ATATGATATTCAACCGACGACAGTAACGGCAGCGGGAGA \\ 
Core 1 Side1 13 &  TGATGAAGGGTAAAGTTAAACGACTTATTAGAGGGAGGG \\ 
Core 1 Side1 14 &  TTAGACGGAATTTGCCACTCAAACTTACCGCCAGCCATT \\ 
Core 1 Side1 15 &  TGGTAGAATATCACCGCAGAGCACTCTCGTCGCTGGCAGC \\ 
Core 1 Side1 16 &  CCGGCACCGCTTCTGGTGCCGGAAAAACGACGTCACCGCC \\ 
Core 1 Side1 17 &  TAAATTGTAAACGTTATGTATAAGCAAATATTGCGGTATG \\ 
Core 1 Side1 18 &  TGATAAATATGAACGGTAATCGTATTGATAATCAGAAAAG \\ 
Core 1 Side1 19 &  TCGTCTTCGCGTCCGTTGCCTAATGAGGGTCACTGTTGCC \\ 
Core 1 Side1 20 &  ATCTTTTCCCGGAACCTGCTGATTAACGTCAGCGTGGTGC \\ 
Core 1 Side1 21 &  GTAAGAATACGTGGCACAGACAATATTAGAGGGTAAGCGCA \\ 
Core 1 Side1 22 &  ACAAAGTCTTTGAATGACCGAGTAAAAGAGTTAGAAGAAGTTACA \\ 
Core 1 Side1 23 &  AACGTCACTTCATTAATTTGGGAATTAGAGCCACCAGAGCGAAAC \\ 
Core 1 Side1 24 &  TATTAAAAATATCCATCTTCAGCAAATCGTTAAGGCCGGAGACAG \\ 
Core 1 Side1 25 &  GAGAATTAGTGAGGCCGCTATTAGTCTTTAATGCGCGAACTGCGGC \\ 
Core 1 Side1 26 &  TTGAGCCAAGGTGAATCACCCTGACACTCAATCCGCCGGGCGCGGTT \\ 
Core 1 Side1 27 &  TTATCCTGAAATAAACAGAGCCGCCAGCCGCCACCCTCAGAACCGCCA \\ 
Core 1 Side1 28 &  GAGGGGACGTTCTAGCTTCATAAACATCCCTTCCCGAACGACAACTCG \\ 
Core 1 Side1 29 &  AAAGATTCGCAAGGATAACGAGCGGAACAATATATCGGCCTTGCTGGT \\ 
Core 1 Side1 30 &  AATCTTATGACAGGGAGTAATAAGCGCCATTCGCCATTCAGGCTGGCCTCTT \\ 
Core 1 Side1 31 &  AAACAGGGAATTGAGCTATTGACGACTTGTAGGCCGTTCCGGCAAACGCGGTCCG \\ 
Core 1 Side1 32 &  GCCAGCATCCAACGCTAAAAATTTTTAGAACCCTCATATATTTTCCCCTACAATT \\ 
Core 1 Side2 1 &  TGAGCAAAAGAAGATTACATTTGGCTCCA \\ 
Core 1 Side2 2 &  GGCCAACGACGCTGAGGTCTGAGAGACTA \\ 
Core 1 Side2 3 &  AATTAAACGTATAAACAGTTAATGCCCCCT \\ 
Core 1 Side2 4 &  GAGCCAGCAGCAAATGAAAAATCAATATAC \\ 
Core 1 Side2 5 &  AGACAAAGAACGCGACAGAACGCGCCTGAT \\ 
Core 1 Side2 6 &  GGTGAATTCATATGGTTTACCATTAGCAAA \\ 
Core 1 Side2 7 &  AAGTTACCAGAAGCGCGGCAGCACCGTCGGT \\ 
Core 1 Side2 8 &  GAAAACATAGCGAGCAAGAAACAATGACCAT \\ 
Core 1 Side2 9 &  ATATTTTAGTTAATTCGACGACAGAGAATAT \\ 
Core 1 Side2 10 &  AGATAGCTTAGATTAAAGAGATAGTCACCAGT \\ 
Core 1 Side2 11 &  ATTTACGAAATACATACATAAAGGACAATCAA \\ 
Core 1 Side2 12 &  GAATCGCCATATTTAAAGTAGGGCTAAAGAAA \\ 
Core 1 Side2 13 &  AATTCTTAAATTTATCAAATGCTGAACCTCAA \\ 
Core 1 Side2 14 &  CTTTTTAAGAAAAGTACTATCTTACCGAAGCC \\ 
Core 1 Side2 15 &  GATCATTTTCGAGCCATTAGTATCTTTCAAAT \\ 
Core 1 Side2 16 &  TATAGAAGGCCAGAATGGAAAGCGCCATCCTA \\ 
Core 1 Side2 17 &  TTAAGCCAATGAAACGTAGAAGCATGTAGAAACTAA \\ 
Core 1 Side2 18 &  CCTTTTTAGCAATACTAACCCTTCTGACCTGAAAGC \\ 
Core 1 Side2 19 &  CACACGATTTACATTGGCAGATACCTACATTTTGACG \\ 
Core 1 Side2 20 &  ATTTTCAGGTTTAACCCAGTATAAAGCCAAAGCCTGT \\ 
Core 1 Side2 21 &  CCACGGAATAAGTTTATTTTGTCTGGCAACATATAAAA \\ 
Core 1 Side2 22 &  GATAGCAAGGATTAGCGGGGTTACAGTGCCAGTAATTC \\ 
Core 1 Side2 23 &  CTCAATCGTATTATAACTATATGGATTTATCAACAATA \\ 
Core 1 Side2 24 &  AAGAAAAATAATATCCAGTCTCTTAAATGCTCAATCGCA \\ 
Core 1 Side2 25 &  GAAAACTTATATGCGTTCTATCAACAGTTGAAAGGAATT \\ 
Core 1 Side2 26 &  GATGAAACGTCAGATGTAAAGCATCACCTAGGGACATTCT \\ 
Core 1 Side2 27 &  GGTTAACATCCTGAACGAAACGCAAAGATTGTATCGGTTT \\ 
Core 1 Side2 28 &  TCGTCACCGTAGCAACAGTTTTGTCACGTTGAACTTTTTC \\ 
Core 1 Side2 29 &  TACCAGGCCACGCCACCACCCTCAGGCTACAGAGGCTTTT \\ 
Core 1 Side2 30 &  CAATAATATAGAAAATAGCAATAGAGCAGATAGCCGAACA \\ 
Core 1 Side2 31 &  GTAATAAATAAACATAACGGGGTCAGTGCCTTGAGTATTGCTCAG \\ 
Core 1 Side2 32 &  TCCCTTAGAATCCTTGCTATTAATTAATTTGCTTCTGTAAATCGTC \\ 
Core 1 Side2 33 &  TCTGTATTTTAATGGAAACAGTACATTTGAATTACCTTTGGGATTT \\ 
Core 1 Side2 34 &  GATGCAATCAAGATTACGATTTTTTGTTTAACCTCCGGCTTAGGTTG \\ 
Core 1 Side2 35 &  TTGCGTAGAGTAACAGAAAAAAAAAACAATTTCATAAATCAATATAT \\ 
Core 1 Side2 36 &  CAACGCCATGCTTGAGGACTAAAGAAATCTCCTACCTTTTCAATTACC \\ 
Core 1 Side2 37 &  GGTCAGTTGGCAAGAAATGGATTACCAGTAATAAAATCATAGAAGAGT \\ 
Core 1 Side2 38 &  GTGAGTGAGCGCCGACAAAGGAGCCTTTAACACAATAGTGAAACATCA \\ 
Core 1 Side2 39 &  AAATCAGACGTCATACATGGCTTTTTACCGTTCCAGTAAGTTAAAATC \\ 
Core1 Side12 1 &  AAACAGCGGTCTCCGTTTTTTGGTGAAGGGATAGCT \\ 
Core1 Side12 2 &  TAAGATTTTTACGCGAGGCGTTGTTGGGA \\ 
Core1 Side12 3 &  GCAACAGGAAAAATTTTTCGCTCATGGAAAT \\ 
Core1 Side12 4 &  TCTCCGTTGAGCGAGTAACAACCCACTCCAG \\ 
Core1 Side12 5 &  ATAGCAGTTTTTCCTTTACAGAGAGAATAACATAA \\ 
Core1 Side12 6 &  TTTGCCGCCAGTTTTTCAGTTGGGCGGGAGACGCAG \\ 
Core1 Side12 7 &  CCCACGCAACCATTTTTGCTTACGGCTGGAGGTGTCC \\ 
Core1 Side12 8 &  TCTTTGATTAGTAATATTTTTACATCACTTGCCTGAG \\ 
Core1 Side12 9 &  TGGTCTGGTCAGCAGCAACCGCAATTTTTGAATGCCAA \\ 
Core1 Side12 10 &  AAGCCGTTTTTATTTTCATTTTTTCGTAGGAATCATTACC \\ 
Core1 Side12 11 &  CTGTCCATCACGCAAATTTTTTTAACCGTTGTAACGTCAAA \\ 
Core1 Side12 12 &  CGGGTATTAAACCAAGTACTTTTTCGCACTCATCGAGAACAAGC \\ 
Core1 Side12 13 &  AAGGTAAAGCTAATATCAGAGAGATAACCCACATTTTTAGAATTGAG \\ 
Core1 Side12 14 &  CCAGCTTTGGCCTCAGGAAGATCGATACTTTTTCAGATGCACAATTCG \\ 
Core1 Side12 15 &  TATTCACTTTTTAAACAAATAAATCCTCATTAAAGCTTATCTAGCAAGC \\ 
Core1 Side12 16 &  ACATAAAAAAATCCCGTAAAAATTTTTAAGCCGCACAGGCGGCCTTTAG \\ 
Core1 Side12 17 &  GTTGCTATTTTTTTTTGCACCCAGGACTTGCGTTTTTGGAGGTTTTGAAGCC \\ 
Core1 Side12 18 &  GCAGGTCACCACCCTCAGCCATATTATTTATCCCAATTTTTTCCAAATAAGAAA \\ 
Core 2 Side3 1 &  GCAAAAACGAAAGAGGCGAGAGAAAGATT \\ 
Core 2 Side3 2 &  CGTTTACACATTCATTCCCAATTCTGCGA \\ 
Core 2 Side3 3 &  CGTACAATACCAAGTTACAAAATGACAGGT \\ 
Core 2 Side3 4 &  AACGTCAAATCATTGTGAATTAGCTCATTC \\ 
Core 2 Side3 5 &  AATCAGAGCCTAACGTGCTTTCCTTCATTG \\ 
Core 2 Side3 6 &  AAAGGAGCGGGCGCTAACCCTAAAGGGAGC \\ 
Core 2 Side3 7 &  AACAAAGTACAACGGAGATTCTATGCATCAG \\ 
Core 2 Side3 8 &  ACACTCATCTTTGACCCCCAGCGACAGGTAG \\ 
Core 2 Side3 9 &  ACCTGCTCCATGTTACTTAGAAGGGGAAGAA \\ 
Core 2 Side3 10 &  AGGCACCAACCTCTCAGTTTTGCAATCCCCC \\ 
Core 2 Side3 11 &  GGTGCCGTAAAGCACTGGACTCCGTTTTTCT \\ 
Core 2 Side3 12 &  ACATACGATGTAGCGGTCACGCTGCGCGTAA \\ 
Core 2 Side3 13 &  TCCAATAAATCATACAGGCAAGGAAGGCTTG \\ 
Core 2 Side3 14 &  CAGAGGGGGTAATAGCCAAAATAGCAAAAGA \\ 
Core 2 Side3 15 &  ATACAGATAAATAAGGGTAAAATATTAGACTG \\ 
Core 2 Side3 16 &  CCCCGATTATCCTGTTGCTGATTGCCCTTAAT \\ 
Core 2 Side3 17 &  GGTCAGGATTAGAGAGCGAAAGACAGTTTCAG \\ 
Core 2 Side3 18 &  ACCGACCGGGAATACCCAGACGACGATAAAAA \\ 
Core 2 Side3 19 &  CTTGAGATGGTTTAATTTCAACTTTAAGGGCGA \\ 
Core 2 Side3 20 &  ACGAACCACCAGCAGTCACAATCGTAATCTGAGAGA \\ 
Core 2 Side3 21 &  CTCGAATTTCCACACAGTGAGACGGGCAACACAAGAGT \\ 
Core 2 Side3 22 &  TTGCTTTGAACACCGCTCTGAATATCGTTAGGGAATTA \\ 
Core 2 Side3 23 &  CGGGAGCTACAGAGGTGAGGCGGTCAGTATTACGAGCA \\ 
Core 2 Side3 24 &  GAACGTGCAGTTCAGAAAACGACATAAATAATGGAAGG \\ 
Core 2 Side3 25 &  ACCGTCCAATACTGCGGAATCGTGAATGACCTTGTATGG \\ 
Core 2 Side3 26 &  CGTATGTGTGAAATTGCCCGCCGCGCTTAATGCGCCGCT \\ 
Core 2 Side3 27 &  CACTACGATCATCATACTAACAACTAATAGATTAGAGCC \\ 
Core 2 Side3 28 &  ATTATACCGCGCCTGTGTCAATAGAACCACCAGAAGGAGC \\ 
Core 2 Side3 29 &  GCTTTTGCAAAGAACATAACGAGATGTGGTGCAGCTGTTT \\ 
Core 2 Side3 30 &  ACGAGTAGATTTAGTTCATCAGTTGAGATTTATGTCTAAA \\ 
Core 2 Side3 31 &  TTAGCAAATTGGGGCGAGAGCATAAAGTGTATCATCGCCT \\ 
Core 2 Side3 32 &  AAGCAAAGCGGATTGCTGACTATTATAGTCAGACAGGGCG \\ 
Core 2 Side3 33 &  AAGATAAAAAACAGGAGGCCGATTTTGCGGGAATGCGGCGCAGT \\ 
Core 2 Side3 34 &  GATAAATAACCTGTTTAGCTATATTTTCATATTCCGGATAGGCTGG \\ 
Core 2 Side3 35 &  CTACCATATGATTGCTATAAATCAAAAATATAGAAAGGAAATATGCA \\ 
Core 2 Side3 36 &  TCACACCGCCTGGCCCATGGTCATTGCGGCCAAACAAAGAATAATACA \\ 
Core 2 Side3 37 &  ACGATCCAGCGGGCCGAAATCTACGTTTAAGAACTGGCTCTTTCACCA \\ 
Core 2 Side3 38 &  GCTTTAAAGCGAGAAAGGAAGGGAAGAAAGCGCCACCACATTATCCGC \\ 
Core 2 Side3 39 &  TAAAATGTCGTAATGCGTTAGAACATTATACTCTGCAACAGTGCCACGCT \\ 
Core 2 Side23 1 &  TTAATTGAGGAAGTTTTTTTTCCATTAAACGG \\ 
Core 2 Side23 2 &  TCAAAATTATTTGTTTTTTACGTAAAACAGAAA \\ 
Core 2 Side23 3 &  ACATCGGGAGAAACTTTTTTATAACGGATTCGCC \\ 
Core 2 Side23 4 &  ATAATATAATGCTGTAGCTTTTTTCAACCAAACTAC \\ 
Core 2 Side23 5 &  GAGGAAGGTTATCTAATTTTTTTTATCTTTAGGAGCA \\ 
Core 2 Side23 6 &  CGGAGTGACGCGCAGAGTTTTTTCGAATTATTCATTT \\ 
Core 2 Side23 7 &  ATATCAAACCCTCAATATAATCCTTTTTTTATTGTTTGG \\ 
Core 2 Side23 8 &  GCCTATTTTTTTTTTGAACCTATTATTCTGAATATAATTCCAAC \\ 
Core 2 Side23 9 &  AGGAATTTTTTTGCGAATAATATGCTAAACAACTTTTTTTTCAAC \\ 
Core 2 Side23 10 &  TGTCCAGATCATCTTCTGACCTAAATTTAATTTTTTGGTTTGAAAT \\ 
Core 2 Side23 11 &  TTCCTGATTATCAGATTTTTTTTTGGCAATTCATCAATCAATATCT \\ 
Core 2 Side23 12 &  GATAGTAAAGTATTTTGCGGATGGCTTTTTTTAGAGCTTAATTGCTGGA \\ 
Core 2 Side23 13 &  TTCAAATATCGCGTTTTAATTTTTTTCGAGCTTCAAAGCGTAAATGAATTT \\ 
Core 2 Side23 14 &  ATGAAAGTATTAAGATTTTTTTCTGAGACTCCTCAAGAGAAGGATTAGCCCAAT \\ 
Core 2 Side23 15 &  CGTTAAATAAGTTTTTTATAAACACCGGAATCATAATTACTAGAAAACGCTCAAC \\ 
Core 2 Side23 16 &  ATTTTTTCGTCTTTCCAGACGTTAGAACCAGACCGGAAGTTTTTTTAACTCCAACA \\ 
Core 2 Side23 17 &  AAAGTACCGACAAAAGGTCGAGGCATAGTAAGAGTTTTTTAACACTATCATAACCCT \\ 
Core 2 Side31 1 &  GAGCCTCCTCACATTTTTTTTGAGGATCCCCGG \\ 
Core 2 Side31 2 &  CCTAAAACATCGCCATTTTTTTTAAAATACCGA \\ 
Core 2 Side31 3 &  TTTGAGGATTTAGAAGTATTATTTTTTTCTTTACAA \\ 
Core 2 Side31 4 &  CGGGTTACCTGCAGCCAGCGGTTTTTTGCCGGTGCCCCC \\ 
Core 2 Side31 5 &  GTCATACCGGGGGTTTTTTTTCTGCCAGCACGCGTGCCTG \\ 
Core 2 Side31 6 &  GCCCGAAGGCCGGAAGCATAAAGTTTTTTGTAAAGCCTGGGG \\ 
Core 2 Side31 7 &  AGCCGTGAGCTAACTCACATTAATTGCGTTTTTTTGCGCTCACT \\ 
Core 2 Side31 8 &  TTATTAATTTTAAAAGTTTTTTTTGAGTAACATTATCATTAAAGGGA \\ 
Core 2 Side31 9 &  ATTATACCAGCTATCAGGTCATTGCCTAGAACGCCATCAGTAAATTG \\ 
Core 2 Side31 10 &  TTTTCACGGTACCGAGTTTTAGACATTTTTTTAACGGTACGCCAGAAT \\ 
Core 2 Side31 11 &  CGAGCTGAAAAGGTGGCATCAATTCTACTTTTTTTATAGTAGTCCAGCTTT \\ 
Core 2 Side31 12 &  AGGACGTTCAATAAAGCCGGTCACGTTGGTGTGATTCCTGTAGAGCATTAA \\ 
Core 2 Side31 13 &  CCCCAAACGCGCGGGGAGAGGCGGTTTTTTTTGCGTATTGGGCGCCAGGGTG \\ 
Core 2 Side31 14 &  CAGTCGGGAAACCTGTCGTGCTTTTTTAGCTGCATTAATGAATCGGCCAAAAC \\ 
Core 2 Side31 15 &  AATCGGTTGTATTTTTTCAAAAACATTATGACCCTGTACGTCGGATGCCAGTTT \\ 
Core 2 Side31 16 &  CATCAACATTAAATGGGGAACAAACGGCGGATTGTTTTTTCCGTAATGGGATATC \\ 
Core 2 Side31 17 &  GGAGAGGGTAGCTATTTTTTTTTTTAGAGATCTACAAAGGTCAGTGAATCAAAGAA \\ 
Core 2 Side31 18 &  ATTTTTGTTAAATCAACGTGAACCTTTTTTTCACCCAAATCAAGTTTTTTGGGGTCGA \\ 
Active Hubble1 1 &  ACTGTAGCCGTTTGCCTTGCCTTTATAGCCCC \\ 
Active Hubble1 2 &  GGGGATGTGCTGCAAGCGCCAGCTTCGGTGCGGCGCAACT \\ 
Active Hubble1 3 &  ACGCCAGGCGCTATTAGCGATTAACCATGTTTGCCTCCCT \\ 
Active Hubble1 4 &  CAGAGCAAAGCCACCAATAATCAAAATCACCGCAATGAAACCATC \\ 
Active Hubble1 5 &  TGTGAGAGATAGACTTATCAAACTTAAGCATTTTCGGTCAGCGTCAG \\ 
Active Hubble1 6 &  GATAGCAGCACCGAGTAGCACAACAATCGACCACCACCAGAGAATCAGAGCCT \\ 
Active Hubble1 7 &  TAACGGAATATTTTTCCCAAAAGAACTGGCCTCGGAATTAGGGCGAGGCGAAAG \\ 
Active Hubble1 8 &  TTAGCGAACCTAAATGCAATGCCTAGGTTGAGGTTTTCCCGTACAGCGGTTGGGTA \\ 
Active Hole1 1 &  CTGGCCTGGGCGCATCGTAACCGTGCATCT \\ 
Active Hole1 2 &  GAGCCGCCACGGGAACCAAGCTTTCAGAGGTG \\ 
Active Hole1 3 &  GCCAGTGCGGATAACCTCACCGGACATTACCATTA \\ 
Active Hole1 4 &  AATAGGAGTCTGGAGCAAACAAGAGAATCGTAATGCC \\ 
Active Hole1 5 &  GCAAGGCCGGATTTTTTCGATCCTCATAACGGAACCGCTTTCG \\ 
Active Hole1 6 &  CCCTGACGAGAAACACTTTTTTTGAACGAGTAAAAATAATTCGCGT \\ 
Active Hole1 7 &  AAAACCGTCTATCAGGGCTTTTTTTTGGCCCACTGCTCATTTTTTAACC \\ 
Active Hubble2 1 &  CCCTCAGACGTTATTCGGTCGCTGAGG \\ 
Active Hubble2 2 &  ATCGCCCACGCATAATTTCTTAAACAGCTTGA \\ 
Active Hubble2 3 &  GAACGAGGCTCAGCAGCGAAAGACAATGACAACAACC \\ 
Active Hubble2 4 &  GGCCGCTTTTGCGGGACTTGCAGGCGATCTAATTTTCAGG \\ 
Active Hubble2 5 &  AGCATCGCGAGGTGAACCGATAACTCAGGAGGTTTAGTACCGC \\ 
Active Hubble2 6 &  CACCCTCAGAACCTGCCGTCGATAAGTTTACATGTTCAGCTAATG \\ 
Active Hubble2 7 &  ATCAGCTACATGTAGGTGTATGGATAAGGCCACCCTCAGAACCGCCA \\ 
Active Hubble2 8 &  AACGCCTGTTTTTTTTCATTCCACAGACAGCCCTCATAGTTAGCGTAAGAGTTAAA \\ 
Active Hole2 1 &  CTAAACGCAACAATCAATAATCGGCTGT \\ 
Active Hole2 2 &  TGATGATACAGGAGTGTACTGGTAAGAGGGTTGAT \\ 
Active Hole2 3 &  TACCGATAGTTATAACCTTAGAAAACAAAATTAAT \\ 
Active Hole2 4 &  ATAAGTATAGCCCGGAATAATTTAGGCAGAGGAAG \\ 
Active Hole2 5 &  CTTTCCTTATCATTCCAAGAAGCGCCCAACGGTATTC \\ 
Active Hole2 6 &  CGAGGCAGCAGTATGGCGCCAAAGACAAATTTTTAGGGCGACATTCAA \\ 
Active Hole2 7 &  CCATGATTAAGACTCCTTATTTTTTTACCGGAAAAATTGTGTACATGAAAC \\ 
Active Hubble3 1 &  GAAATCGGCAAAAGTCCACGCTGGTTTGCCCC \\ 
Active Hubble3 2 &  GCAAGCGTCCCTTATACTGACCAACTTTGAAAGAGG \\ 
Active Hubble3 3 &  GGTCAATCATAAGACAAAGCTCCTTATGCGATTTGCA \\ 
Active Hubble3 4 &  CCACTATTTTGTTCCAGTTTGGAATGATGGTGGTTCC \\ 
Active Hubble3 5 &  TCAACGTAGGAACCGAAAATCAAAAGAATAGCCCGAG \\ 
Active Hubble3 6 &  ATAGGGTTGAGTGAAAGAACGTAAATCGGAGGGCGCTGGCCTTTC \\ 
Active Hubble3 7 &  ACAGATGAACGGTTCATCAAGTGGTCAATTGTGTCGAACATTATT \\ 
Active Hubble3 8 &  GTTAATAACCCAAACTGACCTGTACAGACCAGGCGCAACGAGGCGCAGAC \\ 
Active Hole3 1 &  TGCTCCTTTTGATAAGTAATGTTTTAACAACTAA \\ 
Active Hole3 2 &  TGACCATTAGATACATTTCGCAAAAGTAATCTTGA \\ 
Active Hole3 3 &  AGCAGGCGAAATAGAGCTTGACGGGGAAAGCCGGC \\ 
Active Hole3 4 &  CTTTACCCATCAAAAAGATTAAGAGGAAGCCTACCTTTAAT \\ 
Active Hole3 5 &  TGGAAGTTTCACAGTTGAACTAATGCAGATACATTTTTTTACGCCAAAAGG \\ 
Active Hole3 6 &  CAAGAACCGGATATTCATTAAACGAACAATCCGCGGTCACTGCAAGCGCGA \\ 
Active Hole3 7 &  AGGAACCCATGTACCGTTTTTTTACACTGAGTTTCGTCACCAGAGGTCATCGGTGTC \\ 
Passive Hubble1 1 &  TTTTTACTGTAGCCGTTTGCCTTGCCTTTATAGCCCC \\ 
Passive Hubble1 2 &  TTTTTGGGGATGTGCTGCAAGCGCCAGCTTCGGTGCGGCGCAACT \\ 
Passive Hubble1 3 &  TTTTTACGCCAGGCGCTATTAGCGATTAACCATGTTTGCCTCCCT \\ 
Passive Hubble1 4 &  CAGAGCAAAGCCACCAATAATCAAAATCACCGCAATGAAACCATCTTTTT \\ 
Passive Hubble1 5 &  TGTGAGAGATAGACTTATCAAACTTAAGCATTTTCGGTCAGCGTCAGTTTTT \\ 
Passive Hubble1 6 &  TTTTTGATAGCAGCACCGAGTAGCACAACAATCGACCACCACCAGAGAATCAGAGCCT \\ 
Passive Hubble1 7 &  TAACGGAATATTTTTCCCAAAAGAACTGGCCTCGGAATTAGGGCGAGGCGAAAGTTTTT \\ 
Passive Hubble1 8 &  TTAGCGAACCTAAATGCAATGCCTAGGTTGAGGTTTTCCCGTACAGCGGTTGGGTATTTTT \\ 
Passive Hole1 1 &  TTTTTCTGGCCTGGGCGCATCGTAACCGTGCATCT \\ 
Passive Hole1 2 &  TTTTTGAGCCGCCACGGGAACCAAGCTTTCAGAGGTGTTTTT \\ 
Passive Hole1 3 &  GCCAGTGCGGATAACCTCACCGGACATTACCATTATTTTT \\ 
Passive Hole1 4 &  TTTTTAATAGGAGTCTGGAGCAAACAAGAGAATCGTAATGCC \\ 
Passive Hole1 5 &  TTTTTGCAAGGCCGGATTTTTTCGATCCTCATAACGGAACCGCTTTCG \\ 
Passive Hole1 6 &  CCCTGACGAGAAACACTTTTTTTGAACGAGTAAAAATAATTCGCGTTTTTT \\ 
Passive Hole1 7 &  AAAACCGTCTATCAGGGCTTTTTTTTGGCCCACTGCTCATTTTTTAACCTTTTT \\ 
Passive Hubble2 1 &  TTTTTCCCTCAGACGTTATTCGGTCGCTGAGG \\ 
Passive Hubble2 2 &  TTTTTATCGCCCACGCATAATTTCTTAAACAGCTTGATTTTT \\ 
Passive Hubble2 3 &  GAACGAGGCTCAGCAGCGAAAGACAATGACAACAACCTTTTT \\ 
Passive Hubble2 4 &  TTTTTGGCCGCTTTTGCGGGACTTGCAGGCGATCTAATTTTCAGG \\ 
Passive Hubble2 5 &  AGCATCGCGAGGTGAACCGATAACTCAGGAGGTTTAGTACCGCTTTTT \\ 
Passive Hubble2 6 &  TTTTTCACCCTCAGAACCTGCCGTCGATAAGTTTACATGTTCAGCTAATG \\ 
Passive Hubble2 7 &  ATCAGCTACATGTAGGTGTATGGATAAGGCCACCCTCAGAACCGCCATTTTT \\ 
Passive Hubble2 8 &  AACGCCTGTTTTTTTTCATTCCACAGACAGCCCTCATAGTTAGCGTAAGAGTTAAATTTTT \\ 
Passive Hole2 1 &  CTAAACGCAACAATCAATAATCGGCTGTTTTTT \\ 
Passive Hole2 2 &  TGATGATACAGGAGTGTACTGGTAAGAGGGTTGATTTTTT \\ 
Passive Hole2 3 &  TTTTTTACCGATAGTTATAACCTTAGAAAACAAAATTAAT \\ 
Passive Hole2 4 &  TTTTTATAAGTATAGCCCGGAATAATTTAGGCAGAGGAAG \\ 
Passive Hole2 5 &  TTTTTCTTTCCTTATCATTCCAAGAAGCGCCCAACGGTATTC \\ 
Passive Hole2 6 &  TTTTTCGAGGCAGCAGTATGGCGCCAAAGACAAATTTTTAGGGCGACATTCAA \\ 
Passive Hole2 7 &  CCATGATTAAGACTCCTTATTTTTTTACCGGAAAAATTGTGTACATGAAACTTTTT \\ 
Passive Hubble3 1 &  TTTTTGAAATCGGCAAAAGTCCACGCTGGTTTGCCCCTTTTT \\ 
Passive Hubble3 2 &  GCAAGCGTCCCTTATACTGACCAACTTTGAAAGAGGTTTTT \\ 
Passive Hubble3 3 &  TTTTTGGTCAATCATAAGACAAAGCTCCTTATGCGATTTGCA \\ 
Passive Hubble3 4 &  CCACTATTTTGTTCCAGTTTGGAATGATGGTGGTTCCTTTTT \\ 
Passive Hubble3 5 &  TCAACGTAGGAACCGAAAATCAAAAGAATAGCCCGAGTTTTT \\ 
Passive Hubble3 6 &  TTTTTATAGGGTTGAGTGAAAGAACGTAAATCGGAGGGCGCTGGCCTTTC \\ 
Passive Hubble3 7 &  TTTTTACAGATGAACGGTTCATCAAGTGGTCAATTGTGTCGAACATTATT \\ 
Passive Hubble3 8 &  GTTAATAACCCAAACTGACCTGTACAGACCAGGCGCAACGAGGCGCAGACTTTTT \\ 
Passive Hole3 1 &  TTTTTTGCTCCTTTTGATAAGTAATGTTTTAACAACTAA \\ 
Passive Hole3 2 &  TGACCATTAGATACATTTCGCAAAAGTAATCTTGATTTTT \\ 
Passive Hole3 3 &  TTTTTAGCAGGCGAAATAGAGCTTGACGGGGAAAGCCGGC \\ 
Passive Hole3 4 &  CTTTACCCATCAAAAAGATTAAGAGGAAGCCTACCTTTAATTTTTT \\ 
Passive Hole3 5 &  TTTTTTGGAAGTTTCACAGTTGAACTAATGCAGATACATTTTTTTACGCCAAAAGG \\ 
Passive Hole3 6 &  TTTTTCAAGAACCGGATATTCATTAAACGAACAATCCGCGGTCACTGCAAGCGCGA \\ 
Passive Hole3 7 &  AGGAACCCATGTACCGTTTTTTTACACTGAGTTTCGTCACCAGAGGTCATCGGTGTCTTTTT \\ 
\\
\caption{\textbf{Staple sequences for X-triangle.}}
\label{xseq}
\end{longtable}

\begin{longtable}{c c}
Name &  Sequence \\ 
\hline 
\endfirsthead

Name &  Sequence \\ 
\hline 
\endhead

\endfoot

\endlastfoot
Core 1 Side1 1 &  AGCCGCCGAAACAAATTCAGAACCACCCTCAGAACCGCCATACTCAGGGTCGAGAG \\ 
Core 1 Side1 2 &  GTGCACTCAGGCTGGCCAAGCTTTAACACCGTCGGTGGTGATTTGCCGTTAGCTAT \\ 
Core 1 Side1 3 &  ATAAACAGATCGCCCAAAAGCACTTAATGCGCCGCTACAGGGCGCGTTGCAGGG \\ 
Core 1 Side1 4 &  ATTTGTATGACCTGCTCTGAGGCTACTATGGTTGCTTAAAATCCCGTA \\ 
Core 1 Side1 5 &  TGAGCTAAGTGGTTTTTCTTTTCATGGTTCCGTTTGATGAGTGCCCGT \\ 
Core 1 Side1 6 &  CAACGGAGGGTTGATATTCGGTCGGTGCCTTGCAATACTTCTTTGATA \\ 
Core 1 Side1 7 &  AGTAAGTCAGACGATTATTAGGATTTTAGTTTTAACCTGTCCAGCAGT \\ 
Core 1 Side1 8 &  GTAGATGGGCGCATCGTAACCGTGCATCTGCGAACTGACCAGACGGT \\ 
Core 1 Side1 9 &  CCTACATTAGGGAACCCAGGCCATTGCAACAGGAAAAACGCTATGAG \\ 
Core 1 Side1 10 &  TATTTCAACCGCACAGTGGGCGGTGGAATTTGCAGCGCCATGTTTAC \\ 
Core 1 Side1 11 &  ATCGGCGACAGAGGTGATTACGCCGATCGGTGCGGGCCTGGCCTTGC \\ 
Core 1 Side1 12 &  GGAAATACTGCCAGCACGCGTGCCTGTTCTTTTTTCATAGATTATTT \\ 
Core 1 Side1 13 &  TGCAATGCAACATTATGACCCTGTAATACTTATTTTCATAAGAGGT \\ 
Core 1 Side1 14 &  ATTTACCGTTCCAAGTTAGCGTTTGCCATCCGCGTCCGGTGTAAAG \\ 
Core 1 Side1 15 &  AGAGCTTAAAACACCAGTAATCTTCAGCGCAGTGTCACTCTTCCTG \\ 
Core 1 Side1 16 &  AGTAACATACAGGACAGAATGGCACCGGAACCAGAGCCGGGGGTTT \\ 
Core 1 Side1 17 &  AGAGAAGGGGCCTTGAAGGCGCATACAGATGAACGGTGTCAACTTT \\ 
Core 1 Side1 18 &  CTTCGCTGAGCCGCGACGACAGTATCGGCCTCAGGAAGGAAGGGC \\ 
Core 1 Side1 19 &  TTTGACCCCCAGCGAACCCATGTACCGTAAAGAATAGCCCTGCCT \\ 
Core 1 Side1 20 &  CCTGGGGTGCCTACATCTGAAATGATCAAAATAAAGCGCA \\ 
Core 1 Side1 21 &  ACGTTGGTCGTCTGGCGCGCGCCTCAGATTGGGCTTGAGA \\ 
Core 1 Side1 22 &  GCCAGTGCTGACCTTCCGAGTAGATAGTTGCGGATGGCTT \\ 
Core 1 Side1 23 &  TTGCGGGATTTAGAACCCTCATATATTTTAAACCTTTTGA \\ 
Core 1 Side1 24 &  AAGAACTCAAACTATCAGCTAAACACCTCACCATCGACAT \\ 
Core 1 Side1 25 &  GGTGCCGTCGCATAACAGTTAAAGAGCGCGAAACAAAGTA \\ 
Core 1 Side1 26 &  CGGGCCGTTTTCACGGTCATACCACCACCGGCAATCATA \\ 
Core 1 Side1 27 &  TTTGGGGCTCCCAATTCTGCGAAATCAAGAGAACGAGT \\ 
Core 1 Side1 28 &  AGGAGGCCTTAGAATCAGAGCGGGCAGTCCCTGTGTAC \\ 
Core 1 Side1 29 &  CCATGTTAAAATGACGAGCACGTATAACCGAGTTGTAG \\ 
Core 1 Side1 30 &  AAAAAAGTGTGTCGTTCGCAAATGGTCAAGACCATTA \\ 
Core 1 Side1 31 &  AGCGTCATACATTGGCAGATTCAGCATAAATGAGCCT \\ 
Core 1 Side1 32 &  GGTGTGCTTTCCTCGGATTAAAGGGATTTTTGATTGC \\ 
Core 1 Side1 33 &  CATTTCGGGGTTTTGCTAAAAATTGAAGCCTT \\ 
Core 1 Side1 34 &  GCCGGAAACGTCACCACAGTTTCAGCGGAGTG \\ 
Core 1 Side1 35 &  CAATAGGATTATACCAGCCGCTTTTGCGGGAT \\ 
Core 1 Side1 36 &  GTCTCTGAATTTCGGAACCTATTAGCCAAGCC \\ 
Core 1 Side1 37 &  CATCAAAAATAATTCGTGCGGCCAGGTTGAGG \\ 
Core 1 Side1 38 &  CCAGTGAAATATTACCGCCATTTTGGTAATA \\ 
Core 1 Side1 39 &  GAGGGGACCACGGGAATTTAGGCGACAGACC \\ 
Core 1 Side1 40 &  ATTAAAGCGTGTACTGTTGACGCTCAATCGT \\ 
Core 1 Side1 41 &  GTAATAAGCGGATTAGTAATAACATCACTGG \\ 
Core 1 Side1 42 &  CTTAGCCGAAATCCGCCATCGCCTGATAAAT \\ 
Core 1 Side1 43 &  CGATAGTTGCGCCGCTCAGCAGCACTCATC \\ 
Core 1 Side1 44 &  GGAAACAGATACATGAACGTAACGGGGTCA \\ 
Core 1 Side1 45 &  CGCAAGGATCAGTACCAGGCGGACTCCTCA \\ 
Core 1 Side2 1 &  CGTATTAACTTTACAATATTCCTGCAAGCCGTTATAGAAGGCTTATCCACCCACAA \\ 
Core 1 Side2 2 &  GTGTGATATGGCTATTAGTCTTTAATGCGCGTTTACGAGAATCATAAAGAACAAG \\ 
Core 1 Side2 3 &  CTTTAATTGTATCGGGGCTCCAATGGACTCCGAGTGTTGCGGTCCACGCCCTGAG \\ 
Core 1 Side2 4 &  TGACGGAAAGGAATCAAGATGCTGATGCAAATCCAATCAACAATTCAAGAATA \\ 
Core 1 Side2 5 &  TTCATCGAGCAGCACCGTAATCTTTCTGTAGAATAATAAAACAGCTTGATAC \\ 
Core 1 Side2 6 &  ACATTCAAAGGAGTGAATAACCTTGCTTTCGGTTATATAATCAGATTTTATT \\ 
Core 1 Side2 7 &  AGCGGTCTACAGAGCGAGGTGAGTTAGCCGTTAGTAAATGAATAGTAGCGA \\ 
Core 1 Side2 8 &  CGCTCACAATTCTGGCCAACAGAGAGCCATTTATAGGGTTAACGTTTT \\ 
Core 1 Side2 9 &  CATCCTAAAACTGATATAAAGCATCACCTTGCTGAACACCAACACCGG \\ 
Core 1 Side2 10 &  CCGATTGACAAAGACAAAATTCATATGGTTTAATGAGGAAGTTTCCAT \\ 
Core 1 Side2 11 &  TGCAAATTTAATGGTTTGAAATACCGCTCAAATATCAAAGTATTAGA \\ 
Core 1 Side2 12 &  GGCCAACGCTTTCCAGTCGGGAAACCTGTCGTTAAAGGTGTAAATAT \\ 
Core 1 Side2 13 &  ATAAAAGAAACGCAAGACTTGCGGAGAGATAGGTATTCTTTTTTAAC \\ 
Core 1 Side2 14 &  GTAAAGTCATCCTTGAGCTATTAATTAATTTTGTCTATCCGTGGCG \\ 
Core 1 Side2 15 &  AAACATAGCGATAGCTCATCGCAAGACAAAGTTAATTTGACAACT \\ 
Core 1 Side2 16 &  ACCGTCACCGACTTGATAGAACCCTTTATCGAATCTGACCTGAAA \\ 
Core 1 Side2 17 &  AGGGCGATAAAGGGCGAAAAACCCAGCAAGTTCCAGTATGAATC \\ 
Core 1 Side2 18 &  TTGTTATCACTGCCCGCGCGGGGAGAGGCGGTAGCCCGAG \\ 
Core 1 Side2 19 &  AGCCATATTTGAAGCCTTAAATCAATAAAGGTGGCAACAT \\ 
Core 1 Side2 20 &  ACGTGGCACAGACAATATTTTTGAAAATAAGGATTATTCA \\ 
Core 1 Side2 21 &  GAACGCGAGAAAACTTCCCTTAGACACTATTAAGAAAGGA \\ 
Core 1 Side2 22 &  TTCATCGTCGTTAAATCATCTTCTGACCTCAGATATTTTA \\ 
Core 1 Side2 23 &  AGAGTTGTTTCAAATCTGCATTATTGGAACATTAGGAAG \\ 
Core 1 Side2 24 &  GATTTAGAGCTTGACCTTGCTTTGCTTTGAGTAAACGGG \\ 
Core 1 Side2 25 &  AATAACGGTTTTTAATGGAAACAGTACATAAAAACAAT \\ 
Core 1 Side2 26 &  CAAAATCAAGTAATAAAAGGGACATTCCACTCATGGT \\ 
Core 1 Side2 27 &  TCTTACCGAAGCCCTAATATCAGGAGGTTTATTTAT \\ 
Core 1 Side2 28 &  TCAATAGAAAAGGGCGAGAGCAAGATCAATAT \\ 
Core 1 Side2 29 &  CATAGCTGTTTCCTGTGTGAAAGCGTAAGAAT \\ 
Core 1 Side2 30 &  CACAGACACACTACGACGAGGGTAGCAACGGC \\ 
Core 1 Side2 31 &  GACTAAAGACTTTTTCGGGCGCTACCGGCGAA \\ 
Core 1 Side2 32 &  GGAGAAACGAATTGAGGAAATAGCAATAGCTA \\ 
Core 1 Side2 33 &  CATTAAAACCTGAACAAGAAAAATTATAAAGT \\ 
Core 1 Side2 34 &  AGAAGGAGCGGAATTATTAGATTATAGGTTGG \\ 
Core 1 Side2 35 &  TTAAGCCCAATAATAAGCAAGCAAACTATAT \\ 
Core 1 Side2 36 &  AGACCGCACTCATCGTTACTAGAATTTAGAA \\ 
Core 1 Side2 37 &  AAGAACGAAGGAGCGGGAGCCGCGCCCAAT \\ 
Core 1 Side2 38 &  AAGAACGCTTAAACCAAGTAGAAAATATCC \\ 
Core 1 Side2 39 &  ATGTGGAAGAAAGCGAAAGGAGCCCAGCGC \\ 
Core 2 Side3 1 &  CTTGTAGACCATATCACAATTACCTCATTGCCGAGAGATCTGCCGGAGAGGGTAGC \\ 
Core 2 Side3 2 &  CTAGCTGACCATAAATCAGGCAAGTAAAATGTATTGCAGGCGCTTCTAACGTCAGC \\ 
Core 2 Side3 3 &  CCAACGCTATTTAACAGGATTATAAAAAATTTATCAAAATACAAAATCGAGAATTA \\ 
Core 2 Side3 4 &  TCAGCAGCAGCATCAGGAAAAAGAATCCAATAAATTAAGCTGATATTCAACCGTT \\ 
Core 2 Side3 5 &  GCATTAACGACGCAGATCCGGCAAACGCGGTCCGTTTTTTTAGCAAAATCATA \\ 
Core 2 Side3 6 &  TTTCAGGTGAAATAAAGAAATTGAACATTATAATGGAGCACTAAGGGCTTA \\ 
Core 2 Side3 7 &  ACAGTTGAAGATTAGAGCCGTCAAGATTGTTTACGCCAACAATAGCAG \\ 
Core 2 Side3 8 &  CGATGAACAAAGCTCGTCTTTTATTGCGGCTGGTATGAGCCGGGTCAC \\ 
Core 2 Side3 9 &  AACGGAATACCCAAAAGAGGAAACGCAATAATACAAACATGCGAATTA \\ 
Core 2 Side3 10 &  GAGGCGGTCAGTATTAGTACCCCGGTTGATAAAGCATGTCAATCATAT \\ 
Core 2 Side3 11 &  AAGCAAATATTTAAATTGTAAACAGGACGTTTTACCTGCACCAAAAT \\ 
Core 2 Side3 12 &  CCTTTACATTAGACGGGCGCAGAGCAAGAAAACAAAATTAATTACAT \\ 
Core 2 Side3 13 &  AAACAGGGAAGCGCAGAGAGAATTTGCACGTCTTCTGAATCATTTTG \\ 
Core 2 Side3 14 &  TAAATCGGAAGGCCGGAGACAGTCAAATCACGCTTCAAATAGTAGTA \\ 
Core 2 Side3 15 &  TAAAGGTTTCATAAACATCCCTTACATGGGAACAAACGGTGGTGCCG \\ 
Core 2 Side3 16 &  GTGCCGGAGCGAGTTTTGTTAAAATTCGCATTAAATTCGATTTTA \\ 
Core 2 Side3 17 &  AACGGAACGTGCCGGACGTAGATTTCATTTAAATTATAACATAA \\ 
Core 2 Side3 18 &  AACTCGCACTCAATCCGCCGTTGGCAAAGCGCATCAGGTGTGTT \\ 
Core 2 Side3 19 &  GTTATCTAGGTAATCGGAGCAAACAAGAGAATGTTTTGCC \\ 
Core 2 Side3 20 &  GTTAATATAACAACCCGTCGGATTCTCCGCTGATGCCGGG \\ 
Core 2 Side3 21 &  ATCCTCATGTGGTGCTGGTCTGGAAGGGCGCGGTTGCG \\ 
Core 2 Side3 22 &  CCTCCGGCCAGAACGCGGGGTCTTAGACTGCTTTTG \\ 
Core 2 Side3 23 &  ATAAAGCCTACTAAGCGAACCTACGGTGCGTCATAA \\ 
Core 2 Side3 24 &  GAAACCAGGGTAACGCAGCGAGAGGGATAGCG \\ 
Core 2 Side3 25 &  TGAGAGGGGGTAATAGCTATCAGGTGAGGCAG \\ 
Core 2 Side3 26 &  TTTAAAAGTTTGAGTTGTTGCCCCTTTAGGA \\ 
Core 2 Side3 27 &  AGGGTTAGAGAGTCTGATTGAGAATCGCCAT \\ 
Core 2 Side3 28 &  CAACAGTACAACTAATAAGGAATTGAGGAAG \\ 
Core 2 Side3 29 &  CTCGTTTACCAGAAACTCATTATACCAGTC \\ 
Core 2 Side3 30 &  TCGCTGCAAAAGAAGATGATGAATATTTTT \\ 
Core 2 Side3 31 &  TCTGGAACCCACGACGATAAAAAGCCAGCG \\ 
Core2 Side12 1 &  TAAAATACGTAATGCGCCCTCATAATTTCTTATTTTTTCCACCGAGTTTTTTAAAAGA \\ 
Core2 Side12 2 &  ACATGGCTAAATCGGCATCACGCAACGGTACGCCTTTTTAGAATCCTGAGAAGTGTC \\ 
Core2 Side12 3 &  CAGTACAAACTTTTTTACAACGCCTGTTTGCTAAACTTTTTAACTTTCAAATGAAAC \\ 
Core2 Side12 4 &  CAACAACCTTAATGCCGTCTGTCCAAAATCCCTTATATTTTTAATCAAAAGAAT \\ 
Core2 Side12 5 &  CTGAGTAGTCCAGAACGACGGGCAACAGCTGATTGCTTTTTCCTTCACCGCCTG \\ 
Core2 Side12 6 &  GCTGGTTTGCCCCAGCTTTTTAGGCGAAAATCCTGTTAGACAGGAAATTAACC \\ 
Core2 Side12 7 &  TTGCGTATTTTTTTGGGCGCCAGGCTCACATTTTTTTAATTGCGTTGCGCTC \\ 
Core2 Side12 8 &  CCTCACAGTTGAGGATCCTTTTTCCGGGTACCGAGCTCGAAGCGCGTT \\ 
Core2 Side12 9 &  CACTGAGTTTCGTCACCACTAAAACGAAAGACATTTTTGCATCGGAA \\ 
Core2 Side12 10 &  GGCCCACTACGTGAACCATCTTTTTACCCAAATCAAGTTTTTTGG \\ 
Core2 Side12 11 &  ACGCTGCGCGTATTTTTACCACCACACCCCGTCACCACAATGA \\ 
Core2 Side12 12 &  ACAACATACGAGCTTTTTCGGAACCAGTCATTTTTCACGACC \\ 
Core2 Side12 13 &  AAAGGAACAACTAAAGGAATTTTTTTGCTGGGATTAGCATTC \\ 
Core2 Side12 14 &  AGGCACCAACCTAAAACGAAATTTTTGAGGCAAAAGAATA \\ 
Core2 Side12 15 &  GCCGCGCTAAATCGGAACCCTAAAGTTTTTGGAGCCCCC \\ 
Core2 Side12 16 &  CATCGATGCATTTTTTTTTCGGTCATAGCCCCCTTA \\ 
Core2 Side12 17 &  CCAGTAGCACTTTTTCATTACCATTAGCAGTA \\ 
Core 2 Side23 1 &  GTATGTTAGCAAACGTAGTTTTTTTAATACATACAGATTAGTTTTTTTTTTATTTTGCAC \\ 
Core 2 Side23 2 &  CAAATAAGAAATTTTTTGATTTTTCCAGCCAGCTACAATTTTAATTACGCAGATAGCCG \\ 
Core 2 Side23 3 &  TTAACGTCAGATGAATATTTTTTTTAGTAACAGTACCTTTTACACTGTAAATCGTC \\ 
Core 2 Side23 4 &  TACCAGAACTGAACACCCTGAAAACGTCAACTGAGATTTTTTACTACCCCCAATC \\ 
Core 2 Side23 5 &  CCCTCAATCAATATTTTTTTTGGTCAGTTGGCAAATCAACACCGCCGTATAAAG \\ 
Core 2 Side23 6 &  ACCGACAAAAGGTAAATTTTTTTAATTCTGTATCAACATTTTTTTAGATAAGT \\ 
Core 2 Side23 7 &  TGTTTAGTATCTTTTTTTATGCGTTATACAAATTCTTACCATGCAACA \\ 
Core 2 Side23 8 &  ATTATCAGCTCCGGCTAGACGCTGTTTTTTTAAGAGTCAATAGTGACA \\ 
Core 2 Side23 9 &  ATTCGCCTGATTGCTTTTTTTTGAATACCAAGTTCATAGGTAAATGAA \\ 
Core 2 Side23 10 &  TAGATAATACATTGAGGAAAAGCCAAAAATCGCCCTAAAACATCGC \\ 
Core 2 Side23 11 &  ATCCTTTGTTTTTTTCGAACGTTATTAATCGGAACAAAGAAACCA \\ 
Core 2 Side23 12 &  CCTGTTTCCAGACGACGACCATTTTCTTTTTTAGCCAGTAATA \\ 
Core 2 Side23 13 &  ATACCGAATTTTTTTAACCACCAGCAGAAGATAAAAGAACGCG \\ 
Core 2 Side23 14 &  AACAAAGTTTAACAATTTTTTTTCATTTGAATTACCT \\ 
Core 2 Side23 15 &  AATTGAGCGCTTTTTAAGAAAATTTTTTTAAGCA \\ 
Core 2 Side23 16 &  AGCCTAATTTTTTTTTCCAGTTACAAAATAAAC \\ 
Core 2 Side23 17 &  ATGATGGCAATTCATTTTTTCAATATAATCCT \\ 
Core 2 Side31 1 &  CTGAGTAATGTGTAGGTTTTTTTAAGATTCAAAACCAACAGGTCTTTTTTTGATTAGAGA \\ 
Core 2 Side31 2 &  GTAATGGGGCCATTCGCCATTCAGGCTGCTTTTTTTAACTGTTGGATCGCAC \\ 
Core 2 Side31 3 &  TGGTTTAATTTCAACTAGACGATCGACAAGAACCGTTTTTTATATTCATTA \\ 
Core 2 Side31 4 &  TTGTGAATTTTTTTTCCTTATGTTTGTTTTTTTTAATCAGCTCATTTTT \\ 
Core 2 Side31 5 &  TAACCAATAGGAACGCTAGCCAGCTTTCATCAACATTTTTTTAAATGTG \\ 
Core 2 Side31 6 &  CAGGGTTTAGGTGTCCAACCGCAAGTTTTTTTTGCCAACGGCAGCGTA \\ 
Core 2 Side31 7 &  TGAGAGATAGACTTTCTTTTTTTCGTGGTGAAGGGATAGCTCTCGCAC \\ 
Core 2 Side31 8 &  AGAACTGGTCAACGTAACAAAGCTGCTCATTCTTTTTTTTGAATAAG \\ 
Core 2 Side31 9 &  AACTAAAGAGACCGGAAGCAAACTGGGTGAGATTGTACTTTTTTAAA \\ 
Core 2 Side31 10 &  ATTGCTGACAGTTGATGCGAGCTGAAATTTTTTGGTGGCATCAATT \\ 
Core 2 Side31 11 &  CCATCCCACGCAATTTTTTCAGCTTACGGCTGGTCCCAGTCACGAC \\ 
Core 2 Side31 12 &  GCGGCCTTTAGTGATGAAGGGTAAAGTTATTTTTTACGATGCTG \\ 
Core 2 Side31 13 &  TTGTAAAACGACGTCCAGCCAGCTTTTTTTTTCGGCACCGCTTC \\ 
Core 2 Side31 14 &  GTTTCATTCCATTTTTTTTAAATATAATGTACCTTTAATTGCT \\ 
Core 2 Side31 15 &  ATTGCCGTAACAGCGGATCAAACTTTTTTTAAATTTCTGCTC \\ 
Core 2 Side31 16 &  TGCCCCCTGTTTTTTATCTTAATCAGCTTGCCCTGACGAG \\ 
Core 2 Side31 17 &  AGCTGGCGAAAGGGGGTTTTTTTGTGCTGCAAGGCGATT \\ 
Core 2 Side31 18 &  GCTGTAGCTTTTTTTTACATGTTTTAAATATGC \\ 
Core 2 Side31 19 &  CGGATTGACCGTTTTTTAATGGGATAGGTC \\ 
Active Hubble1 1 &  CTCAGAGCCGCCATGACAGGAGAATGCGGGAAAGAGGTGTGGTGC \\ 
Active Hubble1 2 &  GGGATAGCCGGAATAGGTGTATCACCGCCCTCAGACATGAAAG \\ 
Active Hubble1 3 &  TATTCACCCAGCATCCAGAACCACCACCAGAACCGCCTCCCTC \\ 
Active Hubble1 4 &  CGATATATAAGTATATTCTGAAAGCCACCACCCTCATTTTCA \\ 
Active Hubble1 5 &  TCAGAACCGCCGCCACCCTCAGAGCCACCACC \\ 
Active Hubble1 6 &  AGAGCCGCCACCCAAATCCTCTATTAAGAGGCTGAGA \\ 
Active Hubble1 7 &  TAAGTGCCAGGTTTAGTACCGCCACCC \\ 
Active Hole2 1 &  TTGTCGTCAAAAAAAATTTATCAGGGGGAAAGGGGCGCTGGCAAGTGT \\ 
Active Hole2 2 &  GCGAACCTCCCAGACACCACGGAATAAGTTTATTTTGTCACAA \\ 
Active Hole2 3 &  TATAATCAGTGAGGCACGTTGAGAGCCAGCAGAATCAAGTT \\ 
Active Hole2 4 &  GGGAATTAAAATCTCCTTTCCAGAGTAACGATCTAAAGTT \\ 
Active Hole2 5 &  GCTGTCTTTCCTTATCAACCAATCAATAATCG \\ 
Active Hole2 6 &  CATGTAGAATTCCAAGAACGGGTAGAGGCGTTTTA \\ 
Active Hole2 7 &  TGCCTTTAGCGTCAGACTGTATTCGTAA \\ 
Active Hubble3 1 &  AAAAAGATTAAGATAATTCGACATCAATAAATAAAGCCTCAGAGC \\ 
Active Hubble3 2 &  TACGTTAATAAAATCATAACCATATTCATTCCAATACTGCGGAAT \\ 
Active Hubble3 3 &  TACCACATCAACACTACGAACTAAAGAAGCAAAGCGGATTGCATC \\ 
Active Hubble3 4 &  TGAATCCCTCGCGTTTGGAAGCCCGAAAGACTCAAAAATCAGGTC \\ 
Active Hubble3 5 &  GTAAGAGTCAACTAAGTTGAGATTTAGGAAGGGAAGAAAAATC \\ 
Active Hubble3 6 &  TCAAATACCTCAAATTATAGTCCGGAACAACATTATTACAGGT \\ 
Active Hubble3 7 &  AGAAAGATTCATCATGCAGATGATTGTATCAGCAAAT \\ 
Active Hubble3 8 &  TTTACCCTGACTATGCTTTAAACAGTTCAGAA \\ 
Active Hole3 1 &  TCAGCTAATGCACAGAGGTGTGCCACGCTGAGAGCAGCAGCAAATG \\ 
Active Hole3 2 &  AGGAAACCGAACTGGCATGATTAAGACTCCTTTCCTGAATCTT \\ 
Active Hole3 3 &  AACGAGAATGATAAATTAATACAAAGGGCAAAGAATCGTCTCG \\ 
Active Hole3 4 &  TAAAACTTCAGAAAAGCCCCAAAAACAGGAAACATAACGCCA \\ 
Active Hole3 5 &  ACCAACGCTAACGAGCGTCTTTTGTTTCAAAGTCAGAGGG \\ 
Active Hole3 6 &  AAAGGAATTACGAGGCATACAAAAGAACGTTAACG \\ 
Active Hole3 7 &  ATGTAATTTAGGCAGAGGAATAAACAACATGT \\ 
Passive Hubble1 1 &  TTTTTCTCAGAGCCGCCATGACAGGAGAATGCGGGAAAGAGGTGTGGTGC \\ 
Passive Hubble1 2 &  TTTTTGGGATAGCCGGAATAGGTGTATCACCGCCCTCAGACATGAAAG \\ 
Passive Hubble1 3 &  TATTCACCCAGCATCCAGAACCACCACCAGAACCGCCTCCCTCTTTTT \\ 
Passive Hubble1 4 &  CGATATATAAGTATATTCTGAAAGCCACCACCCTCATTTTCATTTTT \\ 
Passive Hubble1 5 &  TTTTTTCAGAACCGCCGCCACCCTCAGAGCCACCACCTTTTT \\ 
Passive Hubble1 6 &  TTTTTAGAGCCGCCACCCAAATCCTCTATTAAGAGGCTGAGA \\ 
Passive Hubble1 7 &  TAAGTGCCAGGTTTAGTACCGCCACCCTTTTT \\ 
Passive Hole2 1 &  TTTTTTTGTCGTCAAAAAAAATTTATCAGGGGGAAAGGGGCGCTGGCAAGTGT \\ 
Passive Hole2 2 &  TTTTTGCGAACCTCCCAGACACCACGGAATAAGTTTATTTTGTCACAA \\ 
Passive Hole2 3 &  TATAATCAGTGAGGCACGTTGAGAGCCAGCAGAATCAAGTTTTTTT \\ 
Passive Hole2 4 &  GGGAATTAAAATCTCCTTTCCAGAGTAACGATCTAAAGTTTTTTT \\ 
Passive Hole2 5 &  TTTTTGCTGTCTTTCCTTATCAACCAATCAATAATCGTTTTT \\ 
Passive Hole2 6 &  CATGTAGAATTCCAAGAACGGGTAGAGGCGTTTTATTTTT \\ 
Passive Hole2 7 &  TTTTTTGCCTTTAGCGTCAGACTGTATTCGTAA \\ 
Passive Hubble3 1 &  TTTTTAAAAAGATTAAGATAATTCGACATCAATAAATAAAGCCTCAGAGC \\ 
Passive Hubble3 2 &  TTTTTTACGTTAATAAAATCATAACCATATTCATTCCAATACTGCGGAAT \\ 
Passive Hubble3 3 &  TACCACATCAACACTACGAACTAAAGAAGCAAAGCGGATTGCATCTTTTT \\ 
Passive Hubble3 4 &  TGAATCCCTCGCGTTTGGAAGCCCGAAAGACTCAAAAATCAGGTCTTTTT \\ 
Passive Hubble3 5 &  GTAAGAGTCAACTAAGTTGAGATTTAGGAAGGGAAGAAAAATCTTTTT \\ 
Passive Hubble3 6 &  TCAAATACCTCAAATTATAGTCCGGAACAACATTATTACAGGTTTTTT \\ 
Passive Hubble3 7 &  TTTTTAGAAAGATTCATCATGCAGATGATTGTATCAGCAAAT \\ 
Passive Hubble3 8 &  TTTTTTTTACCCTGACTATGCTTTAAACAGTTCAGAATTTTT \\ 
Passive Hole3 1 &  TTTTTTCAGCTAATGCACAGAGGTGTGCCACGCTGAGAGCAGCAGCAAATG \\ 
Passive Hole3 2 &  AGGAAACCGAACTGGCATGATTAAGACTCCTTTCCTGAATCTTTTTTT \\ 
Passive Hole3 3 &  TTTTTAACGAGAATGATAAATTAATACAAAGGGCAAAGAATCGTCTCG \\ 
Passive Hole3 4 &  TAAAACTTCAGAAAAGCCCCAAAAACAGGAAACATAACGCCATTTTT \\ 
Passive Hole3 5 &  TTTTTACCAACGCTAACGAGCGTCTTTTGTTTCAAAGTCAGAGGG \\ 
Passive Hole3 6 &  TTTTTAAAGGAATTACGAGGCATACAAAAGAACGTTAACG \\ 
Passive Hole3 7 &  ATGTAATTTAGGCAGAGGAATAAACAACATGTTTTTT \\ 
Strong3 Hubble1 1 &  TATTCACCCAGCATCCAGAACCACCACCAGAACCGCCTCCCTCGTT \\ 
Strong3 Hubble1 2 &  CGATATATAAGTATATTCTGAAAGCCACCACCCTCATTTTCATTG \\ 
Strong3 Hubble1 3 &  AGAGCCGCCATGACAGGAGAATGCGGGAAAGAGGTGTGGTGC \\ 
Strong3 Hubble1 4 &  ATAGCCGGAATAGGTGTATCACCGCCCTCAGACATGAAAG \\ 
Strong3 Hubble1 5 &  GCCGCCACCCAAATCCTCTATTAAGAGGCTGAGA \\ 
Strong3 Hubble1 6 &  GAACCGCCGCCACCCTCAGAGCCACCACCTTA \\ 
Strong3 Hubble1 7 &  TAAGTGCCAGGTTTAGTACCGCCACCCGAG \\ 
Strong3 Hole2 1 &  TCGTCAAAAAAAATTTATCAGGGGGAAAGGGGCGCTGGCAAGTGT \\ 
Strong3 Hole2 2 &  TATAATCAGTGAGGCACGTTGAGAGCCAGCAGAATCAAGTTATT \\ 
Strong3 Hole2 3 &  GGGAATTAAAATCTCCTTTCCAGAGTAACGATCTAAAGTTAAA \\ 
Strong3 Hole2 4 &  AACCTCCCAGACACCACGGAATAAGTTTATTTTGTCACAA \\ 
Strong3 Hole2 5 &  CATGTAGAATTCCAAGAACGGGTAGAGGCGTTTTACAG \\ 
Strong3 Hole2 6 &  GTCTTTCCTTATCAACCAATCAATAATCGCGC \\ 
Strong3 Hole2 7 &  CTTTAGCGTCAGACTGTATTCGTAA \\ 
Strong3 Hubble3 1 &  TGAATCCCTCGCGTTTGGAAGCCCGAAAGACTCAAAAATCAGGTCAAC \\ 
Strong3 Hubble3 2 &  TACCACATCAACACTACGAACTAAAGAAGCAAAGCGGATTGCATCACC \\ 
Strong3 Hubble3 3 &  TCAAATACCTCAAATTATAGTCCGGAACAACATTATTACAGGTAAA \\ 
Strong3 Hubble3 4 &  GTAAGAGTCAACTAAGTTGAGATTTAGGAAGGGAAGAAAAATCTCA \\ 
Strong3 Hubble3 5 &  GTTAATAAAATCATAACCATATTCATTCCAATACTGCGGAAT \\ 
Strong3 Hubble3 6 &  AAGATTAAGATAATTCGACATCAATAAATAAAGCCTCAGAGC \\ 
Strong3 Hubble3 7 &  AAGATTCATCATGCAGATGATTGTATCAGCAAAT \\ 
Strong3 Hubble3 8 &  ACCCTGACTATGCTTTAAACAGTTCAGAATTT \\ 
Strong3 Hole3 1 &  AGGAAACCGAACTGGCATGATTAAGACTCCTTTCCTGAATCTTAAA \\ 
Strong3 Hole3 2 &  TAAAACTTCAGAAAAGCCCCAAAAACAGGAAACATAACGCCAAGA \\ 
Strong3 Hole3 3 &  GCTAATGCACAGAGGTGTGCCACGCTGAGAGCAGCAGCAAATG \\ 
Strong3 Hole3 4 &  GAGAATGATAAATTAATACAAAGGGCAAAGAATCGTCTCG \\ 
Strong3 Hole3 5 &  AACGCTAACGAGCGTCTTTTGTTTCAAAGTCAGAGGG \\ 
Strong3 Hole3 6 &  ATGTAATTTAGGCAGAGGAATAAACAACATGTTAC \\ 
Strong3 Hole3 7 &  GGAATTACGAGGCATACAAAAGAACGTTAACG \\ 
\\
\caption{\textbf{Staple sequences for A-triangle.}}
\label{aseq}
\end{longtable}

\begin{longtable}{c c}
Name &  Sequence \\ 
\hline 
\endfirsthead

Name &  Sequence \\ 
\hline 
\endhead

\endfoot

\endlastfoot
Core 1 Side1 1 &  GAGGGTTGATATAAGGATAAGTGATTGTGAAGAGTAGTATTCATCAACAGGCGCA \\ 
Core 1 Side1 2 &  ACAAATTCGGTGGTTTTCCCAGTCACGAATTGGAGACAAACGGATAACCTCTCG \\ 
Core 1 Side1 3 &  GTTTTAACCCGTACTCGGATTAGCGGGGTTTACGCAATAATAACGGATGCAACT \\ 
Core 1 Side1 4 &  TTAACCAATAGGAACGTGCATCTGCCAGTTTGAGGGGACAGCCTTATCCCGAC \\ 
Core 1 Side1 5 &  AAGGCTTATCCGGTTACAATTTTAAAAACAGGGAAGCGCATTAGATAAGAGAA \\ 
Core 1 Side1 6 &  TTTAAGAGCTGACCAATTGGGATTGTGTCGAAATCCACAAAGTACAACGGA \\ 
Core 1 Side1 7 &  AAGCTAAATCGGTACAAGAGATAGACTTTCTCGAAAAAGCGTCTGGAG \\ 
Core 1 Side1 8 &  GTGGAGCCACGGCCAGTGCCAAGCGGTGCCGGATTTCAACATACATAA \\ 
Core 1 Side1 9 &  GAAGTTTCGAAGGAAAACCAGAACTTAATGACCATAAATCGATTTAGG \\ 
Core 1 Side1 10 &  CCCAATAAATAACCCAAATTGAGCGCTAATATAACGATCTAAAGTTTT \\ 
Core 1 Side1 11 &  CGAGAAACCCGAGGAATGCTCAGTACCAGGCGTATAGCCCAATACCAC \\ 
Core 1 Side1 12 &  ATTCTGCGAACGAGTAGATTTAGTTTGACCATGTTTACCAATAGCTAT \\ 
Core 1 Side1 13 &  CCCTCATAGTTAGCGTATAGTAAGATTCAACTCATTATACTATGCGAT \\ 
Core 1 Side1 14 &  TCAACCGGAAACAATCGGCGAAACGTTGTACCAAAAACAAAGGGTGA \\ 
Core 1 Side1 15 &  TAAACACCGGAACTTATACCGACCGTGTGATAACCTGTAGAGCATA \\ 
Core 1 Side1 16 &  AAAGTACCCATTAAAAGACTTTTTCATGATAACAGTTCACAATCA \\ 
Core 1 Side1 17 &  GGAATAGGACAGACAGGTCGTCTTTCCAGACGTTAGTAA \\ 
Core 1 Side1 18 &  TAGGCTGTTTCAGAGCCTGATAACTTGAGATTTAGCAAG \\ 
Core 1 Side1 19 &  AAACCAGGTTCCGGCACCGCTTCTACTGGCTAATGCAG \\ 
Core 1 Side1 20 &  GAGTTAAGCGCGCCTCAGGAAGATCGCATTTATTAAGT \\ 
Core 1 Side1 21 &  CAACAGTTATCATATGCGTTATGTCCCGGAATTTGTG \\ 
Core 1 Side1 22 &  CGAACAAAGTTACCAAAGTAAGCAGATAGCAGCCCTT \\ 
Core 1 Side1 23 &  CATCTCCAGCCAGCTCAAAGCGCCATTCGCGTCCGTT \\ 
Core 1 Side1 24 &  ACCCAGCATTCTAATATAAAGACCAATCTTTAACAA \\ 
Core 1 Side1 25 &  ATTACTACGTGGTGAAGGGATAAAGCCTCTTAGCTA \\ 
Core 1 Side1 26 &  AGGTTGGGTTATATAATTACGAGCATGTAGAA \\ 
Core 1 Side1 27 &  TTGCGGGACCAACGCTTGGGTAACGCCAGTTA \\ 
Core 1 Side1 28 &  AATCAAGATTAACTGAACACCCTGAACAAAGT \\ 
Core 1 Side1 29 &  GATTTGTACTTACCGAAAACAATGAAATAGCA \\ 
Core 1 Side1 30 &  CAGAGGGTCAAGAATTGGTTTTGAGACGACAG \\ 
Core 1 Side1 31 &  TTTAATCCCGTCGATAAGACCAGTATAAAG \\ 
Core 1 Side1 32 &  TATCGCAAAAGGAATTACGAGGCCAGAGAG \\ 
Core 1 Side2 1 &  TAGCAATACCGAACGAGCACTCATCGTAGGAATTCATCAACGGATTGACCGTAATG \\ 
Core 1 Side2 2 &  CGTTTTTATTTTCATCGAGAACAACAGGAAGGTCAATCAAGAGGGTACTAGCTGA \\ 
Core 1 Side2 3 &  AGGGCGATCCCGTAAAAAAAGCCGCACAGGCAGAATCCTAGTTTTTTTTCACCAG \\ 
Core 1 Side2 4 &  TAGCTGTTGTCGTGCCAGCTGCATTAATGAAAGAAATAAAGAAATTCATTTGAA \\ 
Core 1 Side2 5 &  ACGCTGAGAAGAGTCTTTCTGCTCATTAAGCCTTAGAATCCTTGAAACATTCTG \\ 
Core 1 Side2 6 &  TGCTTCTGCGTAAAACTCGGCCAACGCGCGTCGGATTCTCCGTGGGAGTGAGTG \\ 
Core 1 Side2 7 &  CAACATGTTTTTGTTTCGCTAACGAGCGTCTTTCCAGAGGCGCCCATGCAGAAC \\ 
Core 1 Side2 8 &  TCATCTTTGATGCAAATCCAATTCCTGAACCATTCCAACAAATCAGATATAG \\ 
Core 1 Side2 9 &  GGCAAGGCGCGGATCACCGAAACTTTTTCAAATATATTTTTCAATTCT \\ 
Core 1 Side2 10 &  TTACCTTTTTTAATGGGGCAAATCTATTTGCAAACCCTTCAAAATCCC \\ 
Core 1 Side2 11 &  CAAATTAAAACTTAAAAATAGTGAACTAGCATATTCGGTGCGGGCCTC \\ 
Core 1 Side2 12 &  AAAGCGTAGCTATTAGAAAGGAATTGAGGAAGAACAATTTGCGTAGAT \\ 
Core 1 Side2 13 &  ACCTGTTTATCAACAATAGATAAGCGCAAGACAGGTCTGAAGAGACGC \\ 
Core 1 Side2 14 &  AAGGATAAGATAGCTTAGATTTGCCGCCTAAATCATACAGGGAGAAG \\ 
Core 1 Side2 15 &  CAAAATAAACAAACGGCATTAAATACGTTAATTAAGCAAATATTTAA \\ 
Core 1 Side2 16 &  AAATTAATTACATTTGTTATCTAATGGAAGGTGGCACAGAATCCTGT \\ 
Core 1 Side2 17 &  TTATAATCGAGACGGGAGGGTGGTTTTTCTTTATTGTAAGTGAGCG \\ 
Core 1 Side2 18 &  ATGCCGGTATGTACCTCATATATTTTAAACCTTTATTTCAACGC \\ 
Core 1 Side2 19 &  TGCAATGCCTGAGTAAATCGTAAAATTTATCACAAGTACC \\ 
Core 1 Side2 20 &  CAGTGCCACGCTGAGACTTGCTGAGGTCAGTTAAACAGTA \\ 
Core 1 Side2 21 &  AGTGTTGTTCCAGTTTTCACCAGTAGAAAAGCAGTAACAA \\ 
Core 1 Side2 22 &  TAAAAGGGTACGTGAAACGTGGACTCCAATTTAGTCCACT \\ 
Core 1 Side2 23 &  GTATTAAATTTAGGAGCACTAACAAAACATCAAGAAAACA \\ 
Core 1 Side2 24 &  ACTGCTCAATCGTCTGTTTGAATGAGAATACGGTTAGAAC \\ 
Core 1 Side2 25 &  ACCTGCAGCCAGCGGTCTTCACCGGGTTCCGACATGGTCA \\ 
Core 1 Side2 26 &  TAAAAAAATGGCCCACAACATAGCAAACGTCAAAGGGCGA \\ 
Core 1 Side2 27 &  TAAATTAGGAACAAGTTAGAACCCCCGGTTGGCCTTTTC \\ 
Core 1 Side2 28 &  TTCCTGTAGCCAGCTTCATTACCCCTAATTTTCCCAATC \\ 
Core 1 Side2 29 &  GTAATGGGTAAAGGTGTTGCCCTGCCCCCTGTCACACGA \\ 
Core 1 Side2 30 &  ACAGCCATATAAAACACCTCAATTCGCCATGCCTGCAA \\ 
Core 1 Side2 31 &  AGTCTGTCCATCACGCAAAGCAGTTGGGCGGTTGTGTA \\ 
Core 1 Side2 32 &  CAACAGCTGATTGCCGCCGAGATAGGGTTGATTAAAGA \\ 
Core 1 Side2 33 &  TAAATCGTCCCGGGGAGAGGCGGTTTGCAATAATCGGC \\ 
Core 1 Side2 34 &  AGAAACAAAAGAATACTAATAGTAGTAGCATTAACATC \\ 
Core 1 Side2 35 &  TATTGCGCAGGAACGGTACGCCGGCCTTTACGTCAGCG \\ 
Core 1 Side2 36 &  CTGAATAAAATATCTCCTTTGACAATATTAAATGGAT \\ 
Core 1 Side2 37 &  TTCGCTAATTAAACAAATCATAAAGAACGCGAGATTG \\ 
Core 1 Side2 38 &  ATAAATCAAAAGAATAGCCCGGTGCGGCTGCCATCAC \\ 
Core 1 Side2 39 &  GTTAAACGTTAAAGGGATTTTAGACGTAACCA \\ 
Core 1 Side2 40 &  GCGCCACCAGCAGAAGATTATTTAGCCAGTTA \\ 
Core 1 Side2 41 &  TGGTGCTGGTCTGATCCCAAATCAGAGAAGTG \\ 
Core 1 Side2 42 &  AAAACCGTCTGTCAGCAGCAACCGGGTCACT \\ 
Core 1 Side2 43 &  CGCTATTAAATAACCTCATAAATCAATATAT \\ 
Core 1 Side2 44 &  GCTTACATTGGCAGAGGGGTCGATATGAGCC \\ 
Core 1 Side2 45 &  CCCAAAAAGCAAGCATTAAAACAGAGATAG \\ 
Core 2 Side3 1 &  TAATTTTTCCACCCTCACCCTCATGTTTGCCAATCAAAAAGATTATCATCATAAAT \\ 
Core 2 Side3 2 &  CGTCACCCAACCTAAAACGAAAGAGGCAATGCGGCTTAGATTAAAGGTCACGTAAT \\ 
Core 2 Side3 3 &  CTTGACGGGGGAAGAATGTGGTGCAGCAGCGGTCCACGCTTTCCACACAGATGATG \\ 
Core 2 Side3 4 &  CGGAATCGCGGTCATAGTGTAAAGTTTATCAGAAAAGGAGAAAATCTCCAAAAAAA \\ 
Core 2 Side3 5 &  CTCAAATGAGCAAAGCACTAACGGACCCTCAGGCCCAATAAGGAATTGCGAATAA \\ 
Core 2 Side3 6 &  GAACCGCCAACAACATAATAGCGAGAGGCTTTTGCAAAAATAGCAAAGCCACC \\ 
Core 2 Side3 7 &  GAGCCTCCCGTGCCTGTTCTTCGCAGCAAAGAATGCGGGGGTCAGGCGAAC \\ 
Core 2 Side3 8 &  CCATCCCAGCTTTCGCACTCAATCGTGCACTCAGCGAAAGCGGAACAA \\ 
Core 2 Side3 9 &  TTTGAATACCAAGTTAGGATTCGCCTGATTGCGCTAACTCAGCCGGAA \\ 
Core 2 Side3 10 &  CAGCTTGATGTCAGGACAGGTCCAAACCAGACCGTTTTAATTCGAGCT \\ 
Core 2 Side3 11 &  ACAATAACCAAAATCGCGCAGAGGCGAATTATTCATTTCACAGTACCT \\ 
Core 2 Side3 12 &  AACTCCAATTAGAGAGTACCTTTAATAAGAATACACTAAGTCAATCA \\ 
Core 2 Side3 13 &  AGTTTCGTCAGCGGAGTGAGAATAGAAAGGATCTGAATTCACCCTCA \\ 
Core 2 Side3 14 &  TTATCATCATATTCCCCAGAAGGTTTCTGCCTGCGGCCATCGTTAAC \\ 
Core 2 Side3 15 &  GTTAAAGGCCGCTTTTGCGGGATATTATTCAGCTTAATTTCAGACTG \\ 
Core 2 Side3 16 &  AGAAACCATGATTATCAACATACGACATTAATTGCGTTGCGCTCACT \\ 
Core 2 Side3 17 &  TAATGCTAGGCACCTCAGCAGCGAAAGACAGCATCGGATTGAGGG \\ 
Core 2 Side3 18 &  TTTAGAGGAAGCCCGAAAGAATCCCGAACTTGCGGATTCCTTTT \\ 
Core 2 Side3 19 &  ATAGCGTCCAATACTGCGTCCGTGCATAAACCGGGGGAGCGGAA \\ 
Core 2 Side3 20 &  GCTGGAGGTACCGATAAGGTGAATTTCTTAAACGGTCATA \\ 
Core 2 Side3 21 &  TAGACTGGATTCATTGAATCCCCCAACTTCAAATATCG \\ 
Core 2 Side3 22 &  GTAATAGTAAAACGAGGATTGCTCTTTTCTTTCATC \\ 
Core 2 Side3 23 &  TAAGGGAAAACGTAACTAGCGCGTATAATCAA \\ 
Core 2 Side3 24 &  CAGTAGCGCGGAACCGAATCACCGGAACCAGA \\ 
Core 2 Side3 25 &  CTTGCCCCTTATTAGCTGTATCGGCCTAGGGG \\ 
Core 2 Side3 26 &  CATCGCCCACGCATAACGACAATGACAACAAC \\ 
Core 2 Side3 27 &  GCGGGCCGGCCTTTCGGTGGCGAGAAAGGAA \\ 
Core 2 Side3 28 &  GGAAAGCCTTGCAGGCCGCAACCAGCTTACG \\ 
Core 2 Side3 29 &  TTACACTGGTGTGTTTCAAAGCGGCATCAGC \\ 
Core 2 Side3 30 &  GCCAGGGGGTGCCTAATGAGTGAAGGCTCC \\ 
Core 2 Side3 31 &  ACAGAATCAAGTAAGTAAATATTGACGGAA \\ 
Core 2 Side3 32 &  AAGAGGCATTTTGCCTTTAGCGGCTGAATA \\ 
Core 2 Side3 33 &  GCCCCCGCCAGAGCCGTCTCGTCGCTGGCA \\ 
Core2 Side12 1 &  CAAATAAGAAACGATTTCAGCTAAATAGCAAGGAACGGGTTTACGCCTTTTTAGCTGGC \\ 
Core2 Side12 2 &  AGTAATTCTGTTTTTTCCAGACGACGAATAATATCCTTTTTCATCCTAATCTATATG \\ 
Core2 Side12 3 &  GAACGCGATCGCCATAGAAAGGGGCAAACAAGAGAATTTTTTCGATGAACGGTA \\ 
Core2 Side12 4 &  GTAAAACGGCCACGGGGTCAAATCACCATCAATATGTTTTTATATTCAACCGTT \\ 
Core2 Side12 5 &  GCTATTTTTGAGAGATTTTTTCTACAAAGGCTATCAGCATTCAGGGCAAGGCG \\ 
Core2 Side12 6 &  TGTGTAGGTTTTTTAAAGATTCAATTATGACCTTTTTCTGTAATACTTTTGC \\ 
Core2 Side12 7 &  TATTTTCATTTGGGGCGCTTTTTGAGCTGAAAAGGTGGCAAGTTAATT \\ 
Core2 Side12 8 &  TATTTTGCCGTAACCGCCATCAAAAATAATTCGTTTTTCGTCTGGCC \\ 
Core2 Side12 9 &  TACCGACAAAAGGTAAGAATAACATATCCTGAATTTTTTCTTACCAA \\ 
Core2 Side12 10 &  ATTTTGTTAAAATTCGCATTTTTTTAAATTTTTGTTAAATCAGCT \\ 
Core2 Side12 11 &  AATAATCGGCTGTCTTTCCTTTTTTTATAAGAAAACAATAAA \\ 
Core2 Side12 12 &  TAGCAAAATTAAGTTTTTCAATAGCTCTCATTTTTCGGAAAA \\ 
Core2 Side12 13 &  GATGTGCTCTGCGCAACTTTTTTGTTGGGAAGGGCGATGTA \\ 
Core2 Side12 14 &  AACGTCAAAAATGAAAATAGCTTTTTAGCCTTTACAGAGA \\ 
Core2 Side12 15 &  TAAATGCCTGATTTTTCCTAAATTTAATGGTTTGAA \\ 
Core2 Side12 16 &  GGATAGGTCACGTTGGTGTTTTTTAGATGGGCGCAT \\ 
Core2 Side12 17 &  GAGACTACCTTTTTTTTTTAACCTCCGGTCATA \\ 
Core 2 Side23 1 &  ATTACCTGAGCAAAAGAATTTTTTTTGATGAAACACTAATAGATTTTTTTAGAGCCGTCA \\ 
Core 2 Side23 2 &  ATGCTGATTTTTTTTCCGTTCCGGCAAACGCGGTCAGAATCAGAGCGGGACTACAGG \\ 
Core 2 Side23 3 &  TCACAGTTGAGGATCCCCTTTTTTTGTACCGAGCTCGAATTCGTGTATTGGGCGCC \\ 
Core 2 Side23 4 &  GGGAGAAGCAATTCATCAATATAACATTATCCCAGCTTTTTTGGCGAACCCGAAC \\ 
Core 2 Side23 5 &  GCAAGAATGCCAACTTTTTTGCAGCACCGTCGGTGGTGGCCTCCGGATTTAGAG \\ 
Core 2 Side23 6 &  ATAACGGACGTGCCGGACTTAAAGCACTAATTTTTTTCGGAACCCTAAAGGGA \\ 
Core 2 Side23 7 &  CCACACCCGCCGCGCTTTTTTTTATGCGCCGGCTAAACTTTTTTTGAGGCCGA \\ 
Core 2 Side23 8 &  GCGCGTACTATGGTTGCGCTAGGGCGCTGGCAAGTTTTTTGTAGCGGTCAC \\ 
Core 2 Side23 9 &  TTTACATCGCCCGCTTTTTTTTTCAGTCGGGAAACCTTCCTGTGTGAAATT \\ 
Core 2 Side23 10 &  CATCAGACTTGATGGTCCTGGCCCTTTTTTTAGAGAGTTGCAGCAACG \\ 
Core 2 Side23 11 &  CGCCGGGCGCGTTGCGGGGTGCCGTGTAGAAGTGATGAAGGGTAAA \\ 
Core 2 Side23 12 &  TTCTTTGCTTTTTTTGTCATAAACATCCCGGCATCAGATGCCGGG \\ 
Core 2 Side23 13 &  TTTCAGGTTTAACGTCAGATGAATATTTTTTTCAGTAA \\ 
Core 2 Side23 14 &  TTAGACTTTTTTTTTTAAACAATTCGACAACTC \\ 
Core 2 Side23 15 &  GATCCAGCGCAGTGTTTTTTCACTGCGCGCCT \\ 
Core 2 Side31 1 &  TCTGTATGGGATTTTGTTTTTTTAAACAACTTTCATGGCTTTTGTTTTTTTGATACAGGA \\ 
Core 2 Side31 2 &  TGAGACTCCTCATTTTTTGAGAATGCCCGTGTACTGGTAATAAATGAATTTAACTACAA \\ 
Core 2 Side31 3 &  GCGACCTGGTACAGACGAGTAATCTTGACAAGTTTTTTTCCGGATATTCATTACC \\ 
Core 2 Side31 4 &  CGGAAGCAGACCAACTTTGAAAGAGGACATTTTTTTTGAACGGTCTCCATGT \\ 
Core 2 Side31 5 &  GGTTTACCAGCGCCAATTTTAAATAATACCCAAAATTTTTTAACTGGCATG \\ 
Core 2 Side31 6 &  GAGGCTTTGAGGACTAACGGGTAAAATACGTAATGTTTTTTCACTACGA \\ 
Core 2 Side31 7 &  AGGGCGACTTTTTTTTCAACCGAACGATTTTTTTGTAGCAACGGCTACA \\ 
Core 2 Side31 8 &  AAAGCTGCTAGTCAGACTTTAAACATTTTTTTTCAGAAAACGAGACCT \\ 
Core 2 Side31 9 &  CAGTCAGGACGTTGGGTTTTTTTGAAAAATCTACGTTAATAAAATGTT \\ 
Core 2 Side31 10 &  AAACATGACCAGTAAGCGTCATACAACAGTTTCACCAGTTTTTTACA \\ 
Core 2 Side31 11 &  AGGGAAGGACTCCTTATTACGCAGTATGTTAGTTTTTTTAACGTAGA \\ 
Core 2 Side31 12 &  AAACAGTTAAGGATTAAGGAGGTTTAGTTTTTTACCGCCACCCTCA \\ 
Core 2 Side31 13 &  AACACTCATCTTTTTTTTGACCCCCAGCGATTATACCAGGAAGTTT \\ 
Core 2 Side31 14 &  AAAAATCAGGTCTTTTTTTTACCCTGACTATTATCATTCAGTGAAT \\ 
Core 2 Side31 15 &  AGGCTTGCCCTGATACTTAGCCGGAATTTTTTGAGGCGCAGACG \\ 
Core 2 Side31 16 &  AGCAACACTATCATAACCCTCGTTTACCATTTTTTACGACGATA \\ 
Core 2 Side31 17 &  AAAACCAATATTACAGGTAGAAATTTTTTATTCATCAGTTGA \\ 
Core 2 Side31 18 &  GTAGCTCAATTTTTTATGAGACAAAAAATACATACATAAA \\ 
Core 2 Side31 19 &  CCTGCCTATTTTTTTTCGGAACCTATTATTCTG \\ 
Active Hole1 1 &  GAAAGGCCGCCTGAGACTGTTTAGAGGGCTTATGTAATTTAGGCA \\ 
Active Hole1 2 &  TTAAATAAGAACGCCAACAATTGAGAAGGCGTTTTAGCGAACC \\ 
Active Hole1 3 &  GTTTATTTTGTGATTCCCATTTAAGAAATTCCATAAGCGCGAA \\ 
Active Hole1 4 &  GGTGTCTGATAGAAAAAAGAAACGCAAAGACACCACGGAATAA \\ 
Active Hole1 5 &  TTGAGTAACAGTGCCCGTATGGTGGCAACATATAATTCATAT \\ 
Active Hole1 6 &  GCGCCATTAGATACATTTCGCAAATGGTCAATAAATAAGGCG \\ 
Active Hole1 7 &  GAGGCATTTTCGAGCCAGTAACGGGAGAATTAGTTGC \\ 
Active Hole1 8 &  CGCCTGTAGCATTCCTGTATCAGGGGTCAGTGCC \\ 
Active Hubble2 1 &  TGACCTGCCAGTAAGAACTCAATTTTTATACGCTCATGGAAATACCTAC \\ 
Active Hubble2 2 &  CTAAAACACAATATCTACCTCAAATATCAAACGAGGTGAGGCGGT \\ 
Active Hubble2 3 &  ATTTTGACATCGGCCTTGCTGGTAATAAGGAAAAAATCAGTGA \\ 
Active Hubble2 4 &  AAATCTAAAGCATCACGCCAGCAGGCGCGAACGAGTAGAA \\ 
Active Hubble2 5 &  CGCCAGCCATTGCAACTCCAGAACACTTGCCTTGATAGCC \\ 
Active Hubble2 6 &  CATCGACAGGCCACCGAGTAAAAGATAACATCAATATTAC \\ 
Active Hubble2 7 &  CTACCATATCAAAATAACAGTTGTCTTTAATCAAATGAA \\ 
Active Hubble2 8 &  CAGTATTAACACCTAAAAATACTTCTTTGATTAGTA \\ 
Active Hubble3 1 &  GGAACCCATGTACCGTAACACTGGAACCGCTACCGTTAAGTATTAGGTCAGA \\ 
Active Hubble3 2 &  GCAGCACCCAGCCAGCATTGACAGGAGTCACAAACGCGCAGTCACAACTAA \\ 
Active Hubble3 3 &  CCTCCCTCAATGGAAAAAATAAATCCTCATTAAGAGCCGCCACCA \\ 
Active Hubble3 4 &  AAGCCAGAGAGCCGAGCCGCCGTAGCACCATTACCATTAGCAA \\ 
Active Hubble3 5 &  GAACCACCACCAGCCACCCTCAGAACCGCCAC \\ 
Active Hubble3 6 &  GGCCGGAAACGTCCACCGTCATGCAGGGAGATAAGAG \\ 
Active Hubble3 7 &  GCAAAATGAATTATACCAATGAAACCATCGATA \\ 
Active Hubble3 8 &  CGATTGGCCTTGATATGTTGAGGCGCCACCAC \\ 
Active Hole3 1 &  TTATCCGCTCACAAGGTTTGCCATTTTGGAGCGGGCTTTGACGAGCAC \\ 
Active Hole3 2 &  GAGGATTTGTTTGAGTAATCCTGATTGTTTTTTTTGGATTATACTT \\ 
Active Hole3 3 &  GTTATTAATTTTTTTTTAAAAAGAAGTAATAGATAATACATTT \\ 
Active Hole3 4 &  CCTCAGAGCCATCACGTTGCCTTTAATTTTCAGGGGAAGTTTT \\ 
Active Hole3 5 &  GTTGCGCCCGATATATTCGGTCGCTGAGGCTCCGACTTGAGC \\ 
Active Hole3 6 &  GTATAACGTGCTTTCCTCGTTCGTTTTTTACATCCTC \\ 
Active Hole3 7 &  CATTTGGGAATTAGAGCCAGGCATTTTGTCATTTT \\ 
Passive Hole1 1 &  GAAAGGCCGCCTGAGACTGTTTAGAGGGCTTATGTAATTTAGGCATTTTT \\ 
Passive Hole1 2 &  TTTTTTTAAATAAGAACGCCAACAATTGAGAAGGCGTTTTAGCGAACC \\ 
Passive Hole1 3 &  TTTTTGTTTATTTTGTGATTCCCATTTAAGAAATTCCATAAGCGCGAA \\ 
Passive Hole1 4 &  GGTGTCTGATAGAAAAAAGAAACGCAAAGACACCACGGAATAATTTTT \\ 
Passive Hole1 5 &  TTTTTTTGAGTAACAGTGCCCGTATGGTGGCAACATATAATTCATAT \\ 
Passive Hole1 6 &  GCGCCATTAGATACATTTCGCAAATGGTCAATAAATAAGGCGTTTTT \\ 
Passive Hole1 7 &  TTTTTGAGGCATTTTCGAGCCAGTAACGGGAGAATTAGTTGC \\ 
Passive Hole1 8 &  CGCCTGTAGCATTCCTGTATCAGGGGTCAGTGCCTTTTT \\ 
Passive Hubble2 1 &  TGACCTGCCAGTAAGAACTCAATTTTTATACGCTCATGGAAATACCTACTTTTT \\ 
Passive Hubble2 2 &  CTAAAACACAATATCTACCTCAAATATCAAACGAGGTGAGGCGGTTTTTT \\ 
Passive Hubble2 3 &  TTTTTATTTTGACATCGGCCTTGCTGGTAATAAGGAAAAAATCAGTGA \\ 
Passive Hubble2 4 &  TTTTTAAATCTAAAGCATCACGCCAGCAGGCGCGAACGAGTAGAA \\ 
Passive Hubble2 5 &  TTTTTCGCCAGCCATTGCAACTCCAGAACACTTGCCTTGATAGCC \\ 
Passive Hubble2 6 &  CATCGACAGGCCACCGAGTAAAAGATAACATCAATATTACTTTTT \\ 
Passive Hubble2 7 &  CTACCATATCAAAATAACAGTTGTCTTTAATCAAATGAATTTTT \\ 
Passive Hubble2 8 &  TTTTTCAGTATTAACACCTAAAAATACTTCTTTGATTAGTA \\ 
Passive Hubble3 1 &  GGAACCCATGTACCGTAACACTGGAACCGCTACCGTTAAGTATTAGGTCAGATTTTT \\ 
Passive Hubble3 2 &  TTTTTGCAGCACCCAGCCAGCATTGACAGGAGTCACAAACGCGCAGTCACAACTAA \\ 
Passive Hubble3 3 &  CCTCCCTCAATGGAAAAAATAAATCCTCATTAAGAGCCGCCACCATTTTT \\ 
Passive Hubble3 4 &  AAGCCAGAGAGCCGAGCCGCCGTAGCACCATTACCATTAGCAATTTTT \\ 
Passive Hubble3 5 &  TTTTTGAACCACCACCAGCCACCCTCAGAACCGCCACTTTTT \\ 
Passive Hubble3 6 &  TTTTTGGCCGGAAACGTCCACCGTCATGCAGGGAGATAAGAG \\ 
Passive Hubble3 7 &  GCAAAATGAATTATACCAATGAAACCATCGATATTTTT \\ 
Passive Hubble3 8 &  TTTTTCGATTGGCCTTGATATGTTGAGGCGCCACCAC \\ 
Passive Hole3 1 &  TTATCCGCTCACAAGGTTTGCCATTTTGGAGCGGGCTTTGACGAGCACTTTTT \\ 
Passive Hole3 2 &  TTTTTGAGGATTTGTTTGAGTAATCCTGATTGTTTTTTTTGGATTATACTT \\ 
Passive Hole3 3 &  GTTATTAATTTTTTTTTAAAAAGAAGTAATAGATAATACATTTTTTTT \\ 
Passive Hole3 4 &  TTTTTCCTCAGAGCCATCACGTTGCCTTTAATTTTCAGGGGAAGTTTT \\ 
Passive Hole3 5 &  GTTGCGCCCGATATATTCGGTCGCTGAGGCTCCGACTTGAGCTTTTT \\ 
Passive Hole3 6 &  TTTTTGTATAACGTGCTTTCCTCGTTCGTTTTTTACATCCTC \\ 
Passive Hole3 7 &  TTTTTCATTTGGGAATTAGAGCCAGGCATTTTGTCATTTT \\ 
Strong3 Hole1 1 &  GAAAGGCCGCCTGAGACTGTTTAGAGGGCTTATGTAATTTAGGCATCA \\ 
Strong3 Hole1 2 &  GGTGTCTGATAGAAAAAAGAAACGCAAAGACACCACGGAATAAAGA \\ 
Strong3 Hole1 3 &  GCGCCATTAGATACATTTCGCAAATGGTCAATAAATAAGGCGCTC \\ 
Strong3 Hole1 4 &  AATAAGAACGCCAACAATTGAGAAGGCGTTTTAGCGAACC \\ 
Strong3 Hole1 5 &  TATTTTGTGATTCCCATTTAAGAAATTCCATAAGCGCGAA \\ 
Strong3 Hole1 6 &  AGTAACAGTGCCCGTATGGTGGCAACATATAATTCATAT \\ 
Strong3 Hole1 7 &  CGCCTGTAGCATTCCTGTATCAGGGGTCAGTGCCGGG \\ 
Strong3 Hole1 8 &  GCATTTTCGAGCCAGTAACGGGAGAATTAGTTGC \\ 
Strong3 Hubble2 1 &  TGACCTGCCAGTAAGAACTCAATTTTTATACGCTCATGGAAATACCTACTGC \\ 
Strong3 Hubble2 2 &  CTAAAACACAATATCTACCTCAAATATCAAACGAGGTGAGGCGGTGCG \\ 
Strong3 Hubble2 3 &  CATCGACAGGCCACCGAGTAAAAGATAACATCAATATTACGCT \\ 
Strong3 Hubble2 4 &  CTACCATATCAAAATAACAGTTGTCTTTAATCAAATGAATTG \\ 
Strong3 Hubble2 5 &  TTGACATCGGCCTTGCTGGTAATAAGGAAAAAATCAGTGA \\ 
Strong3 Hubble2 6 &  TCTAAAGCATCACGCCAGCAGGCGCGAACGAGTAGAA \\ 
Strong3 Hubble2 7 &  CAGCCATTGCAACTCCAGAACACTTGCCTTGATAGCC \\ 
Strong3 Hubble2 8 &  TATTAACACCTAAAAATACTTCTTTGATTAGTA \\ 
Strong3 Hubble3 1 &  GGAACCCATGTACCGTAACACTGGAACCGCTACCGTTAAGTATTAGGTCAGAGAG \\ 
Strong3 Hubble3 2 &  GCACCCAGCCAGCATTGACAGGAGTCACAAACGCGCAGTCACAACTAA \\ 
Strong3 Hubble3 3 &  CCTCCCTCAATGGAAAAAATAAATCCTCATTAAGAGCCGCCACCACCT \\ 
Strong3 Hubble3 4 &  AAGCCAGAGAGCCGAGCCGCCGTAGCACCATTACCATTAGCAACAT \\ 
Strong3 Hubble3 5 &  GCAAAATGAATTATACCAATGAAACCATCGATAGTA \\ 
Strong3 Hubble3 6 &  CGGAAACGTCCACCGTCATGCAGGGAGATAAGAG \\ 
Strong3 Hubble3 7 &  CCACCACCAGCCACCCTCAGAACCGCCACGAA \\ 
Strong3 Hubble3 8 &  TTGGCCTTGATATGTTGAGGCGCCACCAC \\ 
Strong3 Hole3 1 &  TTATCCGCTCACAAGGTTTGCCATTTTGGAGCGGGCTTTGACGAGCACGCA \\ 
Strong3 Hole3 2 &  GTTATTAATTTTTTTTTAAAAAGAAGTAATAGATAATACATTTCGA \\ 
Strong3 Hole3 3 &  GTTGCGCCCGATATATTCGGTCGCTGAGGCTCCGACTTGAGCGGC \\ 
Strong3 Hole3 4 &  GATTTGTTTGAGTAATCCTGATTGTTTTTTTTGGATTATACTT \\ 
Strong3 Hole3 5 &  CAGAGCCATCACGTTGCCTTTAATTTTCAGGGGAAGTTTT \\ 
Strong3 Hole3 6 &  TAACGTGCTTTCCTCGTTCGTTTTTTACATCCTC \\ 
Strong3 Hole3 7 &  TTGGGAATTAGAGCCAGGCATTTTGTCATTTT \\ 
\\
\caption{\textbf{Staple sequences for B-triangle.}}
\label{bseq}
\end{longtable}

\normalsize